\documentclass[11pt]{article}
\usepackage{latexsym}
\usepackage{amssymb}
\usepackage{graphicx}
\usepackage{epsfig}
\usepackage{rotating}
\usepackage{adjustbox}
\usepackage{amsfonts}
\usepackage{amsmath}
\usepackage{hyperref}
\hypersetup{
	colorlinks=true,
	linkcolor=blue,
	filecolor=magenta,
	urlcolor=cyan,
}
\begin{document}

\begin{titlepage}
\begin{flushright}
IRMP-CP3-24-01\\
\end{flushright}

\vspace{5pt}

\begin{center}

{\Large\bf Noncommutativity in Configuration Space}\\

\vspace{7pt}

{\Large\bf Induced by}

\vspace{7pt}

{\Large\bf A Conjugate Magnetic Field in Phase Space}

\vspace{60pt}

Jan Govaerts$^{a,b,c}$

\vspace{30pt}

$^{a}${\sl Centre for Cosmology, Particle Physics and Phenomenology (CP3),\\
Institut de Recherche en Math\'ematique et Physique (IRMP),\\
Universit\'e catholique de Louvain (UCLouvain),\\
2, Chemin du Cyclotron, B-1348 Louvain-la-Neuve, Belgium}\\
E-mail: {\em Jan.Govaerts@uclouvain.be}\\
ORCID: {\tt http://orcid.org/0000-0002-8430-5180}\\
\vspace{15pt}
$^{b}${\sl International Chair in Mathematical Physics and Applications (ICMPA--UNESCO Chair)\\
University of Abomey-Calavi, 072 B.P. 50, Cotonou, Republic of Benin}\\
\vspace{15pt}
$^{c}${\sl Fellow of the Stellenbosch Institute for Advanced Study (STIAS),\\
Stellenbosch, Republic of South Africa}\\

\vspace{10pt}

%\today

\vspace{20pt}

\begin{abstract}
\noindent
As is well known, an external magnetic field in configuration space coupled to a quantum dynamics induces
noncommutativity in its velocity momentum space. By phase space duality, an external vector potential in the conjugate momentum
sector of the system induces noncommutativity in its configuration space. Such a rationale for noncommutativity is explored herein
for an arbitrary configuration space of Euclidean geometry. Ordinary quantum mechanics with a commutative configuration space is revisited first.
Through the introduction of an arbitrary positive definite $*$-product, a one-to-one correspondence between the Hilbert space of abstract quantum states
and that of the enveloping algebra of the position quantum operators is identified. A parallel discussion is then presented when configuration space is
noncommutative, and thoroughly analysed when the conjugate magnetic field is momentum independent and nondegenerate.
Once again the space of quantum states may be
identified with the enveloping algebra of the noncommutative position quantum operators. Furthermore when the positive definite $*$-product is chosen
in accordance with the value of the conjugate magnetic field which determines the commutator algebra of the coordinate operators,
these operators span a Fock algebra of which the canonical coherent states are the localised noncommutative quantum analogues of the sharp and structureless
local points of the associated commutative configuration space geometry. These results generalise and justify {\sl a posteriori} within the context of ordinary
canonical quantisation the heuristic approach to quantum mechanics in the noncommutative Euclidean plane as constructed and developed by F.~G.~Scholtz
and his collaborators.

\end{abstract}

\end{center}

\end{titlepage}

\setcounter{footnote}{0}

\section{Introduction}
\label{Intro}

It is a widely held conviction that even though conceived as a differentiable continuum up to the heretofore experimentally accessible distance and energy scales,
in actual fact as may be revealed by quantum fluctuations the space-time fabric consists of some coarse-grained structure to be physically manifested at the smallest
distance and wavelength, and largest energy scales possible. Yet the nature and characteristics of this coarse-grained structure remains at best speculative,
presumably to be found among a large ensemble of possible complementary ideas and concepts.

Among these and by analogy with the coarse-grained properties of the quantised phase space of a dynamical system,
the possibility of a noncommutative configuration space
with its geometry has been explored starting already seven decades ago along lines largely inspired by the structure of quantum mechanics \cite{Snyder}.
The last twenty-five years or so have seen
a very strong resurgence of interest in such a possibility, from a diversity of vantage points and complementary approaches all sharing
as primary driving motivation nowadays the quest for a complete quantum unification of all fundamental interactions inclusive of relativistic
gravity \cite{Connes,Ahlu,DFR,SW}, beginning with quantum field theories and even more simply quantum mechanical systems constructed
over noncommutative spaces and geometries \cite{Szabo,Girotti}. Some recent introductory reviews may also be found in \cite{LGouba,Pittaway,Liang}.

Among a large body of rather diversified and most interesting works motivated by different approaches and points of view, the cases of two and three dimensional
spaces with flat or curved geometries such as the plane or the sphere have attracted a lot of efforts and developments. Standing out among the latter is the
work and results over the years by F. G. Scholtz and his collaborators
(see for instance \cite{Scholtz1,Scholtz2,Scholtz3,Scholtz4,Scholtz5,Scholtz6,Scholtz7,Scholtz8,Scholtz9,Scholtz10,Scholtz11}).

Based on a heuristic approach conceived within the context of the noncommutative plane and then extended
to three dimensions, a quantum mechanics on such noncommutative configuration spaces has been constructed. This construction \cite{Pittaway,Scholtz3}
involves a so-called classical Hilbert space which provides a representation space for the noncommuting coordinates of configuration space,
to replace and play the role of the ordinary classical coordinates of a commutative configuration space. Then by analogy with the ordinary configuration
space wave functions of quantum states that are functions of the commuting configuration space coordinates in the commutative case, the quantum Hilbert space
of a quantum mechanics and its quantum states defined over the noncommutative configuration space corresponds to the enveloping algebra
of the noncommutative configuration space coordinate operators, namely composite operators of these coordinates that act on the classical Hilbert space.

Even though the construction does not rely on Dirac's correspondence principle with its classical canonical phase space formulation of a dynamics
based on a minimal action principle which is then quantised to identify a corresponding quantum dynamical system, Scholtz's construction of
a quantum mechanics over a noncommutative space and its geometry allows as well for the identification of momenta canonically conjugate
to the noncommuting configuration space coordinates, whose action on the quantum Hilbert space and its states is realised in terms
of the coordinate operators themselves and their commutators.
Furthermore, a path integral representation of such so-called noncommutative quantum mechanics
also becomes feasible, in terms of an action principle over configuration space which is then not local in time on account of space noncommutativity \cite{Scholtz4}.
Yet in the limit of vanishing noncommutativity parameters in configuration space, the usual results of ordinary or so-called commutative quantum mechanics 
are indeed recovered.

Clearly the success of Scholtz's heuristic construction of a noncommutative quantum mechanics and of its specific features calls for an independent understanding
of it, and how it may possibly find its justification as well from within the more traditional and standard approach of Dirac's canonical quantisation
of a classical phase space realisation of a classical dynamics over a classical and commutative configuration space. Addressing and clarifying these questions
is the main purpose of the present work.

As is well known at least in the case of the celebrated Landau problem, coupling a dynamics to a magnetic field in configuration space induces noncommutativity
in the gauge invariant velocity space \cite{Govaerts1}. Likewise but less well known than it would deserve, by phase space duality,
coupling a dynamics to a magnetic field in conjugate momentum space induces noncommutativity in configuration space. This observation serves
as a classical starting point for the exploration of answers to the above
questions, leading {\sl in fine} to the {\sl a posteriori} justification through Dirac's canonical quantisation of Scholtz's construction of a quantum mechanics
over noncommutative configuration spaces, at least when their geometry is Euclidean and even dimensional which is the situation explicitly analysed herein.

The discussion is structured as follows. Section~\ref{Sect2} revisits ordinary quantum mechanics over commutative configuration spaces. A series of
considerations which are usually not to be found in standard quantum mechanics textbooks are addressed and developed, to anticipate on a parallel
ensemble of considerations which prove to be required in the noncommutative case. In particular, a specific $*$-product is introduced already for a commutative
quantum mechanics, based on a specific class of localised and normalisable quantum states that are in one-to-one correspondence with the points
in configuration space, yet that are not sharp position eigenstates. With section~\ref{Sect3} the discussion is then extended to include a generic
conjugate magnetic field in the canonical conjugate momentum sector of phase space, which readily implies a noncommutativity of the configuration space
coordinate operators for the canonically quantised dynamical system. After a series of general considerations, in section~\ref{Sect4} the analysis is restricted
to such a conjugate magnetic field which is simply constant over phase space, thereby leading to noncommuting coordinate operators whose commutation
relations are simply constants. This analysis is then mostly parallel to that of section~\ref{Sect2}, to finally identify, in the case of an even-dimensional
configuration space, how Dirac's canonical quantisation indeed provides an {\sl a posteriori} justification for Scholtz {\sl et al.}'s heuristic quantum mechanics
on noncommutative configuration spaces. Some concluding considerations are outlined in section~\ref{Sect5}.

\section{A ``Star'' (Product) for Ordinary Quantum Mechanics}
\label{Sect2}

\subsection{A classical formulation}

Even though the considerations presented in this work may be extended to more general geometries, for the purposes of the present discussion
let us restrict to any physical system whose configuration space is specifically an Euclidean space of finite dimension $N\ge 1$, of metric $\delta_{ij}$,
and parametrised by real cartesian coordinates $x_i\in\mathbb{R}$ ($i=1,2,\cdots,N $). Within its canonical Hamiltonian formulation
the system's phase space is spanned by coordinates $(x_i,p_i)$, with the variables $p_i\in\mathbb{R}$ ($i=1,2,\cdots,N$) being conjugate momenta
canonically conjugate to $x_i$. Their (equal time) canonical Poisson brackets hence determine the phase space's symplectic structure, with
\begin{equation}
\left\{x_i,x_j\right\}=0,\qquad
\left\{x_i,p_j\right\}=\delta_{ij},\qquad
\left\{p_i,p_j\right\}=0,\qquad i,j=1,2,\cdots,N.
\end{equation}
The system's time evolution, say relative to a physical time coordinate $t\in\mathbb{R}$, is thereby generated from some Hamiltonian function
$H(\vec{x},\vec{p}\,)=H(x_i,p_i)$ -- assumed herein to be time independent, in order to keep the presentation the simplest possible -- with the first-order Hamiltonian
equations of motion
\begin{equation}
\dot{x}_i=\left\{x_i,H\right\}=\frac{\partial H}{\partial p_i},\qquad
\dot{p}_i=\left\{p_i,H\right\}=-\frac{\partial H}{\partial x_i},\qquad \dot{}\equiv\frac{d}{dt},\qquad i=1,2,\cdots,N.
\end{equation}
Within the context of nonrelativistic Newtonian dynamics, for a single massive particle of mass $\mu$ the Hamiltonian is generally of the form
\begin{equation}
H(\vec{x},\vec{p}\,)=\frac{1}{2\mu}\vec{p}\,^2+V(\vec{x}\,),
\label{eq:HV}
\end{equation}
where $V(\vec{x}\,)$ is some mechanical potential energy for the particle.

As a matter of fact, these different structures are encoded in the Hamiltonian first-order action defined over phase space,
\begin{equation}
S_0[x_i,p_i]=\int dt\left(\dot{\vec{x}}\cdot\vec{p} - H(\vec{x},\vec{p}\,)\right)=\int dt\left(\dot{x}_i  p_i - H(x_i,p_i)\right)
\label{eq:action1}
\end{equation}
(with the implicit summation convention over repeated indices being understood throughout).
Of course, the variations of this action relative to $x_i$ and $p_i$ lead back to the above first-order equations of motion.
Note that an arbitrary total time derivative, or time surface term
in the form of $d\Lambda(\vec{x},\vec{p}\,)/dt$ may be added to this action without affecting neither these equations of motion nor the phase space symplectic
structure. Finally in the case of a Hamiltonian in the form of (\ref{eq:HV}), the reduction of the conjugate momenta through the Hamiltonian equations of motion
for $x_i$ leads to the familiar form of the Lagrangian action for this class of systems,
\begin{equation}
S_0[x_i]=\int dt\left(\frac{1}{2}\mu\dot{\vec{x}}\,^2\,-\,V\left(\vec{x}\,\right)\right).
\end{equation}

\subsection{A canonically quantised formulation}

Through the correspondence principle, the canonically quantised system is implicitly specified by the equal time commutation relations which are in correspondence
with the above Poisson brackets. These commutation relations are to be considered at some reference time $t=t_0$ which is not made explicit hereafter
(note that because of the lack of time dependency of the Hamiltonian, invariance of the dynamics under constant translations in time implies that one may always
take $t_0=0$ for convenience). Hence one has the following tensor product over $i,j=1,2,\cdots,N$ of $N$ copies of the Heisenberg algebra with
generators and quantum phase space observables\footnote{Note that throughout all operators are systematically distinguished by a caret above
their symbol.} $(\hat{x}_i,\hat{p}_i,\mathbb{I})$,
\begin{equation}
\left[\hat{x}_i,\hat{x}_j\right]=0,\quad
\left[\hat{x}_i,\hat{p}_j\right]=i\hbar\,\delta_{ij}\,\mathbb{I},\quad
\left[\hat{p}_i,\hat{p}_j\right]=0,\quad
\hat{x}^\dagger_i=\hat{x}_i,\quad
\hat{p}^\dagger_i=\hat{p}_i,\quad i,j=1,2,\cdots,N.
\end{equation}
The abstract Hilbert space of quantum states of the system provides a representation of this Heisenberg algebra. It must be equipped with a positive definite
sesquilinear inner product for which $\hat{x}_i$ and $\hat{p}_i$ must each be self-adjoint (or at least hermitian), as must be the quantum Hamiltonian itself as well,
\begin{equation}
\hat{H}=\hat{H}(\hat{\vec{x}},\hat{\vec{p}}\,),\qquad \hat{H}^\dagger=\hat{H},
\end{equation}
in order for time evolution of the quantised system as generated from this quantum operator to be unitary. As is well known from the Stone-von Neumann theorem,
in the case of a configuration space of trivial fundamental group -- such as applies to the space $\mathbb{R}^N$ --, up to unitary equivalence there exists a single
unitary representation of the Heisenberg algebra \cite{Gov2}, namely the well known canonical representation and its wave function realisations
whether in configuration or in momentum space.

The construction of these $|\vec{x}\,\rangle$- and $|\vec{p}\,\rangle$-representations is standard. Since either all position operators $\hat{x}_i$,
or all momentum operators $\hat{p}_i$, commute among themselves there always exists a basis of eigenstates of each of these two ensembles of operators,
$|\vec{x}\,\rangle$ or $|\vec{p}\,\rangle$, respectively,
\begin{equation}
\hat{x}_i|\vec{x}\,\rangle = x_i |\vec{x}\,\rangle,\qquad
\hat{p}_i|\vec{p}\,\rangle = p_i |\vec{p}\,\rangle,\qquad x_i, p_i\in\mathbb{R},\quad i=1,2,\cdots,N.
\end{equation}
Given the Euclidean metric of configuration space, the normalisation of these states is chosen such that,
\begin{equation}
\langle\vec{x}\,|\vec{x}\,'\rangle=\delta^{(N)}(\vec{x}-\vec{x}\,'),\qquad
\langle\vec{p}\,|\vec{p}\,'\rangle=\delta^{(N)}(\vec{p}-\vec{p}\,'),
\end{equation}
which in turn implies the corresponding completeness relations or representations of the unit operator,
\begin{equation}
\mathbb{I}=\int_{(\infty)}d^N\vec{x}\,|\vec{x}\,\rangle \langle\vec{x}\,|,\qquad
\mathbb{I}=\int_{(\infty)}d^N\vec{p}\,|\vec{p}\,\rangle \langle\vec{p}\,| .
\end{equation}
Note that since these normalisation choices do not impact the pure imaginary phase factors of these position and momentum
eigenstates, these phase factors, local in $\vec{x}$ or in $\vec{p}$, are left unspecified \cite{Gov2}, with the understanding that some choice
albeit implicit has been effected at this stage.

Consequently, in view of the above spectral representations of the unit operator, any abstract quantum state $|\varphi\rangle$ may be decomposed
in either of these two bases in terms of complex valued functions whether in configuration or in momentum space, namely its configuration or momentum
wave function representations $\varphi(\vec{x}\,)=\langle\vec{x}\,|\varphi\rangle$ or $\tilde{\varphi}(\vec{p}\,)=\langle\vec{p}\,|\varphi\rangle$, respectively,
such that,
\begin{equation}
|\varphi\rangle=\int_{(\infty)}d^N\vec{x}\,|\vec{x}\,\rangle\,\varphi(\vec{x}\,),\quad 
|\varphi\rangle=\int_{(\infty)}d^N\vec{p}\,|\vec{p}\,\rangle\,\tilde{\varphi}(\vec{p}\,),\quad
\varphi(\vec{x})=\langle\vec{x}\,|\varphi\rangle,\quad
\tilde{\varphi}(\vec{p}\,)=\langle\vec{p}\,|\varphi\rangle.
\end{equation}
It may be established \cite{Gov2} as well that the operators $\hat{x}_i$ and $\hat{p}_i$ then possess the following differential operator realisations acting on these
wave functions, whether in the $|\vec{x}\,\rangle$-representation,
\begin{equation}
\langle\vec{x}\,|\hat{x}_i|\varphi\rangle=x_i\,\varphi(\vec{x}\,),\qquad
\langle\vec{x}\,|\hat{p}_i|\varphi\rangle=-i\hbar\frac{\partial}{\partial x_i}\varphi(\vec{x}\,),\qquad i=1,2,\cdots,N,
\end{equation}
or in the $|\vec{p}\,\rangle$-representation,
\begin{equation}
\langle\vec{p}\,|\hat{x}_i|\varphi\rangle=i\hbar\frac{\partial}{\partial p_i}\tilde{\varphi}(\vec{p}\,),\qquad
\langle\vec{p}\,|\hat{p}_i|\varphi\rangle=p_i\,\tilde{\varphi}(\vec{p}\,),\qquad i=1,2,\cdots,N.
\end{equation}
Once these choices made for the differential operator representations of $\hat{p}_i$ in the $|\vec{x}\,\rangle$ basis and of $\hat{x}_i$ in the $|\vec{p}\,\rangle$ basis,
it follows that the unitary transformations mapping one of these bases onto the other are realised in terms of the following matrix elements,
\begin{equation}
\langle\vec{x}\,|\vec{p}\,\rangle=\frac{1}{(2\pi\hbar)^{N/2}}\,e^{\frac{i}{\hbar}\vec{x}\cdot\vec{p}},\qquad
\langle\vec{p}\,|\vec{x}\,\rangle=\langle\vec{x}\,|\vec{p}\,\rangle^* =\frac{1}{(2\pi\hbar)^{N/2}}\,e^{-\frac{i}{\hbar}\vec{x}\cdot\vec{p}},
\end{equation}
where an implicit choice for the relative quantum phase factor between these two classes of quantum states is made such that no other pure imaginary
constant phase factor is involved. As a consequence for any abstract state $|\varphi\rangle$ and its position and momentum wave functions
one has the reciprocal (Fourier transformation) relations,
\begin{equation}
\varphi(\vec{x}\,)=\int_{(\infty)}\frac{d^N\vec{p}}{(2\pi\hbar)^{N/2}}\,e^{\frac{i}{\hbar}\vec{x}\cdot\vec{p}}\,\tilde{\varphi}(\vec{p}\,),\qquad
\tilde{\varphi}(\vec{p}\,)=\int_{(\infty)}\frac{d^N\vec{p}}{(2\pi\hbar)^{N/2}}\,e^{-\frac{i}{\hbar}\vec{x}\cdot\vec{p}}\,\varphi(\vec{x}\,).
\end{equation}

\subsection{Alternative quantum operator representations}

Motivated by the spectral decompositions or completeness relations of the unit operator,
\begin{equation}
\mathbb{I}=\int_{(\infty)}d^N\vec{x}\,|\vec{x}\,\rangle \langle\vec{x}\,|,\qquad
\mathbb{I}=\int_{(\infty)}d^N\vec{p}\,|\vec{p}\,\rangle \langle\vec{p}\,|,
\end{equation}
it is also possible to establish similar spectral decomposition representations of the position and momentum operators,
$\hat{x}_i$ and $\hat{p}_i$, whether in the $|\vec{x}\,\rangle$ or the $|\vec{p}\,\rangle$ basis. Such results can also find their own
practical utility, as the case may be. They also extend to the noncommutative configuration spaces considered later on.

Obviously in their respective eigenbases one has,
\begin{equation}
\hat{x}_i=\int_{(\infty)}d^N\vec{x}\,|\vec{x}\,\rangle\,x_i\,\langle\vec{x}\,|,\qquad
\hat{p}_i=\int_{(\infty)}d^N\vec{p}\,|\vec{p}\,\rangle\,p_i\,\langle\vec{p}\,|,\qquad i=1,2,\cdots,N.
\end{equation}
While in the other basis which is not that of their eigenstates, one also finds for $\hat{p}_i$,
\begin{eqnarray}
\hat{p}_i &=& \int_{(\infty)}d^N\vec{x}\,|\vec{x}\,\rangle\left(-\frac{1}{2}i\hbar\left(\frac{\stackrel{\rightarrow}{\partial}}{\partial x_i} -
\frac{\stackrel{\leftarrow}{\partial}}{\partial x_i} \right)\right) \langle\vec{x}\,| \nonumber \\
&=& \int_{(\infty)}d^N\vec{x}\,|\vec{x}\,\rangle\left(-i\hbar\frac{\stackrel{\rightarrow}{\partial}}{\partial x_i}\right) \langle\vec{x}\,|
= \int_{(\infty)}d^N\vec{x}\,|\vec{x}\,\rangle\left(i\hbar\frac{\stackrel{\leftarrow}{\partial}}{\partial x_i} \right) \langle\vec{x}\,|,\qquad i=1,2,\cdots,N,
\end{eqnarray}
and for $\hat{x}_i$,
\begin{eqnarray}
\hat{x}_i &=& \int_{(\infty)}d^N\vec{p}\,|\vec{p}\,\rangle\left(\frac{1}{2}i\hbar\left(\frac{\stackrel{\rightarrow}{\partial}}{\partial p_i} -
\frac{\stackrel{\leftarrow}{\partial}}{\partial p_i} \right)\right) \langle\vec{p}\,| \nonumber \\
&=& \int_{(\infty)}d^N\vec{p}\,|\vec{p}\,\rangle\left(i\hbar\frac{\stackrel{\rightarrow}{\partial}}{\partial p_i}\right) \langle\vec{p}\,|
= \int_{(\infty)}d^N\vec{p}\,|\vec{p}\,\rangle\left(-i\hbar\frac{\stackrel{\leftarrow}{\partial}}{\partial p_i} \right) \langle\vec{p}\,|,\qquad i=1,2,\cdots,N.
\end{eqnarray}
These different relations for the operators $\hat{x}_i$ and $\hat{p}_i$ may easily be confirmed, for example by computing their action on any state $|\varphi\rangle$
and then projecting onto the basis in which their spectral representation is considered.

However it should be emphasised that for the last two sets of identities for $\hat{p}_i$ and for $\hat{x}_i$ to be valid,
it must be understood that these operators are implicitly assumed to be acting
on abstract quantum states $|\varphi\rangle$ whose wave functions $\varphi(\vec{x}\,)$ and $\tilde{\varphi}(\vec{p}\,)$ are normalisable over
$\mathbb{R}^N$, so that no surface contributions at infinity can be generated through integrations by parts.

\subsection{Configuration space localised and normalisable quantum states}

Even though the states $|\vec{x}\,\rangle$ and $|\vec{p}\,\rangle$ are sharp in the sense that they are eigenstates of the position and momentum operators
$\hat{x}_i$ and $\hat{p}_i$, respectively, they are not normalisable since the values of their overlaps within each ensemble are given by the Dirac $\delta^{(N)}$
distribution whether in $\vec{x}$ or in $\vec{p}$ space.

Within the noncommutative configuration space context to be addressed hereafter, the momentum operators $\hat{p}_i$ remain commuting among themselves.
In that case their eigenbasis of non-normalisable though sharp states $|\vec{p}\,\rangle$ still provides a basis of quantum Hilbert space. However since in that context
the configuration space operators $\hat{x}_i$ are no longer all commuting among themselves, they cannot possess a common basis of sharp eigenstates,
as are the present states $|\vec{x}\,\rangle$. Nonetheless, it then remains possible to identify quantum states that are not only localisable in configuration space
at least in a weak sense, namely in terms of expectation values, but these states are then also of finite norm. This is then made possible because
of the existence of length scales intrinsic to the noncommutativity of configuration space.

Likewise within the present context of ordinary quantum physics with its commuting configuration space degrees of freedom $\hat{x}_i$,
let us consider the possibility of constructing similarly localisable quantum states of finite norm, given specific length squared scale factors $\lambda_{ij}$
with $\lambda_{ij}\in\mathbb{R}$ representing a real positive definite symmetric matrix of constant coefficients of which the inverse $(\lambda^{-1})_{ij}$ is thus
also well defined. By definition, these states, to be denoted $|\vec{x};\lambda\rangle$, possess the following representation in the basis of momentum eigenstates,
\begin{equation}
\langle\vec{p}\,|\vec{x};\lambda\rangle=\frac{1}{(2\pi\hbar)^{N/2}}\,e^{-\frac{i}{\hbar}\vec{x}\cdot\vec{p}}\,e^{-\frac{\lambda_{ij}}{4\hbar^2} p_i p_j},\qquad
|\vec{x};\lambda\rangle = \int_{(\infty)}d^N\vec{p}\,|\vec{p}\,\rangle\,\langle\vec{p}\,|\vec{x};\lambda\rangle.
\end{equation}
Obviously one has,
\begin{equation}
\lim_{\lambda_{ij}\rightarrow 0}|\vec{x};\lambda\rangle=|\vec{x}\,\rangle.
\end{equation}
Due to the exponential suppression of the large momentum components of these states $|\vec{x};\lambda\rangle$ induced by
the nonvanishing scale factors $\lambda_{ij}$, the latter quantities are responsible for a spreading or fuzziness on distance scales of the order of $\sqrt{|\lambda_{ij}|}$
for these states' probability density in configuration space, and centered at $\vec{x}$. This is made explicit by the configuration space wave function
of these states,
\begin{equation}
\langle \vec{x}\,|\vec{x}_0;\lambda\rangle=\frac{1}{(\pi^N\,{\rm det}\,\lambda_{ij})^{1/2}}\,e^{-(x-x_0)_i\,(\lambda^{-1})_{ij}\,(x-x_0)_j}
=W_\lambda(\vec{x}-\vec{x}_0)=W_\lambda(\vec{x}_0-\vec{x}),
\end{equation}
where the real Weierstrass kernel $W_\lambda(\vec{x}\,)$ associated to the coefficients $\lambda_{ij}$ is defined by,
\begin{equation}
W_\lambda(\vec{x}\,)\equiv \int_{(\infty)}\frac{d^N\vec{p}}{(2\pi\hbar)^N}\,e^{-\frac{\lambda_{ij}}{4\hbar^2}p_i p_j}\,e^{\frac{i}{\hbar}\vec{x}\cdot\vec{p}}
=\frac{1}{(\pi^N\,{\rm det}\,\lambda_{ij})^{1/2}}\,e^{-x_i (\lambda^{-1})_{ij} x_j},\qquad
W^*_\lambda(\vec{x}\,)=W_\lambda(\vec{x}\,).
\end{equation}
Note that one may consider that the parameters $\lambda_{ij}$ jointly with the choice of localisable states $|\vec{x};\lambda\rangle$ provide a gaussian regularisation
of the sharp but non-normalisable position eigenstates $|\vec{x}\,\rangle$, that leads as well to an ensemble of localised and normalisable quantum states attached
specifically to all points in configuration space which ought to share with the states $|\vec{x}\,\rangle$ most of their properties but in a weak sense
through matrix elements and expectation values.

Even though the above notation for the states $|\vec{x};\lambda\rangle$ is strictly reserved for real values of the parameters $x_i\in\mathbb{R}$,
since $\lambda_{ij}$ defines a real positive definite symmetric matrix
the above definition of these states extends to a larger ensemble of quantum states by allowing the parameters $x_i$ to rather take complex values
$z_i\in\mathbb{C}$. This extended set of  quantum states is then to be denoted specifically as $|z_i;\lambda\rangle$ with,
\begin{equation}
|z_i;\lambda\rangle = \int_{(\infty)}d^N\vec{p}\,|\vec{p}\,\rangle\,\langle\vec{p}\,|z_i;\lambda\rangle,\qquad
\langle\vec{p}\,|z_i;\lambda\rangle=\frac{1}{(2\pi\hbar)^{N/2}}\,e^{-\frac{i}{\hbar} z_i p_i}\,e^{-\frac{\lambda_{ij}}{4\hbar^2} p_i p_j},
\label{eq:zi0}
\end{equation}
while obviously $|z_i=x_i\in\mathbb{R};\lambda\rangle=|x_i;\lambda\rangle=|\vec{x};\lambda\rangle$. Of course, the states $|z_i;\lambda\rangle$
remain normalisable for $z_i\in\mathbb{C}$, with,
\begin{equation}
\langle z_{1,i};\lambda | z_{2,i};\lambda\rangle
= \frac{1}{\left((2\pi)^N\,{\rm det}\,\lambda_{ij}\right)^{1/2}}\,e^{-\frac{1}{2}(z^*_{1,i}-z_{2,i})\,\left(\lambda^{-1}\right)_{ij}\,(z^*_{1,j}-z_{2,j})},
\end{equation}
and in particular,
\begin{equation}
\langle z_i;\lambda | z_i;\lambda\rangle = \frac{1}{\left((2\pi)^N\,{\rm det}\,\lambda_{ij}\right)^{1/2}}\,e^{2\,{\rm Im}\,z_i\,\left(\lambda^{-1}\right)_{ij}\,{\rm Im}\,z_j}.
\end{equation}

\subsection{Properties of Weierstrass kernels and their $*_\lambda$-products}

From the above $\vec{p}\,$-integral representation of the Weierstrass kernel associated to the coefficients $\lambda_{ij}$ one readily sees that,
\begin{equation}
\lim_{\lambda_{ij}\rightarrow 0} W_\lambda(\vec{x}\,)=\delta^{(N)}(\vec{x}\,),\qquad
\int_{(\infty)}d^N\vec{x}\,W_\lambda(\vec{x}\,)=1,
\end{equation}
while of course,
\begin{equation}
\int_{(\infty)}d^N\vec{x}\,\delta^{(N)}(\vec{x}\,)=1.
\end{equation}
The Weierstrass kernel $W_\lambda(\vec{x}\,)$ thus provides a specific gaussian regularisation of the Dirac $\delta^{(N)}(\vec{x}\,)$ distribution.

By using the Weierstrass kernel as a normalised weight distribution in configuration space, direct evaluations of some expectation values in configuration space find,
\begin{equation}
\int_{(\infty)}d^N\vec{x}\,W_\lambda(\vec{x}\,)\,x_i=0,\quad
\int_{(\infty)}d^N\vec{x}\,W_\lambda(\vec{x}\,)\,x_i x_j=\frac{1}{2}\lambda_{ij},\quad
\int_{(\infty)}d^N\vec{x}\,W_\lambda(\vec{x}\,)\,\vec{x}\,^2=\frac{1}{2}{\rm Tr}\,\lambda_{ij}.
\end{equation}
It may readily be checked that this kernel also obeys the following elliptic differential equation,
\begin{equation}
\left(-\lambda_{ij}\frac{\partial}{\partial x_i}\frac{\partial}{\partial x_j}\,+\, 4\, x_i\, (\lambda^{-1})_{ij}\, x_j \,-\, 2N\right)\,W_\lambda(\vec{x}\,)=0.
\end{equation}

In order to display one last most noteworthy property of the Weierstrass kernel $W_\lambda(\vec{x}\,)$,
let us finally introduce a $*_\lambda$-product which will prove to be most
relevant hereafter. This $*_\lambda$-product involves the scale parameters $\lambda_{ij}$, is given as the exponentiated series of powers of the
$\lambda_{ij}$ contracted product of two configuration space gradients acting towards the left and the right on functions of $\vec{x}$
that are multiplied through the $*_\lambda$-product, and is defined by,
\begin{equation}
*_\lambda \equiv e^{\frac{1}{2}\lambda_{ij}\stackrel{\leftarrow}{\frac{\partial}{\partial x_i}}\stackrel{\rightarrow}{\frac{\partial}{\partial x_j}}} \equiv *_{\vec{x},\lambda}.
\label{eq:star-product}
\end{equation}
The very last notation may be used when the variables on which the $*_\lambda$-product acts are to be made explicit if confusion could arise otherwise.
Note that in the limit of vanishing scale coefficients $\lambda_{ij}$ the $*_\lambda$-product reduces to the ordinary commutative product of functions over
configuration space,
\begin{equation}
\lim_{\lambda_{ij}\rightarrow 0}*_\lambda=1,\qquad
\lim_{\lambda_{ij}\rightarrow 0}\,f(\vec{x}\,)\,*_\lambda\,g(\vec{x}\,)=f(\vec{x}\,)\,g(\vec{x}\,).
\end{equation}

A direct evaluation using for the Weierstrass kernel its $\vec{p}\,$-integral representation then finds the following identity valid specifically for the $*_\lambda$-product,
\begin{equation}
W_\lambda(\vec{x}-\vec{a}\,)*_\lambda W_\lambda(\vec{x}-\vec{b}\,)=W_\lambda(\vec{x}-\vec{a}\,)\,\delta^{(N)}(\vec{a}-\vec{b}\,).
\end{equation}
Once again, this result is to be considered on a par with the similar identity valid for the Dirac $\delta^{(N)}(\vec{x}\,)$ distribution,
to which furthermore the above relation reduces in the limit $\lambda_{ij}\rightarrow 0$,
\begin{equation}
\delta^{(N)}(\vec{x}-\vec{a}\,)\,\delta^{(N)}(\vec{x}-\vec{b}\,)=\delta^{(N)}(\vec{x}-\vec{a}\,)\,\delta^{(N)}(\vec{a}-\vec{b}\,).
\end{equation}

\subsection{Properties of the localisable states $|\vec{x};\lambda\rangle$}

Given any quantum operator $\hat{\cal O}$ acting on Hilbert space, let us denote its normalised expectation values for the localised states
$|\vec{x};\lambda\rangle$ as
\begin{equation}
\langle\hat{\cal O}\rangle_\lambda(\vec{x}\,) \equiv \frac{\langle\vec{x};\lambda|\hat{\cal O}|\vec{x};\lambda\rangle}
{\langle\vec{x};\lambda | \vec{x};\lambda\rangle}.
\end{equation}
A simple evaluation finds that the states $|\vec{x};\lambda\rangle$ are indeed of finite norm for nonvanishing scale factors $\lambda_{ij}$,
with a value independent of $\vec{x}$ given as,
\begin{equation}
\langle\vec{x};\lambda | \vec{x};\lambda\rangle=\frac{1}{\left((2\pi)^N\,{\rm det}\,\lambda_{ij}\right)^{1/2}}=W_{2\lambda}(\vec{0}\,),
\end{equation}
where $W_{2\lambda}(\vec{x}\,)$ denotes the Weierstrass kernel now associated to the coefficients $2\lambda_{ij}$, namely,
\begin{equation}
W_{2\lambda}(\vec{x}\,)\equiv \int_{(\infty)}\frac{d^N\vec{p}}{(2\pi\hbar)^N}\,e^{-\frac{2\lambda_{ij}}{4\hbar^2}p_i p_j}\,e^{\frac{i}{\hbar}\vec{x}\cdot\vec{p}}
=\frac{1}{((2\pi)^N\,{\rm det}\,\lambda_{ij})^{1/2}}\,e^{-\frac{1}{2}x_i (\lambda^{-1})_{ij} x_j}.
\end{equation}

Direct evaluations find for the operators $\hat{x}_i$, $\hat{x}_i\hat{x}_j$ and $\hat{\vec{x}}\,^2$ the following expectations values
for the states $|\vec{x};\lambda\rangle$,
\begin{equation}
\langle\hat{x}_i\rangle_\lambda(\vec{x}\,)=x_i,\qquad
\langle\hat{x}_i\hat{x}_j\rangle_\lambda(\vec{x}\,)=x_i x_j+\frac{1}{4}\lambda_{ij},\qquad
\langle\hat{\vec{x}}\,^2\rangle_\lambda(\vec{x}\,)=\vec{x}\,^2+\frac{1}{4}{\rm Tr}\,\lambda_{ij}.
\end{equation}
And likewise for the operators $\hat{p}_i$, $\hat{p}_i\hat{p}_j$ and $\hat{\vec{p}}\,^2$,
\begin{equation}
\langle\hat{p}_i\rangle_\lambda(\vec{x}\,)=0,\qquad
\langle\hat{p}_i \hat{p}_j\rangle_\lambda(\vec{x}\,)=\hbar^2\,\left(\lambda^{-1}\right)_{ij},\qquad
\langle\hat{\vec{p}}\,^2\rangle_\lambda(\vec{x}\,)=\hbar^2\,{\rm Tr}\,\left(\lambda^{-1}\right)_{ij}.
\end{equation}

Consequently for the associated uncertainty relations, one finds,
\begin{equation}
\Delta^{(\hat{\vec{x}})}_{ij}(\vec{x}\,)
\equiv\langle\left(\hat{x}_i-\langle\hat{x}_i\rangle_\lambda\right)\left(\hat{x}_j-\langle\hat{x}_j\rangle_\lambda\right)\rangle_\lambda(\vec{x}\,)
=\frac{1}{4}\lambda_{ij},
\end{equation}
as well as,
\begin{equation}
\Delta^{(\hat{\vec{p}}\,)}_{ij}(\vec{x}\,)
\equiv\langle\left(\hat{p}_i-\langle\hat{p}_i\rangle_\lambda\right)\left(\hat{p}_j-\langle\hat{p}_j\rangle_\lambda\right)\rangle_\lambda(\vec{x}\,)
=\hbar^2\,\left(\lambda^{-1}\right)_{ij},
\end{equation}
so that finally,
\begin{equation}
\Delta^{(\hat{\vec{x}})}_{ij}(\vec{x}\,)\,\Delta^{(\hat{\vec{p}}\,)}_{k\ell}(\vec{x}\,) = \frac{1}{4}\hbar^2\,\lambda_{ij}\,\left(\lambda^{-1}\right)_{k\ell}.
\end{equation}
In other words the states $|\vec{x};\lambda\rangle$ are saturating states for the mixed uncertainty relations of the position and momentum quantum operators,
as is characteristic as well of usual canonical coherent states for the Heisenberg algebra. Indeed it will be seen hereafter that these localised
and normalised quantum states $|\vec{x};\lambda\rangle$ are indeed a specific subclass of such coherent quantum states.

This observation also reflects the fact that of course, the states $|\vec{x};\lambda\rangle$ cannot be sharp position eigenstates of $\hat{x}_i$, of which the action
on these states includes as well contributions from the action of the momentum operators $\hat{p}_i$ that directly involve the coefficients $\lambda_{ij}$.
Indeed, by inserting the spectral decomposition of the unit operator in the $|\vec{p}\,\rangle$ eigenbasis to the left of $\hat{x}_i|\vec{x};\lambda\rangle$
and using the matrix elements $\langle\vec{p}\,|\hat{x}_i|\vec{x};\lambda\rangle$ one establishes the identity,
\begin{equation}
\hat{x}_i\,|\vec{x};\lambda\rangle=x_i\,|\vec{x};\lambda\rangle\,-\,\frac{i}{2\hbar}\lambda_{ij}\,\hat{p}_j\,|\vec{x};\lambda\rangle,\qquad
\left(\hat{x}_i+\frac{i}{2\hbar}\lambda_{ij}\hat{p}_j\right)|\vec{x};\lambda\rangle = x_i\, |\vec{x};\lambda\rangle.
\label{eq:twisted}
\end{equation}
Hence, the localised states $|\vec{x};\lambda\rangle$
are eigenstates of the ``momentum twisted'' position operators $(\hat{x}_i+\frac{i\lambda_{ij}}{2\hbar}\hat{p}_j)$ with eigenvalues $x_i$,
rather than of the position operators $\hat{x}_i$ themselves. Further consequences of this observation are to be addressed hereafter.

Similar considerations may be extended to matrix elements of the operators $\mathbb{I}$, $\hat{x}_i$, $\hat{x}_i\hat{x}_j$, $\hat{\vec{x}}\,^2$,
$\hat{p}_i$, $\hat{p}_i \hat{p}_j$ and $\hat{\vec{p}}\,^2$ but now for two distinct external localised states. Through direct evaluations one finds the following
explicit results. First for their overlaps,
\begin{equation}
\langle\vec{x}_1;\lambda|\vec{x}_2;\lambda\rangle=\int_{(\infty)}\frac{d^N\vec{p}}{(2\pi\hbar)^N}\,e^{\frac{i}{\hbar}\vec{p}\cdot(\vec{x}_1-\vec{x}_2)}
e^{-\frac{\lambda_{ij}}{2\hbar^2}p_i p_j} =W_{2\lambda}(\vec{x}_1-\vec{x}_2),
\end{equation}
an expression which shows that any of these states has a finite nonvanishing overlap with any other such state.
Next for the operators $\hat{x}_i$, $\hat{x}_i\hat{x}_j$ and $\hat{\vec{x}}\,^2$,
\begin{equation}
\langle\vec{x}_1;\lambda|\hat{x}_i|\vec{x}_2;\lambda\rangle=\frac{1}{2}\left(x_{1}+x_{2}\right)_i\,W_{2\lambda}(\vec{x}_1-\vec{x}_2),
\label{eq:matrix1}
\end{equation}
\begin{equation}
\langle\vec{x}_1;\lambda| \hat{x}_i\hat{x}_j |\vec{x}_2;\lambda\rangle
=\left[\frac{1}{4}\left(x_1+x_2\right)_i \left(x_1+x_2\right)_j+\frac{1}{4}\lambda_{ij}\right]\,W_{2\lambda}(\vec{x}_1-\vec{x}_2),
\label{eq:matrix2}
\end{equation}
and,
\begin{equation}
\langle\vec{x}_1;\lambda|\hat{\vec{x}}\,^2|\vec{x}_2;\lambda\rangle
=\left[\left(\frac{\vec{x}_1+\vec{x}_2}{2}\right)^2+\frac{1}{4}{\rm Tr}\,\lambda_{ij}\right]\,W_{2\lambda}(\vec{x}_1-\vec{x}_2).
\label{eq:matrix3}
\end{equation}
And finally for the operators $\hat{p}_i$, $\hat{p}_i \hat{p}_j$ and $\hat{\vec{p}}\,^2$ as well,
\begin{equation}
\langle\vec{x}_1;\lambda|\hat{p}_i|\vec{x}_2;\lambda\rangle=-i\hbar\frac{\partial}{\partial x_{1i}}\,W_{2\lambda}(\vec{x}_1-\vec{x}_2),
\label{eq:matrix4}
\end{equation}
\begin{equation}
\langle\vec{x}_1;\lambda|\hat{p}_i \hat{p}_j |\vec{x}_2;\lambda\rangle=
\left(-i\hbar\frac{\partial}{\partial x_{1i}}\right)\left(-i\hbar\frac{\partial}{\partial x_{1j}}\right)\,W_{2\lambda}(\vec{x}_1-\vec{x}_2),
\label{eq:matrix5}
\end{equation}
and,
\begin{equation}
\langle\vec{x}_1;\lambda|\hat{\vec{p}}\,^2|\vec{x}_2;\lambda\rangle=\left(-i\hbar\vec{\nabla}_{\vec{x}_1}\right)^2\,W_{2\lambda}(\vec{x}_1-\vec{x}_2).
\label{eq:matrix6}
\end{equation}

\subsection{Displacement operators and canonical quantum coherent states}

As is well known the momentum operators $\hat{p}_i$ are the generators of displacements in configuration space, with a finite displacement
by $\vec{a}$ being generated through the action of the unitary operator $e^{-\frac{i}{\hbar}\vec{a}\cdot\hat{\vec{p}}}$.
Explicitly one finds\footnote{By using the Baker-Campbell-Haussdorff (BCH) formula
$e^{\hat{A}}\,e^{\hat{B}}=e^{\hat{A}+\hat{B}+\frac{1}{2}[\hat{A},\hat{B}]}$ which is valid for any two operators $\hat{A}$ and $\hat{B}$
that commute with their commutator $[\hat{A},\hat{B}]$.},
\begin{equation}
e^{\frac{i}{\hbar}\vec{a}\cdot\hat{\vec{p}}}\,\hat{x}_i\,e^{-\frac{i}{\hbar}\vec{a}\cdot\hat{\vec{p}}} = \hat{x}_i\,+\,a_i\,\mathbb{I},\qquad
e^{\frac{i}{\hbar}\vec{a}\cdot\hat{\vec{p}}}\,\hat{p}_i\,e^{-\frac{i}{\hbar}\vec{a}\cdot\hat{\vec{p}}}  = \hat{p}_i,
\end{equation}
while,
\begin{equation}
e^{-\frac{i}{\hbar}\vec{a}\cdot\hat{\vec{p}}}\,|\vec{x}\,\rangle = |\vec{x}+\vec{a}\,\rangle,\qquad
e^{-\frac{i}{\hbar}\vec{a}\cdot\hat{\vec{p}}}\,|\vec{p}\,\rangle=e^{\frac{i}{\hbar}\vec{a}\cdot\vec{p}}\,|\vec{p}\,\rangle.
\end{equation}

In a likewise manner the position operators $\hat{x}_i$ are the generators of displacements in momentum space, with a finite displacement
by $\vec{p}_0$ being generated through the action of the unitary operator $e^{\frac{i}{\hbar}\vec{p}_0\cdot\hat{\vec{x}}}$. One has this time,
\begin{equation}
e^{-\frac{i}{\hbar}\vec{p}_0\cdot\hat{\vec{x}}}\,\hat{x}_i\,e^{\frac{i}{\hbar}\vec{p}_0\cdot\hat{\vec{x}}} = \hat{x}_i,\qquad
e^{-\frac{i}{\hbar}\vec{p}_0\cdot\hat{\vec{x}}}\,\hat{p}_i\,e^{\frac{i}{\hbar}\vec{p}_0\cdot\hat{\vec{x}}}=\hat{p}_i\,+\,p_{0i}\,\mathbb{I},
\end{equation}
while,
\begin{equation}
e^{\frac{i}{\hbar}\vec{p}_0\cdot\hat{\vec{x}}}\,|\vec{x}\,\rangle=e^{\frac{i}{\hbar}\vec{p}_0\cdot\vec{x}}\,|\vec{x}\,\rangle,\qquad
e^{\frac{i}{\hbar}\vec{p}_0\cdot\hat{\vec{x}}}\,|\vec{p}\,\rangle=|\vec{p}+\vec{p}_0\rangle.
\label{eq:displa-p}
\end{equation}

Let us now reconsider the result in (\ref{eq:twisted}). Combined with the fact that
\begin{equation}
\hat{p}_i\,|\vec{x};\lambda\rangle=i\hbar\,\frac{\partial}{\partial x_i}\,|\vec{x};\lambda\rangle,
\end{equation}
which is readily established from the knowledge of the momentum wave functions $\langle\vec{p}\,|\vec{x};\lambda\rangle$, one thus also obtains
the following identities,
\begin{equation}
\hat{x}_i\,|\vec{x};\lambda\rangle = \left(x_i+\frac{1}{2}\lambda_{ij}\frac{\partial}{\partial x_j}\right)\,|\vec{x};\lambda\rangle,\qquad
\hat{p}_i\,|\vec{x};\lambda\rangle=2i\hbar\left(\lambda^{-1}\right)_{ij}\,\left(\hat{x}_j - x_j\mathbb{I}\right)\,|\vec{x};\lambda\rangle.
\end{equation}

Given that the localised states $|\vec{x};\lambda\rangle$ are eigenstates with eigenvalues $x_i$ of the momentum twisted coordinates
$\left(\hat{x}_i+i\frac{\lambda_{ij}}{2\hbar}\hat{p}_j\right)$, for any conjugate momentum value $\vec{p}_0$ one also has the property that,
\begin{equation}
e^{\frac{i}{\hbar}p_{0i}\left(\hat{x}_i+\frac{i\lambda_{ij}}{2\hbar}\hat{p}_j\right)}\,|\vec{x};\lambda\rangle
= e^{\frac{i}{\hbar}\vec{p}_0\cdot\vec{x}}\,|\vec{x};\lambda\rangle.
\end{equation}
Using the BCH formula this last identity may also be brought into the following form,
\begin{equation}
e^{\frac{i}{\hbar}\vec{p}_0\cdot\hat{\vec{x}}}\,|\vec{x};\lambda\rangle
=e^{-\frac{\lambda_{ij}}{4\hbar^2}p_{0i}p_{0j}}\,e^{\frac{i}{\hbar}\vec{x}\cdot\vec{p}_0}\,e^{\frac{\lambda_{ij}}{2\hbar^2}p_{0i}\hat{p}_j}\,|\vec{x};\lambda\rangle,
\label{eq:intermediate0}
\end{equation}
a result which is to be compared to that of the action of the same momentum displacement operator acting on
the sharp states $|\vec{x}\,\rangle$ (see (\ref{eq:displa-p})). Note that the momentum wave functions of these transformed localised states are thus simply,
\begin{equation}
\langle\vec{p}\, | e^{\frac{i}{\hbar}\vec{p}_0\cdot\hat{\vec{x}}}\,|\vec{x};\lambda\rangle = \frac{1}{(2\pi\hbar)^{N/2}}
\,e^{-\frac{\lambda_{ij}}{4\hbar^2}\left(p-p_0\,\right)_i\left(p-p_0\right)_j}\,e^{-\frac{i}{\hbar}\left(\vec{p}-\vec{p}_0\,\right)\cdot\vec{x}}
=\langle \vec{p}-\vec{p}_0|\vec{x};\lambda\rangle,
\label{eq:matrix-elt}
\end{equation}
which display the expected shift $\vec{p}\rightarrow (\vec{p}-\vec{p}_0)$ in their momentum dependency as generated by the
momentum displacement operator ${\rm exp}(i\vec{p}_0\cdot\hat{\vec{x}}/\hbar)$.

Since $e^{-\frac{i}{\hbar}\vec{a}\cdot\hat{\vec{p}}}\,|\vec{x}\,\rangle=|\vec{x}+\vec{a}\,\rangle$, the result in (\ref{eq:intermediate0}) may also be
expressed in terms of the extended quantum states $|z_i;\lambda\rangle$, with
\begin{equation}
e^{\frac{i}{\hbar}\vec{p}_0\cdot\hat{\vec{x}}}\,|\vec{x};\lambda\rangle = e^{-\frac{\lambda_{ij}}{4\hbar^2}p_{0i} p_{0j} }\,e^{\frac{i}{\hbar}\vec{x}\cdot\vec{p}_0}\,
\Big|z_i=x_i+\frac{1}{2}i\lambda_{ij} p_{0j}; \lambda\Big\rangle .
\end{equation}

Incidentally the matrix elements (\ref{eq:matrix-elt}) also imply the following simple representation of momentum eigenstates in terms of a superposition
of the transformed localised states, whatever the values for $\lambda_{ij}$. Indeed one has, using (\ref{eq:matrix-elt}),
\begin{equation}\int_{(\infty)}d^N\vec{x}\,e^{\frac{i}{\hbar}\vec{p}\cdot\hat{\vec{x}}}\,|\vec{x};\lambda\rangle
=\int_{(\infty)}d^N\vec{x}\int_{(\infty)}d^N\vec{p}_1\,|\vec{p}_1\rangle\,\langle\vec{p}_1|e^{\frac{i}{\hbar}\vec{p}\cdot\hat{\vec{x}}}|\vec{x};\lambda\rangle
=(2\pi\hbar)^{N/2}\,|\vec{p}\,\rangle,
\end{equation}
so that,
\begin{equation}
|\vec{p}\,\rangle=\int_{(\infty)}\frac{d^N\vec{x}}{(2\pi\hbar)^{N/2}}\,e^{\frac{i}{\hbar}\vec{p}\cdot\hat{\vec{x}}}\,|\vec{x};\lambda\rangle.
\label{eq:p-int}
\end{equation}
This observation remains valid as well for $\lambda_{ij}=0$ and the sharp position eigenstates $|\vec{x}\,\rangle$, since
\begin{equation}
\int_{(\infty)}d^N\vec{x}\,e^{\frac{i}{\hbar}\vec{p}\cdot\hat{\vec{x}}}\,|\vec{x}\,\rangle
=\int_{(\infty)}d^N\vec{x}\,e^{\frac{i}{\hbar}\vec{p}\cdot\vec{x}}\int_{(\infty)}d^N\vec{p}_1\, |\vec{p}_1\rangle \langle\vec{p}_1|\vec{x}\,\rangle
=(2\pi\hbar)^{N/2}\,|\vec{p}\,\rangle,
\end{equation}
where in the last equality the value $\langle\vec{p}_1|\vec{x}\,\rangle={\rm exp}(-i\vec{p}_1\cdot\vec{x}/\hbar)/(2\pi\hbar)^{N/2}$ is applied.
So that indeed,
\begin{equation}
|\vec{p}\,\rangle=\int_{(\infty)}\frac{d^N\vec{x}}{(2\pi\hbar)^{N/2}}\,e^{\frac{i}{\hbar}\vec{p}\cdot\hat{\vec{x}}}\,|\vec{x}\,\rangle.
\label{eq:star-p-int}
\end{equation}

Finally, let us revisit the issue of the coherent state status of the states $|\vec{x};\lambda\rangle$.
First consider the momentum twisted coordinates in the following form,
\begin{equation}
\hat{a}_i=\hat{x}_i+\frac{i}{2\hbar}\lambda_{ij}\,\hat{p}_j,\qquad
\hat{a}^\dagger_i=\hat{x}_i-\frac{i}{2\hbar}\lambda_{ij}\,\hat{p}_j ,
\label{eq:Fock1}
\end{equation}
to which one may associate the complex valued phase space quantities
\begin{equation}
z_i=x_i+\frac{i}{2\hbar}\,\lambda_{ij}\,p_j,\qquad
z^*_i=x_i-\frac{i}{2\hbar}\,\lambda_{ij}\,p_j.
\label{eq:zi}
\end{equation}
A direct evaluation then finds for their algebra of commutation relations,
\begin{equation}
\left[\hat{a}_i^{},\hat{a}_j^{}\right]=0,\qquad
\left[\hat{a}_i,\hat{a}^\dagger_j\right]=\lambda_{ij}\,\mathbb{I},\qquad
\left[\hat{a}^\dagger_i,\hat{a}^\dagger_j\right]=0.
\end{equation}
Since $\lambda_{ij}$ is real, symmetric and definite positive, these operators $(\hat{a}_i,\hat{a}^\dagger_i)$ thus in fact span a $N$-dimensional Fock algebra
of which the Fock vacuum $|\Omega_\lambda\rangle$ is such that
\begin{equation}
\hat{a}_i\,|\Omega_\lambda\rangle = 0,
\end{equation}
with a non standard choice of normalisation adapted to the present context, namely,
\begin{equation}
\langle\Omega_\lambda|\Omega_\lambda\rangle=W_{2\lambda}(\vec{0}\,)=\frac{1}{\left((2\pi)^N {\rm det}\,\lambda_{ij}\right)^{1/2}}.
\end{equation}
It follows that the momentum wave function representation of the Fock vacuum, together with a trivial choice of phase factor, is given as
\begin{equation}
\langle\vec{p}\,|\Omega_\lambda\rangle=\frac{1}{(2\pi\hbar)^{N/2}}\,e^{-\frac{\lambda_{ij}}{4\hbar^2} p_i p_j}.
\label{eq:Omega}
\end{equation}
Consequently,
\begin{equation}
e^{-\frac{i}{\hbar}\vec{x}\cdot\hat{\vec{p}}}\,|\Omega_\lambda\rangle
=\int_{(\infty)}d^N\vec{p}\,e^{-\frac{i}{\hbar}\vec{x}\cdot\hat{\vec{p}}}\,|\vec{p}\,\rangle\,\langle\vec{p}\,|\Omega_\lambda\rangle = |\vec{x};\lambda\rangle,
\end{equation}
thus proving that the localised and normalised state $|\vec{x};\lambda\rangle$ is simply the Fock vacuum $|\Omega_\lambda\rangle$ displaced in
configuration space by the translation vector $\vec{x}$, with in particular the identification
\begin{equation}
|\Omega_\lambda\rangle = |\vec{x}=\vec{0};\lambda\rangle.
\end{equation}
The states $|\vec{x};\lambda\rangle$ are thus indeed a particular class of canonical quantum coherent states for the Fock algebra $(\hat{a}_i,\hat{a}^\dagger_i)$,
but specifically associated to configuration space alone and to the choice of symmetric definite positive matrix $\lambda_{ij}$.
Indeed given the complex parameters $(z_i,z^*_i)$ defined in (\ref{eq:zi}), the complete set of canonical quantum phase space coherent states associated
to that Fock algebra are constructed as
\begin{equation}
|z_i;\Omega_\lambda\rangle \equiv e^{z_i(\lambda^{-1})_{ij} \hat{a}^\dagger_j - z^*_i(\lambda^{-1})_{ij} \hat{a}_j}\,|\Omega_\lambda\rangle
=e^{-\frac{i}{\hbar}\left(x_i\hat{p}_i - p_i \hat{x}_i\right)}\,|\Omega_\lambda\rangle \equiv |(x_i,p_i);\Omega_\lambda\rangle,
\end{equation}
with thus in particular,
\begin{equation}
|\vec{x};\lambda\rangle=|z_i=x_i;\Omega_\lambda\rangle = |(x_i,p_i=0);\Omega_\lambda\rangle=e^{-\frac{i}{\hbar}x_i\hat{p}_i}\,|\Omega_\lambda\rangle,
\end{equation}
namely with $p_i=0$.

\subsection{The completeness relation of localised states and the $*_\lambda$-representation}
\label{Subsection.2.8}

Even though the localised states $|\vec{x};\lambda\rangle$ cannot provide a basis in quantum Hilbert space for having nonvanishing overlaps among themselves
and thus not being linearly independent, they do generate the entire quantum Hilbert space and are therefore overcomplete. Indeed, while one has the identity
\begin{equation}
\int_{(\infty)}d^N\vec{x}\,|\vec{x};\lambda\rangle \,\langle\vec{x};\lambda|
=\int_{(\infty)}d^N\vec{p}\,|\vec{p}\,\rangle\,e^{-\frac{\lambda_{ij}}{2\hbar^2}p_i p_j}\,\langle\vec{p}\,|
=e^{-\frac{\lambda_{ij}}{2\hbar^2}\,\hat{p}_i\hat{p}_j},
\end{equation}
which is not of much relevance as such except for the charateristic property of the states $|\vec{x};\lambda\rangle$ that it represents (namely a suppression
of their components at large momenta), given the $*_\lambda$-product defined in (\ref{eq:star-product}) one has the following actual completeness relation
this time in terms of the $*_\lambda$-product,
\begin{equation}
\int_{(\infty)}d^N\vec{x}\,|\vec{x};\lambda\rangle *_\lambda\,\langle\vec{x};\lambda|=\mathbb{I}.
\label{eq:complete1}
\end{equation}
Both these identifies may readily be established by computing the matrix elements of the operator on the l.h.s.~of each expression whether in the $|\vec{x}\,\rangle$
or in the $|\vec{p}\,\rangle$ basis, to check that the considered operator indeed provides yet another integral representation of the operator on the r.h.s.,
in particular in the second case the unit operator for the quantum Hilbert space
representation of the Heisenberg algebra generated by the operators $\mathbb{I}$, $\hat{x}_i$ and $\hat{p}_i$ ($i=1,2,\cdots,N$).

Consequently, besides the configuration and momentum space wave functions representations of any abstract quantum state $|\varphi\rangle$, 
namely $\varphi(\vec{x}\,)=\langle\vec{x}\,|\varphi\rangle$ and $\tilde{\varphi}(\vec{p}\,)=\langle\vec{p}\,|\varphi\rangle$, respectively,
a third type of overcomplete wave function representation in $\vec{x}$\,-space arises whatever the values of the spatial scale parameters $\lambda_{ij}\ne 0$,
namely the complex valued function of $\vec{x}$ defined by
$\langle\vec{x};\lambda|\varphi\rangle\equiv\varphi_\lambda(\vec{x}\,)$. Let us call it the $*_\lambda$-representation (in configuration space\footnote{Note that
this $*_\lambda$-representation is simply the ordinary canonical coherent state wave function representation of the Heisenberg algebra for the choice
of coefficients $\lambda_{ij}$, restricted however to configuration space as is warranted by the completeness relation (\ref{eq:complete1}) that applies thanks to
the existence of the $*_\lambda$-product and its specific properties.}) of the abstract quantum Hilbert space that defines a representation of the Heisenberg
algebra $(\hat{x}_i,\hat{p}_i,\mathbb{I})$.

More explicitly, in terms of the $*_\lambda$-product any abstract quantum state $|\varphi\rangle$ is decomposed as a linear combination of the localised
states $|\vec{x};\lambda\rangle$ according to the relation,
\begin{equation}
|\varphi\rangle=\int_{(\infty)}d^N\vec{x}\,|\vec{x};\lambda\rangle *_\lambda \varphi_\lambda(\vec{x}\,),\qquad
\varphi_\lambda(\vec{x}\,) \equiv \langle\vec{x};\lambda|\varphi\rangle.
\end{equation}
In turn the wave function $\varphi_\lambda(\vec{x}\,)$ possesses an integral representation in terms either of the configuration space or of the momentum
wave function of the state $|\varphi\rangle$,
\begin{equation}
\varphi_\lambda(\vec{x}\,)=\int_{(\infty)}\frac{d^N\vec{p}}{(2\pi\hbar)^{N/2}}
\,e^{\frac{i}{\hbar}\vec{x}\cdot\vec{p}}\,e^{-\frac{\lambda_{ij}}{4\hbar^2}p_i p_j}\,\tilde{\varphi}(\vec{p}\,)
=\int_{(\infty)}d^N\vec{x}\,'\,W_\lambda(\vec{x}-\vec{x}\,')\,\varphi(\vec{x}\,').
\end{equation}
Inverse relations are also available for $\varphi(\vec{x}\,)$ and $\tilde{\varphi}(\vec{p}\,)$ in terms of $\varphi_\lambda(\vec{x}\,)$,
but involving now the $*_\lambda$-product, in the form of,
\begin{equation}
\varphi(\vec{x}_1)=\int_{(\infty)}d^N\vec{x}\ W_\lambda(\vec{x}-\vec{x}_1) *_\lambda \varphi_\lambda(\vec{x}\,),
\end{equation}
as well as,
\begin{equation}
\tilde{\varphi}(\vec{p}\,)=\int_{(\infty)}\frac{d^N\vec{x}}{(2\pi\hbar)^{N/2}}\,e^{-\frac{i}{\hbar}\vec{x}\cdot\vec{p}}\,e^{-\frac{\lambda_{ij}}{4\hbar^2}p_i p_j}
*_\lambda \varphi_\lambda(\vec{x}\,).
\end{equation}

In contradistinction to the operators $|\vec{x}\,\rangle\langle\vec{x}\,|$ which are projection operators (more correctly, a projection operator density)
acting on quantum Hilbert space,
the operators $|\vec{x};\lambda\rangle *_\lambda \langle\vec{x};\lambda|$ are not such a projection operator density. Yet since they are the operator density of the
completeness relation integrated over configuration space, they arise continuously when working in the $*_\lambda$-representation,
in particular through their commutators with the position and momentum operators $\hat{x}_i$ and $\hat{p}_i$.
Thus one may also establish the following fundamental commutation relations,
beginning with the operators $\hat{x}_i$,
\begin{equation}
\Big[\,|\vec{x};\lambda\rangle *_\lambda\langle\vec{x};\lambda|\, , \, \hat{x}_i\,\Big] = 0 ,
\label{eq:com-1}
\end{equation}
next with the operators $\hat{p}_i$,
\begin{equation}
\Big[\,|\vec{x};\lambda\rangle *_\lambda\langle\vec{x};\lambda|\, , \, \hat{p}_i\,\Big]
=-i\hbar\frac{\partial}{\partial x_i}\Big[\,|\vec{x};\lambda\rangle *_\lambda \langle\vec{x};\lambda|\,\Big],
\label{eq:com-2.1}
\end{equation}
and with the operators $\hat{p}_i\hat{p}_j$,
\begin{eqnarray}
\Big[\,|\vec{x};\lambda\rangle *_\lambda\langle\vec{x};\lambda|\, , \, \hat{p}_i\hat{p}_j\,\Big] &=&
\ \ \ \ \left(-i\hbar\frac{\partial}{\partial x_j}\right)\left(\,|\vec{x};\lambda\rangle *_\lambda \left(-i\hbar\frac{\partial}{\partial x_i}\langle\vec{x};\lambda | \right)\,\right)
\nonumber \\
&& \,-\,\left(-i\hbar\frac{\partial}{\partial x_i}\right)\left(\,\left(-i\hbar\frac{\partial}{\partial x_j}|\vec{x};\lambda\rangle\right) *_\lambda \langle\vec{x};\lambda | \right),
\label{eq:com-2.2}
\end{eqnarray}
and finally with the operator $\hat{\vec{p}}\,^2$,
\begin{equation}
\Big[\,|\vec{x};\lambda\rangle *_\lambda\langle\vec{x};\lambda|\, , \, \hat{\vec{p}}\,^2\,\Big]
=-i\hbar\vec{\nabla}_{\vec{x}}\cdot\Big[\,|\vec{x};\lambda\rangle*_\lambda\left(-i\hbar\vec{\nabla}_{\vec{x}}\langle\vec{x};\lambda|\right)\,-\,
\left(-i\hbar\vec{\nabla}_{\vec{x}}|\vec{x};\lambda\rangle\right) *_\lambda\langle\vec{x};\lambda|\,\Big].
\label{eq:com-3}
\end{equation}

\subsection{Quantum probability conservation in the $*_\lambda$-representation}
\label{Subsection.2.9}

The abstract Schr\"odinger equation of ordinary quantum dynamics in the Schr\"odinger picture for a single nonrelativistic particle reads,
\begin{equation}
i\hbar\frac{d}{dt}|\psi,t\rangle = \hat{H}\,|\psi,t\rangle=\left[\frac{1}{2\mu}\hat{\vec{p}}\,^2+\hat{V}(\hat{\vec{x}}\,)\right]\,|\psi,t\rangle.
\end{equation}
Hence in the $|\vec{x}\,\rangle$-representation one has the ordinary Schr\"odinger wave equation for $\psi(\vec{x},t)\equiv \langle\vec{x}\,|\psi,t\rangle$,
\begin{equation}
i\hbar\frac{\partial}{\partial t}\psi(\vec{x},t)=\left[-\frac{\hbar^2}{2\mu}\vec{\nabla}^2_{\vec{x}} + V(\vec{x}\,)\right]\,\psi(\vec{x},t).
\end{equation}
Since quantum evolution is unitary (for $\hat{H}$ self-adjoint, or at least hermitian), the normalisation of any initial state is constant under time evolution,
reflecting quantum probability conservation. Thus for an initial state normalised to unity, one has,
\begin{equation}
1=\langle\psi,t|\psi,t\rangle=\int_{(\infty)}d^N\vec{x}\,\psi^*(\vec{x},t)\,\psi(\vec{x},t)=\int_{(\infty)}d^N\vec{x}\,\rho(\vec{x},t),\qquad
\rho(\vec{x},t) \equiv \psi^*(\vec{x},t)\,\psi(\vec{x},t),
\end{equation}
a property which indicates that the density $\rho(\vec{x},t)$ represents a probability density. As is well known, this probability density
obeys the following continuity equation for quantum probability conservation,
\begin{equation}
\frac{\partial}{\partial t}\rho + \vec{\nabla}_{\vec{x}}\cdot\vec{J}=0,\qquad
\vec{J}=\frac{\hbar}{2i\mu}\left(\psi^*\,(\vec{\nabla}_{\vec{x}}\,\psi)\,-\,(\vec{\nabla}_{\vec{x}}\,\psi^*) \,\psi\right),
\end{equation}
as may easily be established based on the above Schr\"odinger wave equation, while $\vec{J}(\vec{x},t)$ thus represents the probability current density.

Let us develop the same considerations now in the $*_\lambda$-representation, in which case the Schr\"odinger equation reads
as follows for the wave function $\psi_\lambda(\vec{x},t)\equiv \langle\vec{x};\lambda | \psi,t\rangle$,
\begin{equation}
i\hbar\frac{\partial}{\partial t}\psi_\lambda(\vec{x},t)=\langle\vec{x};\lambda|\left[\frac{1}{2\mu}\hat{\vec{p}}\,^2+\hat{V}(\hat{\vec{x}}\,)\right]|\psi,t\rangle.
\label{eq:Schro-star}
\end{equation}
Note how the r.h.s.~of this equation may be expressed in terms of the wave function $\psi_\lambda(\vec{x},t)$, a convolution integral,
the Weierstrass kernel $W_{2\lambda}(\vec{x}\,)$, and the $*_\lambda$-product. First insert to the left of $|\psi,t\rangle$
the completeness relation for the unit operator in terms of the $*_\lambda$-product and the states $|\vec{x};\lambda\rangle$.
And then use the matrix elements in (\ref{eq:matrix1}) and (\ref{eq:matrix6}) given in terms of the Weierstrass kernel $W_{2\lambda}(\vec{x}_1-\vec{x}_2)$.

Once again since the normalisation of any quantum state is conserved under quantum time evolution, when the initial quantum state $|\psi,t\rangle$ is normalised
to unity one has for $\psi_\lambda(\vec{x},t)$,
\begin{equation}
1=\langle\psi,t|\psi,t\rangle=\int_{(\infty)}d^N\vec{x}\,\psi^*_\lambda(\vec{x},t) *_\lambda \psi_\lambda(\vec{x},t)=\int_{(\infty)}d^N\vec{x}\,\rho_\lambda(\vec{x},t),
\end{equation}
a property which indicates that in the $*_\lambda$-representation the probability density in configuration space, $\rho_\lambda(\vec{x},t)$,
is given this time by the $*_\lambda$-product of the wave function $\psi_\lambda(\vec{x},t)$ with its complex conjugate,
\begin{equation}
\rho_\lambda(\vec{x},t) \equiv \langle\psi,t|\vec{x};\lambda\rangle *_\lambda \langle\vec{x};\lambda|\psi,t\rangle
=\psi^*_\lambda(\vec{x},t) *_\lambda \psi_\lambda(\vec{x},t).
\end{equation}
Using now the Schr\"odinger wave equation in the $*_\lambda$-representation (\ref{eq:Schro-star}), it readily follows that the time evolution
of this density is governed by the equation,
\begin{equation}
\frac{\partial}{\partial t}\rho_\lambda(\vec{x},t)
=-\frac{i}{\hbar}\langle\psi,t|\,\left[\,|\vec{x};\lambda\rangle *_\lambda\langle\vec{x};\lambda|\, , \, \hat{H}\,\right]\,|\psi,t\rangle.
\end{equation}
The detailed evaluation of the commutator $\left[|\vec{x};\lambda\rangle *_\lambda \langle\vec{x};\lambda|,\hat{H}\right]$ directly derives from (\ref{eq:com-1}-\ref{eq:com-3}).
This calculation leads to the following continuity equation for probability conservation in the $*_\lambda$-repre\-sen\-ta\-tion,
\begin{equation}
\frac{\partial}{\partial t}\rho_\lambda(\vec{x},t)+\vec{\nabla}_{\vec{x}}\cdot\vec{J}_\lambda(\vec{x},t)=0,
\end{equation}
with the following probability current density $\vec{J}_\lambda(\vec{x},t)$, itself also given in terms of the $*_\lambda$-product
of the wave function and its configuration space gradient in the $*_\lambda$-representation,
\begin{equation}
\vec{J}_\lambda(\vec{x},t)=\frac{\hbar}{2i\mu}\left(\psi^*_\lambda(\vec{x},t) *_\lambda(\vec{\nabla}_{\vec{x}}\,\psi_\lambda(\vec{x},t))\,-\,
(\vec{\nabla}_{\vec{x}}\,\psi^*_\lambda(\vec{x},t)) *_\lambda\psi_\lambda(\vec{x},t)\right).
\end{equation}

\subsection{An enveloping algebra representation in configuration space}
\label{Subsection.2.10}

With this last subsection dedicated to ordinary quantum mechanics, one last perspective complementary to the usual discussion
of representations of the Heisenberg algebra is to be addressed. Even though this perspective does not actually provide new useful insight into
that issue in the case of commuting configuration space coordinate operators, this discussion has its exact parallel in the context
of the noncommutative configuration spaces to be considered hereafter where it does indeed provide new powerful and useful insight.

Given an abstract quantum Hilbert space representation of the Heisenberg algebra, quantum states of the physical system are represented by abstract
states in Hilbert space, $|\varphi\rangle$. By identifying a basis or an overcomplete basis in Hilbert space, as are the states $|\vec{x}\,\rangle$,
$|\vec{p}\,\rangle$ or $|\vec{x};\lambda\rangle$, the abstract states are represented by complex quantities that are functions of the discrete or
continuous labels that distinguish the basis or overcomplete basis vectors, {\sl viz}, the wave functions $\varphi(\vec{x}\,)$, $\tilde{\varphi}(\vec{p}\,)$
or $\varphi_\lambda(\vec{x}\,)$ considered above.

The construction to be described presently provides a representation of physical quantum states not in terms of vectors in the abstract Hilbert space, {\sl i.e.},
the abstract states $|\varphi\rangle$, but rather in terms of operators that belong to the enveloping algebra
of the configuration space coordinate operators $\hat{x}_i$ ($i=1,2,\cdots,N)$, namely all the composite operators that may be built out of the $\hat{x}_i$'s,
{\sl i.e.}, $\hat{\varphi}(\hat{\vec{x}}\,)$. To any given abstract state $|\varphi\rangle$ there corresponds a unique element $\hat{\varphi}(\hat{\vec{x}}\,)$
of that enveloping algebra, and vice-versa, This correspondence defines a one-to-one map between states in the abstract quantum Hilbert space
and operators in the enveloping algebra of the $\hat{x}_i$'s that act on that same quantum Hilbert space which spans all possible abstract quantum physical states
of the considered physical system.

The construction relies on the existence of the momentum eigenbasis $|\vec{p}\,\rangle$ of the commuting conjugate momentum operators $\hat{p}_i$
($i=1,2\cdots,N$). When the position operators $\hat{x}_i$ ($i=1,2,\cdots,N$)  all commute, as is the case in this subsection, the construction also goes
through by using the position eigenbasis $|\vec{x}\,\rangle$. However, when configuration space is noncommutative only the conjugate momentum path
remains available with commuting momentum operators $\hat{p}_i$, as is to be discussed when considering noncommutative configuration spaces hereafter.
This is the reason why the conjugate momentum path is taken here as well.

The existence of such a one-to-one map follows from the representations of momentum eigenstates expressed in (\ref{eq:p-int}) and (\ref{eq:star-p-int}),
\begin{equation}
|\vec{p}\,\rangle=\int_{(\infty)}\frac{d^N\vec{x}}{(2\pi\hbar)^{N/2}}\,e^{\frac{i}{\hbar}\vec{p}\cdot\hat{\vec{x}}}\,|\vec{x};\lambda\rangle,\qquad
|\vec{p}\,\rangle=\int_{(\infty)}\frac{d^N\vec{x}}{(2\pi\hbar)^{N/2}}\,e^{\frac{i}{\hbar}\vec{p}\cdot\hat{\vec{x}}}\,|\vec{x}\,\rangle.
\end{equation}
Note that in particular for $\vec{p}=\vec{0}$ these relations imply,
\begin{equation}
\int_{(\infty)}d^N\vec{x}\,|\vec{x};\lambda\rangle = (2\pi\hbar)^{N/2}\,|\vec{p}=\vec{0}\rangle,\qquad
\int_{(\infty)}d^N\vec{x}\,|\vec{x}\,\rangle = (2\pi\hbar)^{N/2}\,|\vec{p}=\vec{0}\rangle.
\end{equation}
Hence any given abstract quantum state $|\varphi\rangle$ in Hilbert space may be represented according to,
with $\tilde{\varphi}(\vec{p}\,)=\langle\vec{p}\,|\varphi\rangle$,
\begin{eqnarray}
|\varphi\rangle &=& \int_{(\infty)}d^N\vec{p}\,|\vec{p}\,\rangle\,\tilde{\varphi}(\vec{p}\,)
=\int_{(\infty)}d^N\vec{p}\int_{(\infty)}\frac{d^N\vec{x}}{(2\pi\hbar)^{N/2}}\,e^{\frac{i}{\hbar}\vec{p}\cdot\hat{\vec{x}}}\,|\vec{x};\lambda\rangle\,\tilde{\varphi}(\vec{p}\,)
\nonumber \\
&=&\int_{(\infty)}d^N\vec{x}\left(\int_{(\infty)}\frac{d^N\vec{p}}{(2\pi\hbar)^{N/2}}\,e^{\frac{i}{\hbar}\vec{p}\cdot\hat{\vec{x}}}\,\tilde{\varphi}(\vec{p}\,)\right)
\,|\vec{x};\lambda\rangle
=\int_{(\infty)}d^N\vec{x}\,\hat{\varphi}(\hat{\vec{x}}\,)\,|\vec{x};\lambda\rangle,
\end{eqnarray}
or equivalently, through a series of identical manipulations in the case that one uses the above second representation of $|\vec{p}\,\rangle$ with $\lambda_{ij}=0$,
\begin{equation}
|\varphi\rangle = \int_{(\infty)}d^N\vec{x}\,\hat{\varphi}(\hat{\vec{x}}\,)\,|\vec{x}\,\rangle,
\label{eq:corres2}
\end{equation}
where in each case the operator $\hat{\varphi}(\hat{\vec{x}}\,)$ is defined by
\begin{equation}
\hat{\varphi}(\hat{\vec{x}}\,) \equiv \int_{(\infty)}\frac{d^N\vec{p}}{(2\pi\hbar)^{N/2}}\,e^{\frac{i}{\hbar}\hat{\vec{x}}\cdot\vec{p}}\,\tilde{\varphi}(\vec{p}\,)
=\int_{(\infty)}\frac{d^N\vec{p}}{(2\pi\hbar)^{N/2}}\,e^{\frac{i}{\hbar}\vec{p}\cdot\hat{\vec{x}}}\,\langle\vec{p}\,|\varphi\rangle.
\end{equation}
In other words, any abstract quantum state $|\varphi\rangle$ defines a unique operator $\hat{\varphi}(\hat{\vec{x}}\,)$ in the enveloping algebra
of the coordinate operators $\hat{x}_i$ through the latter correspondence. And conversely, any operator $\hat{\varphi}(\hat{\vec{x}}\,)$ in that
enveloping algebra defines a unique abstract quantum state $|\varphi\rangle$ through
\begin{equation}
|\varphi\rangle=\int_{(\infty)}d^N\vec{x}\,\hat{\varphi}(\hat{\vec{x}}\,)\,|\vec{x};\lambda\rangle
=\int_{(\infty)}d^N\vec{x}\,\hat{\varphi}(\hat{\vec{x}}\,)\,|\vec{x}\,\rangle=(2\pi\hbar)^{N/2}\,\hat{\varphi}(\hat{\vec{x}}\,)\,|\vec{p}=\vec{0}\,\rangle.
\end{equation}
Note that when the coordinate operators all commute among themselves the position eigenbasis $|\vec{x}\,\rangle$ exists, as is the case in the present subsection,
so that one has right away for any such operator in the enveloping algebra,
\begin{equation}
\hat{\varphi}(\hat{\vec{x}}\,)\,|\vec{x}\,\rangle=\varphi(\vec{x}\,)|\vec{x}\,\rangle=|\vec{x}\,\rangle\,\langle\vec{x}\,|\varphi\rangle,
\end{equation}
hence leading once again to the result that,
\begin{equation}
\int_{(\infty)}d^N\vec{x}\,\hat{\varphi}(\hat{\vec{x}}\,)\,|\vec{x}\,\rangle=\int_{(\infty)}d^N\vec{x}\,|\vec{x}\,\rangle\,\langle\vec{x}\,|\varphi\rangle=|\varphi\rangle.
\end{equation}
There also exists the following alternative representation of the operator $\hat{\varphi}(\hat{\vec{x}}\,)$ associated to the quantum state $|\varphi\rangle$,
valid again when the coordinates $\hat{x}_i$ ($i=1,2,\cdots,N$) are all commuting among themselves,
\begin{equation}
\hat{\varphi}(\hat{\vec{x}}\,)=\int_{(\infty)}d^N\vec{x}_1\,d^N\vec{x}_2\,|\vec{x}_1\rangle\langle\vec{x}_1|\hat{\varphi}(\hat{\vec{x}}\,)|\vec{x}_2\rangle\langle\vec{x}_2|
=\int_{(\infty)}d^N\vec{x}\,|\vec{x}\,\rangle\,\varphi(\vec{x}\,)\,\langle\vec{x}\,|.
\end{equation}
Under such a circumstance this representation of $\hat{\varphi}(\hat{\vec{x}}\,)$ could be a most natural alternative definition of it as well.

The enveloping algebra representation of quantum states in terms of the operators $\hat{\varphi}(\hat{\vec{x}}\,)$ is also equipped with a sesquilinear positive definite
inner product as is required of a quantum mechanical representation of the abstract Heisenberg algebra. This inner product over the enveloping algebra of
the position operators $\hat{x}_i$ is simply identified from the abstract inner product $\langle\varphi_1|\varphi_2\rangle$ of any two abstract quantum states
$|\varphi_1\rangle$ and $|\varphi_2\rangle$, using the above one-to-one map between $|\varphi\rangle$ and $\hat{\varphi}(\hat{\vec{x}}\,)$ involving
the states $|\vec{x};\lambda\rangle$, namely,
\begin{equation}
\langle\varphi_1|\varphi_2\rangle = \int_{(\infty)}d^N\vec{x}_1\,d^N\vec{x}_2\,\langle\vec{x}_1;\lambda|\hat{\varphi}^\dagger_1(\hat{\vec{x}}\,)\,
\hat{\varphi}_2(\hat{\vec{x}}\,)|\vec{x}_2;\lambda\rangle.
\label{eq:inner}
\end{equation}
It is obvious that the r.h.s.~of this relation defines a sesquilinear inner product over the enveloping algebra, which is furthermore positive definite since
\begin{equation}
\int_{(\infty)}d^N\vec{x}_1\,d^N\vec{x}_2\,\langle\vec{x}_1;\lambda|\,\hat{\varphi}^\dagger(\hat{\vec{x}}\,)\,\hat{\varphi}(\hat{\vec{x}}\,)\,|\vec{x}_2;\lambda\rangle
= \langle\varphi|\varphi\rangle \ge 0,
\end{equation}
for any operator $\hat{\varphi}(\hat{\vec{x}}\,)$ and the corresponding abstract quantum state $|\varphi\rangle$.

By considering rather the one-to-one map (\ref{eq:corres2}) between $|\varphi\rangle$ and $\hat{\varphi}(\hat{\vec{x}}\,)$
involving this time the sharp position eigenstates $|\vec{x}\,\rangle$, one realises that the relation (\ref{eq:inner}) extends to the localised and normalisable
states $|\vec{x};\lambda\rangle$ the usual definition of the inner product for the abstract Hilbert space representation, since,
\begin{eqnarray}
\int_{(\infty)}d^N\vec{x}_1\,d^N\vec{x}_2\,\langle\vec{x}_1|\hat{\varphi}^\dagger_1(\hat{\vec{x}}\,)\,\hat{\varphi}_2(\hat{\vec{x}}\,)|\vec{x}_2\rangle
&=& \int_{(\infty)}d^N\vec{x}_1\,d^N\vec{x}_2\,\varphi^*_1(\vec{x}_1)\,\varphi_2(\vec{x}_2)\,\langle\vec{x}_1|\vec{x}_2\rangle \nonumber \\
&=& \int_{(\infty)}d^N\vec{x}\,\varphi^*_1(\vec{x}\,)\,\varphi_2(\vec{x}\,) = \langle\varphi_1 | \varphi_2\rangle.
\end{eqnarray}

Now that the one-to-one map between the abstract quantum Hilbert space and the enveloping algebra of the coordinate operators has been identified,
one may enquire how the action on that Hilbert space of the phase space operators $(\hat{x}_i,\hat{p}_i)$ ($i=1,2,\cdots,N$) is
represented as linear operators acting on that enveloping algebra, to be denoted $(\hat{X}_i,\hat{P}_i$) ($i=1,2,\cdots,N$), respectively.
Working once again in the momentum eigenbasis the action of the operators $(\hat{x}_i,\hat{p}_i)$ on Hilbert space is represented as, respectively,
\begin{equation}
\langle\vec{p}\,|\hat{x}_i|\varphi\rangle=i\hbar\frac{\partial}{\partial p_i}\,\tilde{\varphi}(\vec{p}\,),\qquad
\langle\vec{p}\,|\hat{p}_i|\varphi\rangle=p_i\,\tilde{\varphi}(\vec{p}\,).
\end{equation}
Hence under the map to the enveloping algebra, these operators are now realised as,
\begin{equation}
\hat{X}_i\left(\hat{\varphi}(\hat{\vec{x}}\,)\right)=\int_{(\infty)}\frac{d^N\vec{p}}{(2\pi\hbar)^{N/2}}\,e^{\frac{i}{\hbar}\hat{\vec{x}}\cdot\vec{p}}
\,i\hbar\frac{\partial}{\partial p_i}\tilde{\varphi}(\vec{p}\,)=\hat{x}_i\,\hat{\varphi}(\hat{\vec{x}}\,),
\label{eq:X-1}
\end{equation}
\begin{equation}
\hat{P}_i\left(\hat{\varphi}(\hat{\vec{x}}\,)\right)=\int_{(\infty)}\frac{d^N\vec{p}}{(2\pi\hbar)^{N/2}}\,e^{\frac{i}{\hbar}\hat{\vec{x}}\cdot\vec{p}}
p_i \tilde{\varphi}(\vec{p}\,)=-i\hbar\frac{\partial}{\partial\hat{x}_i}\,\hat{\varphi}(\hat{\vec{x}}\,).
\label{eq:P-1}
\end{equation}
Note that since the coordinate operators $\hat{x}_i$ all commute with one another, whether the multiplication by $\hat{x}_i$ or the
derivative relative to $\hat{x}_i$ is taken from the left or from the right is irrelevant in the present situation. No ordering issue arises.
In particular, one thus has indeed as operators acting on the considered enveloping algebra the following commutation relations,
\begin{equation}
\left[\hat{X}_i,\hat{X}_j\right]=0,\qquad
\left[\hat{X}_i,\hat{P}_j\right]=i\hbar\,\delta_{ij}\,\mathbb{I}_q,\qquad
\left[\hat{P}_i,\hat{P}_j\right]=0,
\end{equation}
which is indeed once again the Heisenberg algebra defining the quantised dynamical system (with $\mathbb{I}_q$ standing for the identity operator
on the enveloping algebra). Note as well that for the inner product over the enveloping algebra of
the coordinate operators $\hat{x}_i$ defined in (\ref{eq:inner}), the operators $\hat{X}_i$ and $\hat{P}_i$ are indeed at least hermitean if not
formally self-adjoint, as it should be.

In particular given now the enveloping algebra representation of the abstract quantum system, the Schr\"odinger equation then reads,
\begin{equation}
i\hbar\frac{\partial}{\partial t}\hat{\psi}(\hat{\vec{x}},t)=\hat{H}\left(\hat{\vec{x}},-i\hbar\vec{\nabla}_{\hat{\vec{x}}}\right)\,\hat{\psi}(\hat{\vec{x}},t),
\end{equation}
with for instance,
\begin{equation}
\hat{H}\left(\hat{\vec{x}}, -i\hbar\vec{\nabla}_{\hat{\vec{x}}}\right) = -\frac{\hbar^2}{2\mu}\vec{\nabla}_{\hat{\vec{x}}}\cdot\vec{\nabla}_{\hat{\vec{x}}}\,
+\,\hat{V}(\hat{\vec{x}}\,).
\end{equation}
Clearly, whether working in terms of the commuting coordinate operators $\hat{x}_i$ and the enveloping algebra operators $\hat{\psi}(\hat{\vec{x}},t)$,
or in terms of their eigenvalues $x_i$ in the position eigenbasis $|\vec{x}\,\rangle$ and thus the wave functions $\psi(\vec{x},t)$, is totally irrelevant
from the algebraic point of view. The same differential equations ensue. One is simply dealing once more with different realisations of
a same and unique abstract Hilbert space in which the same abstract Heisenberg algebra is represented.

When extending hereafter the discussion now to noncommutative configuration spaces of a certain general class, all the above considerations
will find their rightful place in an entirely parallel analysis, based once again on a Hamiltonian formulation and its canonical quantisation.

\section{A Conjugate Magnetic Field in Phase Space}
\label{Sect3}

\subsection{Classical dynamics considerations}

Let us now deform the action of the system given in the form of (\ref{eq:action1}) together with its implicit symplectic structure by adding to it a real non-exact
1-form in phase space, specifically restricted to the conjugate momentum sector of phase space coordinates $p_i$ in the form of a local vector field
$\bar{A}_i(p_i)$ ($i=1,2,\cdots,N$) and the 1-form $dp_i \bar{A}_i(\vec{p}\,)$. One then has a new Hamiltonian first-order action which reads,
\begin{equation}
S_1[x_i,p_i]=\int dt\left(\dot{\vec{x}}\cdot\vec{p}+\dot{\vec{p}}\cdot\vec{\bar{A}}(\vec{p}\,)-H(\vec{x},\vec{p}\,)\right)
=\int dt\left(\dot{x}_i p_i + \dot{p}_i\bar{A}_i(p_i)-H(x_i,p_i)\right).
\label{eq:action2}
\end{equation}
The Hamiltonian equations of motion are then deformed into,
\begin{equation}
\dot{x}_i=\frac{\partial H(\vec{x},\vec{p}\,)}{\partial p_i}+\bar{A}_{ij}(\vec{p}\,)\frac{\partial H(\vec{x},\vec{p}\,)}{\partial x_j},\qquad
\dot{p}_i=-\frac{\partial H(\vec{x},\vec{p}\,)}{\partial x_i},\qquad i,j=1,2,\cdots,N,
\end{equation}
where $\bar{A}_{ij}(\vec{p}\,)$ is the nonvanishing exterior derivative or curl of the vector field $\bar{A}_i(\vec{p}\,)$ in momentum space,
\begin{equation}
\bar{A}_{ij}(\vec{p}\,)=\frac{\partial \bar{A}_j(\vec{p}\,)}{\partial p_i}\,-\,\frac{\partial \bar{A}_i(\vec{p}\,)}{\partial p_j},\qquad i,j=1,2,\cdots,N.
\end{equation}

Consequently, in presence of the non integrable vector field $\bar{A}_i(\vec{p}\,)$ the symplectic structure of phase space is deformed in such a way that 
the phase space parametrisation in terms of the variables $(x_i,p_i)$ now possesses the following Poisson brackets,
\begin{equation}
\left\{x_i,x_j\right\}=\bar{A}_{ij}(\vec{p}\,),\qquad
\left\{x_i,p_j\right\}=\delta_{ij},\qquad
\left\{p_i,p_j\right\}=0,\qquad i,j=1,2,\cdots,N.
\end{equation}
Note how the Poisson brackets of the configuration space coordinates are no longer vanishing. For the canonically quantised system,
its configuration space coordinate operators have become noncommutative, with $\left[\hat{x}_i,\hat{x}_j\right]=i\hbar\,\hat{\bar{A}}_{ij}(\hat{\vec{p}}\,)$.

Clearly, by analogy with what happens for genuine physical magnetic fields and their interactions in configuration space, the vector field $\bar{A}_i(\vec{p}\,)$
may be given the interpretation of a conjugate magnetic vector potential with its conjugate magnetic field $\bar{A}_{ij}(\vec{p}\,)$ present in the conjugate momentum
sector of phase space. Thus, any nontrivial conjugate magnetic field in conjugate momentum space induces noncommutativity in configuration space.

This interpretation, which resonates with a physicist's intuition, is corroborated as well by the following fact.
Any dynamical action is defined up to a total derivative in time, without affecting its equations of motion.
By redefining the vector field through what is in effect a local gauge transformation in conjugate momentum space,
\begin{equation}
\bar{A}_i(\vec{p}\,) \longrightarrow \bar{A}_i(\vec{p}\,) + \frac{\partial\chi(\vec{p}\,)}{\partial p_i},
\end{equation}
where $\chi(\vec{p}\,)$ is an arbitrary function over conjugate momentum space, the action (\ref{eq:action2}) sees its value modified by a total time derivative
contribution, thereby not affecting the equations of motion. Incidentally, this observation also explains why the equations of motion must involve
the field strength $\bar{A}_{ij}(\vec{p}\,)$ rather than directly the vector potential $\bar{A}_i(\vec{p}\,)$,
namely the conjugate magnetic field only. Indeed, this magnetic field is the local gauge invariant field that may be
constructed out of the vector potential $\bar{A}_i(\vec{p}\,)$.

Even though such gauge transformations do not modify the classical equations of motion, they may affect the identification of canonically conjugate
phase space coordinates, with still further more subtle consequences for the canonically quantised system in terms of induced passive unitary
transformations of the unitary representation of the Heisenberg algebra used to identify the actual quantum physical system \cite{Govaerts1}.
Since no such transformation is to be contemplated in this work, we shall not have to concern ourselves with these subtle issues here, even though
they could be of interest in their own right.

Finally, the following phase space reparametrisation in terms of the new set of variables $(y_i,p_i)$ ($i=1,2,\cdots,N$) defined as,
\begin{equation}
y_i=x_i-\bar{A}_i(\vec{p}\,),\qquad p_i=p_i,\qquad i=1,2,\cdots,N,
\end{equation}
consists in a Darboux transformation to canonically conjugate phase space coordinates, with the then canonical Poisson brackets,
\begin{equation}
\left\{y_i,y_j\right\}=0,\qquad
\left\{y_i,p_j\right\}=\delta_{ij},\qquad
\left\{p_i,p_j\right\}=0,\qquad i,j=1,2,\cdots,N.
\end{equation}
Note that in contradistinction to $(x_i,p_i)$ that are invariant under the above gauge transformations of $A_i(\vec{p}\,)$, the coordinates $y_i$ canonically conjugate
to $p_i$ are not gauge invariant under such transformations \cite{Govaerts1}.

\subsection{Quantum dynamics considerations}

The canonically quantised system is thus characterised by the (equal time) commutation relations,
\begin{equation}
\left[\hat{x}_i,\hat{x}_j\right]=i\hbar\,\hat{\bar{A}}_{ij}(\hat{\vec{p}}\,),\qquad
\left[\hat{x}_i,\hat{p}_j\right]=i\hbar\,\delta_{ij}\,\mathbb{I},\qquad
\left[\hat{p}_i,\hat{p}_j\right]=0,\qquad i,j=1,2,\cdots,N,
\end{equation}
or equivalently,
\begin{equation}
\left[\hat{y}_i,\hat{y}_j\right]=0,\qquad
\left[\hat{y}_i,\hat{p}_j\right]=i\hbar\,\delta_{ij}\,\mathbb{I},\qquad
\left[\hat{p}_i,\hat{p}_j\right]=0,\qquad i,j=1,2,\cdots,N,
\end{equation}
where
\begin{equation}
\hat{x}_i=\hat{y}_i+\hat{\bar{A}}_i(\hat{\vec{p}}\,),\qquad i=1,2,\cdots,N.
\end{equation}
Hence the operators $(\hat{y}_i,\hat{p}_i,\mathbb{I})$ ($i=1,2,\cdots,N$) generate a Heisenberg algebra, of which an abstract Hilbert space representation
provides an abstract Hilbert space representation of the algebra of the original operators $(\hat{x}_i,\hat{p}_i,\mathbb{I})$ ($i=1,2,\cdots,N$),
which include the operators $\hat{x}_i$
that are identified with the cartesian coordinates of the noncommutative configuration space. Note that all the operators $\hat{x}_i$, $\hat{p}_i$
and $\hat{y}_i$ need to be self-adjoint (or at least hermitian) as well.

Since all conjugate momentum operators $\hat{p}_i$ commute among themselves, one may choose as basis of Hilbert space all sharp momentum eigenstates
$|\vec{p}\,\rangle$, with precisely all the same properties and conventions as discussed already in the absence of the vector potential $\vec{\bar{A}}(\vec{p}\,)$.
Hence in particular in that $|\vec{p}\,\rangle$-representation, we have for the operators $\hat{y}_i$ and $\hat{x}_i$, given any abstract quantum state
$|\varphi\rangle$ and its momentum space wave function representation $\tilde{\varphi}(\vec{p}\,)=\langle\vec{p}\,|\varphi\rangle$,
\begin{equation}
\langle\vec{p}\,|\hat{y}_i|\varphi\rangle=i\hbar\frac{\partial}{\partial p_i}\tilde{\varphi}(\vec{p}\,),\qquad
\langle\vec{p}\,|\hat{x}_i|\varphi\rangle=\left(i\hbar\frac{\partial}{\partial p_i}+\bar{A}_i(\vec{p}\,)\right)\,\tilde{\varphi}(\vec{p}\,),\qquad i=1,2,\cdots,N.
\end{equation}
In the last of these two expressions of course one recognises the covariant derivative for the conjugate vector potential $\vec{\bar{A}}(\vec{p}\,)$
in conjugate momentum space.

However, since the configuration space coordinate operators $\hat{x}_i$ do not commute among themselves, there does not exist a basis of sharp position eigenstates
for the complete set of these operators, that could have been labelled by all values of $x_i\in\mathbb{R}$. Nevertheless, there exists a basis of
eigenstates $|\vec{y}\,\rangle$ for the twisted configuration space operators $\hat{y}_i$, since these operators all commute with one another,
$\hat{y}_i |\vec{y}\,\rangle =y_i |\vec{y}\,\rangle$. The states $|\vec{y}\,\rangle$ possess all the same properties
as the position eigenstates $|\vec{x}\,\rangle$ considered above in the commutative case, with for instance the change of basis matrix elements,
\begin{equation}
\langle\vec{p}\,|\vec{y}\,\rangle=\frac{1}{(2\pi\hbar)^{N/2}}\,e^{-\frac{i}{\hbar}\vec{y}\cdot\vec{p}}.
\end{equation}
The existence of the basis $|\vec{y}\,\rangle$ and different available representations for the twisted configuration space operators $\hat{y}_i$
may sometimes be put to good use in intermediate steps of calculations for a noncommutative configuration space.
For example as an illustration, one has,
\begin{equation}
\hat{x}_i = \int_{(\infty)}d^N\vec{p}\,|\vec{p}\,\rangle\left(\frac{1}{2}i\hbar\left(\frac{\stackrel{\rightarrow}{\partial}}{\partial p_i} -
\frac{\stackrel{\leftarrow}{\partial}}{\partial p_i} +\bar{A}_i(\vec{p}\,)\right)\right) \langle\vec{p}\,| ,\qquad i=1,2,\cdots,N.
\end{equation}

By analogy with the previous construction of localised but not sharp quantum states associated to positions $\vec{x}$ in configuration space
in the commutative case, one could contemplate as well now in the present noncommutative context the study of localised quantum states
in configuration space labelled by all points $\vec{x}$ in configuration space, in the form of,
\begin{equation}
\langle\vec{p}\,|\vec{x};B\rangle \equiv \frac{1}{(2\pi\hbar)^{N/2}}\,e^{-\frac{i}{\hbar}\vec{x}\cdot\vec{p}}\,e^{-B(\vec{p}\,)},\qquad
|\vec{x};B\rangle=\int_{(\infty)}d^N\vec{p}\,|\vec{p}\,\rangle \langle\vec{p}\,|\vec{x};B\rangle,
\end{equation}
where $B(\vec{p}\,)$ is some specifically chosen function of $\vec{p}$ such that $\lim_{|\vec{p}\,|\rightarrow\infty}B(\vec{p}\,)=+\infty$,
so that these states would be normalisable and with an exponential suppression in their large momentum components.
Possible choices could be once again $B(\vec{p}\,)=\lambda_{ij} p_i p_j/(4\hbar^2)$ with $\lambda_{ij}$ a symmetric positive definite matrix of real spatial
scale factors, or also for instance $B(\vec{p}\,)=\alpha\vec{\bar{A}}^2(\vec{p}\,)/2$ or $B(\vec{p}\,)=\beta \bar{A}_{ij}(\vec{p}\,)\bar{A}_{ij}(\vec{p}\,)/4$
with $\alpha,\beta>0$ and of the proper physical dimension, provided of course these fields grow infinite for infinite $|\vec{p}\,|$.

Some properties valid for the states $|\vec{x};B\rangle$ analogous to those presented in the commutative case for the states $|\vec{x};\lambda\rangle$
may then be established here as well, but not to the same degree of completeness nor generality
unless the choice for $B(\vec{p}\,)$ is quadratic in $\vec{p}$ and the conjugate magnetic field $\bar{A}_{ij}(\vec{p}\,)$
is constant. Indeed, when $\bar{A}_{ij}(\vec{p}\,)$ is independent of $\vec{p}$, the nonvanishing noncommutativity of the configuration space
operators $\hat{x}_i$ is then decoupled from the conjugate momentum operators $\hat{p}_i$, which themselves commute among one another
and have constant Heisenberg-like commutators with the $\hat{x}_i$. Rather than try to pursue a general analysis for an arbitrary choice of
vector potential $\vec{\bar{A}}(\vec{p}\,)$ with its conjugate magnetic field $\bar{A}_{ij}(\vec{p}\,)$, the remainder of the discussion henceforth
focusses now on the specific case of a constant conjugate magnetic field $\bar{A}_{ij}(\vec{p}\,)$.

\section{A Uniform Conjugate Magnetic Field in Phase Space}
\label{Sect4}

\subsection{A canonical quantum representation}

From now on, let us thus assume that the conjugate magnetic field $\bar{A}_{ij}(\vec{p}\,)$ is constant over phase space, $\bar{A}_{ij}(\vec{p}\,)=\bar{A}_{ij}$
with $\bar{A}_{ji}=-\bar{A}_{ij}$ being the constant entries of a real antisymmetric matrix. A convenient choice of gauge for the vector potential $\bar{A}_i(\vec{p}\,)$
is then that of the so-called symmetric gauge with
\begin{equation}
\bar{A}_i(\vec{p}\,)=-\frac{1}{2}\bar{A}_{ij}\,p_j.
\end{equation}
For ease of notation for the quantised system, let us also introduce the quantities $A_i(\vec{p}\,)$ and $A_{ij}$ in the form of,
\begin{equation}
A_{ij}=\hbar\,\bar{A}_{ij},\qquad
A_i(\vec{p}\,)=\hbar\,\bar{A}_i(\vec{p}\,)=-\frac{1}{2}A_{ij}\,p_j,\qquad
\bar{A}_i(\vec{p}\,)=\frac{1}{\hbar} A_i(\vec{p}\,)=-\frac{1}{2\hbar}\,A_{ij}\,p_j .
\end{equation}
Consequently the quantised system is now characterised by the commutation relations
\begin{equation}
\left[\hat{x}_i,\hat{x}_j\right]=i\,A_{ij}\,\mathbb{I},\qquad \left[\hat{x}_i,\hat{p}_j\right]=i\hbar\,\delta_{ij}\,\mathbb{I},\qquad
\left[\hat{p}_i,\hat{p}_j\right]=0,
\end{equation}
with in particular noncommutative configuration space coordinate operators of constant commutation relations specified through the real antisymmetric matrix $A_{ij}$.

The canonical parametrisation of classical phase space in terms of the variables $(y_i,p_i)$ is then provided by the following changes of variables
\begin{equation}
y_i=x_i+\frac{1}{2\hbar}\,A_{ij}\,p_j,\qquad
x_i=y_i-\frac{1}{2\hbar}\,A_{ij}\,p_j,
\end{equation}
while for the canonically quantised system one has the following operator redefinitions
\begin{equation}
\hat{y}_i=\hat{x}_i-\hat{A}_i(\hat{\vec{p}}\,)=\hat{x}_i+\frac{1}{2\hbar} A_{ij}\,\hat{p}_j,\qquad
\hat{x}_i=\hat{y}_i-\frac{1}{2\hbar}A_{ij}\,\hat{p}_j.
\end{equation}
A direct evaluation then finds that these operators $(\hat{y}_i,\hat{p}_i,\mathbb{I})$ indeed generate a Heisenberg algebra for $N$-dimensional Euclidean space,
\begin{equation}
\left[\hat{y}_i,\hat{y}_j\right]=0,\qquad
\left[\hat{y}_i,\hat{p}_j\right]=i\hbar\,\delta_{ij}\,\mathbb{I},\qquad
\left[\hat{p}_i,\hat{p}_j\right]=0.
\end{equation}

Since all conjugate momenta operators $\hat{p}_i$ do commute among themselves, the $|\vec{p}\,\rangle$\,-represen\-ta\-tion certainly still exists
with its basis of sharp and non-normalisable momentum eigenstates $|\vec{p}\,\rangle$, and all the same properties as those recalled above in the
commutative configuration space case. Thus abstract quantum states $|\varphi\rangle$ may be represented in terms of their momentum wave
function $\tilde{\varphi}(\vec{p}\,)=\langle\vec{p}\,|\varphi\rangle$, while the operators $\hat{y}_i$, $\hat{x}_i$ and $\hat{p}_i$ then have the following
differential operator representations,
\begin{equation}
\langle\vec{p}\,|\hat{y}_i|\varphi\rangle=i\hbar\frac{\partial}{\partial p_i}\,\tilde{\varphi}(\vec{p}\,),\qquad
\langle\vec{p}\,|\hat{x}_i|\varphi\rangle=\left(i\hbar\frac{\partial}{\partial p_i}\,-\,\frac{1}{2\hbar}A_{ij}\,p_j\right)\tilde{\varphi}(\vec{p}\,),\qquad
\langle\vec{p}\,|\hat{p}_i|\varphi\rangle = p_i\,\tilde{\varphi}(\vec{p}\,).
\end{equation}

\subsection{Configuration space localised and normalisable quantum states}

In the same way as in the previous discussion of the commutative Euclidean configuration space, let us reconsider the length-squared scale factors defined by some
real positive definite symmetric matrix of entries $\lambda_{ij}$, and introduce once again the localised and normalisable quantum states
$|\vec{x};\lambda\rangle$ labelled by all possible vector values for $\vec{x}$ parametrising now the noncommutative configuration space.
These states coincide with those in the commutative case, and are defined again in the form of,
\begin{equation}
\langle\vec{p}\,|\vec{x};\lambda\rangle=\frac{1}{(2\pi\hbar)^{N/2}}\,e^{-\frac{i}{\hbar}\vec{x}\cdot\vec{p}}\,e^{-\frac{\lambda_{ij}}{4\hbar^2}\,p_i p_j},\qquad
|\vec{x};\lambda\rangle=\int_{(\infty)}d^N\vec{p}\,|\vec{p}\,\rangle\,\langle\vec{p}\,|\vec{x};\lambda\rangle.
\end{equation}

Related to this choice of states, one has again the same definition of the $W_\lambda(\vec{x}\,)$ and $W_{2\lambda}(\vec{x}\,)$ Weierstrass kernels
with all their same properties as discussed previously already, say in the case of $W_\lambda(\vec{x}\,)$,
\begin{equation}
W_\lambda(\vec{x}\,)=\int_{(\infty)}\frac{d^N\vec{p}}{(2\pi\hbar)^N}\,e^{\frac{i}{\hbar}\vec{x}\cdot\vec{p}}\,e^{-\frac{\lambda_{ij}}{4\hbar^2}\,p_i p_j}
=\frac{1}{\left(\pi^N{\rm det}\,\lambda_{ij}\right)^{1/2}}\,e^{-x_i(\lambda^{-1})_{ij} x_j},
\end{equation}
and the same definition of a $*_\lambda$-product acting on pairs of functions of configuration space para\-me\-tri\-sed by $\vec{x}$,
\begin{equation}
*_\lambda \equiv e^{\frac{1}{2}\stackrel{\leftarrow}{\frac{\partial}{\partial x_i}}\,\lambda_{ij}\,\stackrel{\rightarrow}{\frac{\partial}{\partial x_j}}}.
\end{equation}
Hence for instance one finds again for the finite normalisation of these states $|\vec{x};\lambda\rangle$,
\begin{equation}
\langle\vec{x};\lambda|\vec{x};\lambda\rangle=\frac{1}{\left((2\pi)^N{\rm det}\,\lambda_{ij}\right)^{1/2}}=W_{2\lambda}(\vec{x}=\vec{0}\,).
\end{equation}

Furthermore, note that like in the commutative case, it is possible to extend the ensemble of states $|\vec{x};\lambda\rangle=|x_i;\lambda\rangle$
with real values for the parameters $x_i$, to the larger set of states $|z_i;\lambda\rangle$ with $z_i\in\mathbb{C}$ defined in the present noncommutative case
exactly as in (\ref{eq:zi0}).

More specifically, as a matter of fact it should be remarked that the states $|\vec{x};\lambda\rangle$ and $|z_i;\lambda\rangle$ introduced here in the
case of the noncommutative configuration space coincide exactly still with those introduced above in the case of a commutative configuration space,
then in the absence of the conjugate magnetic field.

In contradistinction however, the expectation values for these states of operators involving the configuration space coordinate operators $\hat{x}_i$
(rather than $\hat{y}_i$) take different values now, on account of additional contributions due to the conjugate magnetic field $A_{ij}$.
Thus ready evaluations find the following results, for $\hat{x}_i$,
\begin{equation}
\langle\hat{x}_i\rangle_\lambda(\vec{x}\,)=x_i,
\end{equation}
and for $\hat{x}_i\,\hat{x}_j$,
\begin{equation}
\langle\hat{x}_i\hat{x}_j\rangle_\lambda(\vec{x}\,)=x_i x_j + \frac{1}{4}\lambda_{ij}+\frac{1}{2}iA_{ij}-\frac{1}{4}A_{ik}(\lambda^{-1})_{k\ell} A_{\ell j}.
\end{equation}
On the other hand, operators only involving the momentum operators $\hat{p}_i$ retain the same expectation values as in the absence of the conjugate
magnetic field, such as $\hat{p}_i$ and $\hat{p}_i\,\hat{p}_j$,
\begin{equation}
\langle\hat{p}_i\rangle_\lambda(\vec{x}\,)=0,\qquad
\langle\hat{p}_i \hat{p}_j\rangle_\lambda(\vec{x}\,)=\hbar^2\,(\lambda^{-1})_{ij}.
\end{equation}

As a consequence, in the noncommutative case the uncertainties in position and momentum for the states $|\vec{x};\lambda\rangle$ now read, respectively,
\begin{equation}
\Delta^{(\hat{\vec{x}})}_{ij}(\vec{x}\,)
\equiv \langle\left(\hat{x}_i-\langle\hat{x}_i\rangle_\lambda\right)\left(\hat{x}_j-\langle\hat{x}_j\rangle_\lambda\right)\rangle_\lambda(\vec {x}\,)
= \frac{1}{4}\lambda_{ij}+\frac{1}{2}iA_{ij}-\frac{1}{4}A_{ik}(\lambda^{-1})_{k\ell} A_{\ell j},
\end{equation}
\begin{equation}
\Delta^{(\hat{\vec{p}}\,)}_{ij}(\vec{x}\,)
\equiv \langle\left(\hat{p}_i-\langle\hat{p}_i\rangle_\lambda\right)\left(\hat{p}_j-\langle\hat{p}_j\rangle_\lambda\right)\rangle_\lambda(\vec {x}\,)
=\hbar^2\,(\lambda^{-1})_{ij},
\end{equation}
thus leading to the following uncertainly relations,
\begin{equation}
\Delta^{(\hat{\vec{x}})}_{ij}(\vec{x}\,)\,\Delta^{(\hat{\vec{p}}\,)}_{k\ell}(\vec{x}\,)
=\frac{1}{4}\hbar^2\,(\lambda^{-1})_{k\ell}\left(\lambda_{ij}+2iA_{ij}-A_{im}(\lambda^{-1})_{mn} A_{nj}\right).
\end{equation}
Hence once again the states $|\vec{x};\lambda\rangle$ are saturating states, in the character of quantum coherent states.

For completeness, let us also list matrix elements of some operators involving $\hat{x}_i$ and $\hat{p}_i$ for two distinct external states belonging to the class
$|\vec{x};\lambda\rangle$, beginning with the overlaps of such states,
\begin{equation}
\langle\vec{x}_1;\lambda | \vec{x}_2;\lambda\rangle = W_{2\lambda}(\vec{x}_1 - \vec{x}_2).
\end{equation}
For the operators $\hat{x}_i$ and $\hat{x}_i \hat{x}_j$ one finds, respectively,
\begin{equation}
\langle\vec{x}_1;\lambda | \hat{x}_i | \vec{x}_2;\lambda\rangle
= \left(\frac{1}{2}(x_1+x_2)_i-\frac{1}{2}i A_{ij} (\lambda^{-1})_{jk} (x_1 - x_2)_k\right)\,W_{2\lambda}(\vec{x}_1 - \vec{x}_2),
\end{equation}
and,
\begin{eqnarray}
&& \langle\vec{x}_1;\lambda | \hat{x}_i \hat{x}_j | \vec{x}_2;\lambda\rangle = \nonumber \\
&=& \left\{\frac{1}{4}\left[\left(x_1+x_2\right)_i-i\left(A\cdot \lambda^{-1}\right)_{ik}\left(x_1-x_2\right)_k\right]
\left[\left(x_1+x_2\right)_j-i\left(A\cdot \lambda^{-1}\right)_{j\ell}\left(x_1-x_2\right)_\ell\right] \right. \nonumber \\
&& \left. \ \ \ + \frac{1}{4}\lambda_{ij}+\frac{1}{2}i A_{ij}-\frac{1}{4}\left(A\cdot \lambda^{-1}\cdot A\right)_{ij}\right\}\,W_{2\lambda}(\vec{x}_1 - \vec{x}_2).
\end{eqnarray}
While for operators linear and quadratic in $\hat{p}_i$,
\begin{equation}
\langle\vec{x}_1;\lambda | \hat{p}_i | \vec{x}_2;\lambda\rangle = -i\hbar \frac{\partial}{\partial x_{1i}}\,W_{2\lambda}(\vec{x}_1 - \vec{x}_2),
\end{equation}
\begin{equation}
\langle\vec{x}_1;\lambda | \hat{p}_i \hat{p}_j | \vec{x}_2;\lambda\rangle
= \left(-i\hbar \frac{\partial}{\partial x_{1i}}\right)\left(-i\hbar\frac{\partial}{\partial x_{1j}}\right)\,W_{2\lambda}(\vec{x}_1 - \vec{x}_2),
\end{equation}
and finally,
\begin{equation}
\langle\vec{x}_1;\lambda | \hat{\vec{p}}\,^2 | \vec{x}_2;\lambda\rangle
= -\hbar^2\,\vec{\nabla}^2_{\vec{x}_1}\,W_{2\lambda}(\vec{x}_1 - \vec{x}_2).
\end{equation}

\subsection{Displacement operators and canonical quantum coherent states}

Since the commutation relations of the momentum operators $\hat{p}_i$ with either the configuration space or momentum operators,
$\hat{x}_i$ and $\hat{p}_i$, respectively, have not been modified in the noncommutative configuration space case as compared to the commutative one,
the status of the momentum operators $\hat{p}_i$ as generators of displacements in configuration space through the unitary transformations
$e^{-\frac{i}{\hbar}\vec{a}\cdot\hat{\vec{p}}}$ has not be modified either. One still has of course,
\begin{equation}
e^{\frac{i}{\hbar}\vec{a}\cdot\hat{\vec{p}}}\,\hat{x}_i\,e^{-\frac{i}{\hbar}\vec{a}\cdot\hat{\vec{p}}}=\hat{x}_i\,+\,a_i\,\mathbb{I},\qquad
e^{\frac{i}{\hbar}\vec{a}\cdot\hat{\vec{p}}}\,\hat{p}_i\,e^{-\frac{i}{\hbar}\vec{a}\cdot\hat{\vec{p}}}=\hat{p}_i.
\end{equation}
When acting with that operator on the states $|\vec{x};\lambda\rangle$ and $|\vec{p}\,\rangle$ and by inserting then to the left of it the completeness relation
of the unit operator in the $|\vec{p}\,\rangle$-representation, one readily finds again for the states $|\vec{x};\lambda\rangle$ and $|\vec{p}\,\rangle$,
\begin{equation}
e^{-\frac{i}{\hbar}\vec{a}\cdot\hat{\vec{p}}}\,|\vec{x};\lambda\rangle = |\vec{x}+\vec{a};\lambda\rangle,\qquad
e^{-\frac{i}{\hbar}\vec{a}\cdot\hat{\vec{p}}}\,|\vec{p}\,\rangle=e^{-\frac{i}{\hbar}\vec{a}\cdot\vec{p}}\,|\vec{p}\,\rangle.
\end{equation}
In particular, note the following simple realisation of any state $|\vec{x};\lambda\rangle$ in terms of the displacement of the state at the origin of configuration space,
namely $|\vec{x}=\vec{0};\lambda\rangle$,
\begin{equation}
|\vec{x};\lambda\rangle=e^{-\frac{i}{\hbar}\vec{x}\cdot\hat{\vec{p}}}\,|\vec{x}=\vec{0};\lambda\rangle.
\end{equation}
We shall come back to this observation from the perspective of quantum coherent states once again.

In contradistinction, the situation regarding the position operators $\hat{x}_i$ is certainly much different, on account of the now nonvanishing
commutation relations of these operators among themselves determined by the conjugate magnetic field $A_{ij}$. In other words, which types of transformation
in phase space, and of quantum states do the unitary operators $e^{\frac{i}{\hbar}\vec{p}_0\cdot\hat{\vec{x}}}$ generate in the noncommutative case?

Using the BCH formula and the relevant commutation relations, it is straightforward to establish that one has,
\begin{equation}
e^{-\frac{i}{\hbar}\vec{p}_0\cdot\hat{\vec{x}}}\,\hat{x}_i\,e^{\frac{i}{\hbar}\vec{p}_0\cdot\hat{\vec{x}}} = \hat{x}_i\,-\,\frac{1}{\hbar} A_{ij}\,p_{0j}\,\mathbb{I},\qquad
e^{-\frac{i}{\hbar}\vec{p}_0\cdot\hat{\vec{x}}}\,\hat{p}_i\,e^{\frac{i}{\hbar}\vec{p}_0\cdot\hat{\vec{x}}} = \hat{p}_i\,+\,p_{0i}\,\mathbb{I},
\label{eq:corres}
\end{equation}
thus showing that the operators $e^{\frac{i}{\hbar}\vec{p}_0\cdot\hat{\vec{x}}}$ not only generate displacements in momentum space as they do in the commutative
case, but that because of the noncommutativity of the $\hat{x}_i$ operators among themselves, the same unitary transformations do also generate as well displacements in configuration space by a quantity dependent both on the conjugate magnetic field and the transformation parameters $p_{0i}$.
In other words, in the case of configuration space noncommutativity, displacements in configuration space may be generated by the coordinate operators as well,
and not only by the momentum operators as is usually the case. This specific feature is to be explored further hereafter.

However, determining now the action of these unitary operators on the quantum states $|\vec{x};\lambda\rangle$ and $|\vec{p}\,\rangle$ is another matter
altogether. And this even by attempting to use the completeness relation of the unit operator whether in the $|\vec{p}\,\rangle$ or in the $|\vec{y}\,\rangle$ eigenbases,
that correspond to the canonical phase space parametrisation in terms of the variables $(y_i,p_i)$. Another path is to be followed.

First, by considering in the $|\vec{p}\,\rangle$-representation the action of the operators $\hat{x}_i$ and $\hat{p}_i$ on the states $|\vec{x};\lambda\rangle$
one finds the representations,
\begin{equation}
\hat{x}_i\,|\vec{x};\lambda\rangle = x_i\,|\vec{x};\lambda\rangle\,-\,\frac{i}{2\hbar}\left(\lambda_{ij} - i A_{ij}\right)\,\hat{p}_j\,|\vec{x};\lambda\rangle,\qquad
\hat{p}_i\,|\vec{x};\lambda\rangle = i\hbar\frac{\partial}{\partial x_i}\,|\vec{x};\lambda\rangle.
\end{equation}
Hence one may write for the action of $\hat{x}_i$ on the states $|\vec{x};\lambda\rangle$,
\begin{equation}
\hat{x}_i\,|\vec{x};\lambda\rangle = \left(x_i+\frac{1}{2}\left(\lambda_{ij}-i A_{ij}\right)\frac{\partial}{\partial x_j}\right)\,|\vec{x};\lambda\rangle,
\end{equation}
which includes an explicit contribution of the conjugate magnetic field $A_{ij}$. On the other hand, since one also observes that\footnote{Note that
$\hat{x}_i+\frac{i}{2\hbar}\left(\lambda_{ij}-i A_{ij}\right)\,\hat{p}_j=\hat{y}_i+\frac{i}{2\hbar}\lambda_{ij}\,\hat{p}_j$, to be compared to the Fock
algebra operators $(a_i,a^\dagger_i)$ introduced in (\ref{eq:Fock1}) in the commutative case.}
\begin{equation}
\left(\hat{x}_i+\frac{i}{2\hbar}\left(\lambda_{ij}-i A_{ij}\right)\,\hat{p}_j\right)\,|\vec{x};\lambda\rangle = x_i\,|\vec{x};\lambda\rangle,
\end{equation}
it follows that the states $|\vec{x};\lambda\rangle$ are eigenstates with eigenvalues $x_i$ of the momentum twisted coordinate operators now defined by
\begin{equation}
\hat{x}_i+\frac{i}{2\hbar}\left(\lambda_{ij}-i A_{ij}\right)\,\hat{p}_j,
\end{equation}
thus extending in the presence of the conjugate magnetic field $A_{ij}$ a parallel result already observed for the same states $|\vec{x};\lambda\rangle$ in the case
of a commutative configuration space.

Consequently, one finds that
\begin{equation}
e^{\frac{i}{\hbar}p_{0i}\left(\hat{x}_i+\frac{i}{2\hbar}\left(\lambda_{ij}-i A_{ij}\right)\,\hat{p}_j\right)}\,|\vec{x};\lambda\rangle
= e^{\frac{i}{\hbar}\vec{p}_0\cdot\vec{x}}\,|\vec{x};\lambda\rangle.
\end{equation}
Using once again the BCH formula in the l.h.s.~of this identity, at this intermediary step the action of the unitary operator
$e^{\frac{i}{\hbar}\vec{p}_0\cdot\hat{\vec{x}}}$ on the states $|\vec{x};\lambda\rangle$ may thus be represented in the form of,
\begin{equation}
e^{\frac{i}{\hbar}p_{0i}\hat{x}_i}\,|\vec{x};\lambda\rangle = e^{\frac{i}{\hbar}\vec{p}_0\cdot\vec{x}}\,e^{-\frac{\lambda_{ij}}{4\hbar^2} p_{0i} p_{0j}}\,
e^{\frac{1}{2\hbar^2}\left(\lambda_{ij}-i A_{ij}\right) p_{0i} \hat{p}_j}\,|\vec{x};\lambda\rangle.
\label{eq:intermediate1}
\end{equation}
One first conclusion that may be drawn from this result is the value of the following matrix element,
\begin{equation}
\langle\vec{p}\,| e^{\frac{i}{\hbar}\vec{p}_0\cdot\hat{\vec{x}}} |\vec{x};\lambda\rangle = e^{-\frac{i}{2\hbar^2} A_{ij} p_{0i} p_j}\,
\langle\vec{p}-\vec{p}_0|\vec{x};\lambda\rangle.
\end{equation}
This expression confirms that the unitary operator $e^{\frac{i}{\hbar}\vec{p}_0\cdot\hat{\vec{x}}}$ indeed generates a displacement in momentum space
as already established above in operator form.

Furthermore from the knowledge of the result for the action of $e^{-\frac{i}{\hbar}\vec{a}\cdot\hat{\vec{p}}}$ on the states $|\vec{x};\lambda\rangle$,
in terms of the states $|z_i\in\mathbb{C};\lambda\rangle$ the result in (\ref{eq:intermediate1}) may also be given in the form of,
\begin{equation}
e^{\frac{i}{\hbar}p_{0i}\hat{x}_i}\,|\vec{x};\lambda\rangle = e^{\frac{i}{\hbar}\vec{p}_0\cdot\vec{x}}\,e^{-\frac{\lambda_{ij}}{4\hbar^2} p_{0i} p_{0j}}\,
\Big|z_i=x_i-\frac{1}{2\hbar}A_{ij}\,p_{0j}+\frac{i}{2\hbar} \lambda_{ij}\, p_{0j};\lambda\Big\rangle.
\end{equation}
Or else in the following form still in terms of the states $|\vec{x};\lambda\rangle$,
\begin{equation}
e^{\frac{i}{\hbar}p_{0i}\hat{x}_i}\,|\vec{x};\lambda\rangle = e^{\frac{i}{\hbar}\vec{p}_0\cdot\vec{x}}\,e^{-\frac{\lambda_{ij}}{4\hbar^2} p_{0i} p_{0j}}\,
e^{\frac{\lambda_{ij}}{2\hbar^2} p_{0i} \hat{p}_j}\,\Big|x_i-\frac{1}{2\hbar} A_{ij}\, p_{0j};\lambda\Big\rangle,
\end{equation}
thus showing that $e^{\frac{i}{\hbar}p_{0i}\hat{x}_i}$ indeed generates as well displacements in configuration space. Alternatively by using the fact that
$|\vec{x};\lambda\rangle=e^{\frac{i}{\hbar}\vec{a}\cdot\hat{\vec{p}}}\,|\vec{x}+\vec{a};\lambda\rangle$ and that the momentum operators $\hat{p}_i$
all commute with one another, one also establishes that
\begin{equation}
e^{\frac{i}{\hbar}p_{0i}\hat{x}_i}\,|\vec{x};\lambda\rangle = e^{\frac{i}{\hbar}\vec{p}_0\cdot\vec{x}}\,e^{-\frac{\lambda_{ij}}{4\hbar^2} p_{0i} p_{0j}}\,
e^{\frac{1}{2\hbar^2}\left(\lambda_{ij}+i A_{ij}\right) p_{0i} \hat{p}_j}\,\Big| x_i -\frac{1}{\hbar}\,A_{ij}\,p_{0j};\lambda \Big\rangle,
\end{equation}
where the choice of $|\vec{x};\lambda\rangle$ state on the r.h.s.~of this identity is motivated by the result in (\ref{eq:corres}) for the transformation
under the operator $e^{\frac{i}{\hbar}\vec{p}_0\cdot\hat{\vec{x}}}$ of the coordinate operators $\hat{x}_i$.

By inserting to the left of (\ref{eq:intermediate1}) the completeness relation for the unit operator in the $|\vec{p}\,\rangle$ eigenbasis,
another conclusion that may be drawn from that relation is the following representation of momentum eigenstates in terms of the localised
and normalisable states $|\vec{x};\lambda\rangle$,
\begin{equation}
\int_{(\infty)}d^N\vec{x}\,e^{\frac{i}{\hbar}\vec{p}_0\cdot\hat{\vec{x}}}\,|\vec{x};\lambda\rangle = \left(2\pi\hbar\right)^{N/2}\,|\vec{p}_0\rangle,\qquad
|\vec{p}\,\rangle=\int_{(\infty)}\frac{d^N\vec{x}}{(2\pi\hbar)^{N/2}}\,e^{\frac{i}{\hbar}\vec{p}\cdot\hat{\vec{x}}}\,|\vec{x};\lambda\rangle.
\label{eq:p-identity}
\end{equation}
Hence finally based on this result as well as the identity (\ref{eq:intermediate1}), one finds for the action
of the operators $e^{\frac{i}{\hbar}\vec{p}_0\cdot{\hat{\vec{x}}}}$ on the momentum eigenstates,
\begin{equation}
e^{\frac{i}{\hbar}\vec{p}_0\cdot\hat{\vec{x}}}\,|\vec{p}\,\rangle = e^{-\frac{i}{2\hbar^2} A_{ij} p_{0i} p_j}\,|\vec{p}+\vec{p}_0\rangle.
\end{equation}
It is to be pointed out that this result is confirmed by inserting to the left of the l.h.s.~of this identity the completeness relation for the unit operator
in the $|\vec{y}\,\rangle$ eigenbasis of the coordinate twisted operators $\hat{y}_i$ and then using the values for the matrix elements $\langle\vec{y}\,|\vec{p}\,\rangle$.

In the case that the configuration space is even dimensional with $N=2n$ and $n\ge 1$, and that the real antisymmetric matrix is invertible
(and thus with no vanishing eigenvalue), it is also possible to express the results for the action of the exponentiated coordinate operators $\hat{x}_i$
on the states $|\vec{x};\lambda\rangle$ in a form which makes more explicit their character as generators of displacements in noncommutative
configuration space. For this purpose it suffices to consider the correspondence (see (\ref{eq:corres})),
\begin{equation}
a_i=-\frac{1}{\hbar} A_{ij}\,p_{0j},\qquad
p_{0i}=-\hbar\,(A^{-1})_{ij}\,a_j=\hbar\,a_j\left(A^{-1}\right)_{ji}.
\end{equation}
One thus finds for the transformation of the phase space operators,
\begin{equation}
e^{-ia_i(A^{-1})_{ij}\hat{x}_j}\,\hat{x}_i\,e^{ia_i(A^{-1})_{ij}\hat{x}_j} = \hat{x}_i + a_i\,\mathbb{I},\qquad
e^{-ia_i(A^{-1})_{ij}\hat{x}_j}\,\hat{p}_i\,e^{ia_i(A^{-1})_{ij}\hat{x}_j} = \hat{p}_i - \hbar (A^{-1})_{ij} a_j\,\mathbb{I},
\end{equation}
while for the quantum states $|\vec{x};\lambda\rangle$, by using the fact that
$|\vec{x};\lambda\rangle=e^{\frac{i}{\hbar}\vec{a}\cdot\hat{\vec{p}}}\,|\vec{x}+\vec{a};\lambda\rangle$,
\begin{equation}
e^{i a_i (A^{-1})_{ij} \hat{x}_j}\,|\vec{x};\lambda\rangle = 
e^{i a_i(A^{-1})_{ij} x_j}\,e^{-\frac{\lambda_{ij}}{4}(A^{-1})_{ik} a_k (A^{-1})_{j\ell} a_\ell}\,
e^{\frac{i}{2\hbar}a_i\left(\delta_{ij}-i(A^{-1})_{ik}\lambda_{kj}\right)\hat{p}_j}\,|\vec{x}+\vec{a};\lambda\rangle,
\end{equation}
and finally for the momentum eigenstates $|\vec{p}\,\rangle$,
\begin{equation}
e^{i a_i (A^{-1})_{ij} \hat{x}_j}\,|\vec{p}\,\rangle = e^{-\frac{i}{2\hbar} a_i p_i}\,\Big|p_i-\hbar\left(A^{-1}\right)_{ij}\,a_j\Big\rangle.
\end{equation}

Let us now come back to the issue of the states $|\vec{x};\lambda\rangle$ as quantum coherent states, along lines parallel to those discussed
in the case of a commutative configuration space, and this again independently of whether $N$ is even or odd.
Thus consider the momentum twisted coordinate operators defined this time by,
\begin{equation}
\hat{A}_i=\hat{x}_i+\frac{i}{2\hbar}\left(\lambda_{ij}-iA_{ij}\right)\,\hat{p}_j,\qquad
\hat{A}^\dagger_i=\hat{x}_i-\frac{i}{2\hbar}\left(\lambda_{ij}+iA_{ij}\right)\,\hat{p}_j,
\label{eq:Fock-2}
\end{equation}
jointly with the associated complex valued phase space variables,
\begin{equation}
Z_i=x_i+\frac{i}{2\hbar}\left(\lambda_{ij}-iA_{ij}\right)\,p_j,\qquad
Z^*_i=x_i-\frac{i}{2\hbar}\left(\lambda_{ij}+i A_{ij}\right)\,p_j.
\label{eq:Zi}
\end{equation}
By construction the states $|\vec{x};\lambda\rangle$ are eigenstates with eigenvalues $x_i$ of the operators $\hat{A}_i$,
\begin{equation}
\hat{A}_i\,|\vec{x};\lambda\rangle = x_i\,|\vec{x};\lambda\rangle.
\end{equation}
Note as well that one has,
\begin{equation}
\hat{A}_i=\hat{y}_i+\frac{i}{2\hbar}\,\lambda_{ij}\,\hat{p}_j,\qquad
\hat{A}^\dagger_i=\hat{y}_i-\frac{i}{2\hbar}\,\lambda_{ij}\,\hat{p}_j,
\end{equation}
together with,
\begin{equation}
Z_i=y_i+\frac{i}{2\hbar}\,\lambda_{ij}\,p_j,\qquad
Z^*_i=y_i-\frac{i}{2\hbar}\,\lambda_{ij}\,p_j,\qquad
y_i=x_i+\frac{1}{2\hbar}\,A_{ij}\,p_j.
\end{equation}

A direct evaluation finds that the operators $(\hat{A}_i,\hat{A}^\dagger_i)$ again span a $N$ dimensional Fock algebra since the matrix $\lambda_{ij}$
is real, symmetric and positive definite, with the commutation relations,
\begin{equation}
\Big[\,\hat{A}_i, \hat{A}_j\,\Big]=0,\qquad
\Big[ \,\hat{A}_i, \hat{A}^\dagger_j\,\Big]=\lambda_{ij}\,\mathbb{I},\qquad
\Big[\,\hat{A}^\dagger_i, \hat{A}^\dagger_j\,\Big]=0.
\end{equation}
By substitution one also checks that the displacement operator in phase space for the associated canonical quantum coherent states is given as,
\begin{equation}
Z_i\left(\lambda^{-1}\right)_{ij}\,\hat{A}^\dagger_j\,-\,Z^*_i\left(\lambda^{-1}\right)_{ij}\,\hat{A}_j
= -\frac{i}{\hbar} x_i \hat{p}_i + \frac{i}{\hbar} p_i \hat{x}_i + \frac{i}{\hbar^2} p_i\,A_{ij}\,\hat{p}_j
= -\frac{i}{\hbar} y_i \hat{p}_i + \frac{i}{\hbar} p_i \hat{y}_i,
\end{equation}
as may have been expected for its  very last expression in terms of the momentum twisted operators $\hat{y}_i$ and their eigenvalues $y_i$.

Consider then the associated Fock vacuum $|\Omega_\lambda\rangle$ again with a non standard choice of normalisation, such that,
\begin{equation}
\hat{A}_i\,| \Omega_\lambda \rangle = 0,\qquad
\langle \Omega_\lambda \,|\,\Omega_\lambda\rangle = \frac{1}{\left((2\pi)^N\,{\rm det}\,\lambda_{ij}\right)^{1/2}}.
\end{equation}
Since in view of the above expression for $\hat{A}_i$ in terms of $\hat{y}_i$, the $|\vec{p}\,\rangle$ momentum space wave function of this Fock vacuum
$|\Omega_\lambda\rangle$ is such that,
\begin{equation}
\left(i\hbar\frac{\partial}{\partial p_i}+\frac{i}{2\hbar}\,\lambda_{ij}\,p_j\right)\,\langle\vec{p}\,|\Omega_\lambda\rangle=0,
\end{equation}
this wave function $\langle\vec{p}\,|\Omega_\lambda\rangle$ is identical to that of the Fock vacuum $|\Omega_\lambda\rangle$ introduced earlier already
in (\ref{eq:Omega}) in the case of a commutative configuration space. Hence these two Fock states are a same and identical quantum state irrespective
of the conjugate magnetic field $A_{ij}$, which justifies the same notation used here for both without any risk of confusion. And one has as well that
$|\Omega_\lambda\rangle=|\vec{x}=\vec{0};\lambda\rangle$.

The canonical phase space coherent states related to the Fock algebra $(\hat{A}_i,\hat{A}^\dagger_i)$ are thus constructed as,
\begin{eqnarray}
\Big| Z_i;\Omega_\lambda \Big\rangle & \equiv &
e^{Z_i\left(\lambda^{-1}\right)_{ij}\,\hat{A}^\dagger_j\,-\,Z^*_i\left(\lambda^{-1}\right)_{ij}\,\hat{A}_j}\,\Big| \Omega_\lambda \Big\rangle \nonumber \\
&=&  e^{-\frac{i}{\hbar}\left(x_i \hat{p}_i - p_i \hat{x}_i\right)+\frac{i}{\hbar^2} p_i A_{ij} \hat{p}_j}\,\Big| \Omega_\lambda \Big\rangle
\equiv \Big| (x_i,p_i;A_{ij});\Omega_\lambda \Big\rangle,
\end{eqnarray}
and are the eigenstates of the Fock operators $\hat{A}_i$ with complex eigenvalues $Z_i$,
\begin{equation}
\hat{A}_i\,\Big| Z_i;\Omega_\lambda \Big\rangle = Z_i\,\Big| Z_i;\Omega_\lambda \Big\rangle,\qquad
Z_i=y_i+\frac{i}{2\hbar}\,\lambda_{ij}\,p_j = x_i +\frac{i}{2\hbar}\left(\lambda_{ij}-i A_{ij}\right)\,p_j.
\end{equation}
Hence the localised and normalisable quantum states $|\vec{x};\lambda\rangle$ belong to the ensemble of such canonical coherent states
for the $(\hat{A}_i,\hat{A}^\dagger_i)$ Fock algebra, restricted however solely to configuration space with $Z_i=x_i$ or $p_i=0$, namely,
\begin{equation}
|\vec{x};\lambda\rangle = \Big| (x_i,p_i=0;A_{ij});\Omega_\lambda \Big\rangle = \Big| Z_i=x_i;\Omega_\lambda \Big\rangle.
\end{equation}

With the obvious identification now that
\begin{equation}
|\Omega_\lambda\rangle=|\vec{x}=\vec{0};\lambda\rangle,
\end{equation}
one thus also recovers the general relation observed earlier already, namely,
\begin{equation}
|\vec{x};\lambda\rangle = e^{-\frac{i}{\hbar}\vec{x}\cdot\hat{\vec{p}}}\,|\Omega_\lambda\rangle
= e^{-\frac{i}{\hbar}\vec{x}\cdot\hat{\vec{p}}}\,|\vec{x}=\vec{0};\lambda\rangle.
\end{equation}

Based on this observation, together with the facts that the momentum operators $\hat{p}_i$ and Fock vacua $|\Omega_\lambda\rangle$ for the Fock algebras
$(\hat{a}_i,\hat{a}^\dagger_i)$ and $(\hat{A}_i,\hat{A}^\dagger_i)$ remain identical irrespective of whether configuration space is commutative or noncommutative,
one also concludes that the entire ensemble of localised and normalisable quantum states $|\vec{x};\lambda\rangle$ labelled by all points $\vec{x}$ in configuration
space are as well the same and identical quantum states irrespective of the value of the conjugate magnetic field $A_{ij}$, for any given choice
of gaussian length squared scale parameters $\lambda_{ij}$.

\subsection{The $*_\lambda$-representation}

Consequently, the completeness relation (\ref{eq:complete1}) established in section~\ref{Subsection.2.8} remains valid, namely
\begin{equation}
\int_{(\infty)}d^N\vec{x}\,|\vec{x};\lambda\rangle *_\lambda \langle\vec{x};\lambda|=\mathbb{I},
\end{equation}
thereby allowing for the $*_\lambda$-representation of any abstract quantum state $|\varphi\rangle$ in terms of a $*_\lambda$-wave function
$\varphi_\lambda(\vec{x}\,)$, such that,
\begin{equation}
|\varphi\rangle = \int_{(\infty)}d^N\vec{x}\,|\vec{x};\lambda\rangle *_\lambda \varphi_\lambda(\vec{x}\,),\qquad
\varphi_\lambda(\vec{x}\,)=\langle\vec{x};\lambda|\varphi\rangle,
\end{equation}
while this $*_\lambda$-wave function $\varphi_\lambda(\vec{x}\,)$ and the momentum wave function $\tilde{\varphi}(\vec{p}\,)$
of the same state are transformed into one another according to the following expressions involving the $*_\lambda$-product,
\begin{equation}
\varphi_\lambda(\vec{x}\,)=\int_{(\infty)}\frac{d^N\vec{p}}{(2\pi\hbar)^{N/2}}\,e^{-\frac{i}{\hbar}\vec{x}\cdot\vec{p}}\,
e^{-\frac{\lambda_{ij}}{4\hbar^2} p_i p_j} \tilde{\varphi}(\vec{p}\,),
\end{equation}
and
\begin{equation}
\tilde{\varphi}(\vec{p}\,)=\langle\vec{p}\,|\varphi\rangle = \int_{(\infty)}\frac{d^N\vec{x}}{(2\pi\hbar)^{N/2}}\,
e^{-\frac{i}{\hbar}\vec{x}\cdot\vec{p}}\,e^{-\frac{\lambda_{ij}}{4\hbar^2} p_i p_j}*_\lambda \varphi_\lambda(\vec{x}\,).
\end{equation}

All other similar considerations as they have already been discussed in sections~\ref{Subsection.2.8} and~\ref{Subsection.2.9} in the commutative
configuration space case apply here as well without modification, except of course for those that involve the now noncommuting configuration space
coordinates $\hat{x}_i$, and the configuration space wave functions $\varphi(\vec{x}\,)$ which only exist in the commutative case. Thus for instance
the commutators of the operator $|\vec{x};\lambda\rangle *_\lambda \langle\vec{x};\lambda|$ with the operators $\hat{p}_i$, $\hat{p}_i \hat{p}_j$,
and $\hat{\vec{p}}\,^2$ remain as given in (\ref{eq:com-2.1}), (\ref{eq:com-2.2}) and (\ref{eq:com-3}), respectively.

However, the commutators of that same operator $|\vec{x};\lambda\rangle *_\lambda \langle\vec{x};\lambda|$
with the coordinate operators $\hat{x}_i$ are now dependent on the conjugate magnetic field $A_{ij}$, with values given according to,
\begin{equation}
\Big[\,|\vec{x};\lambda\rangle *_\lambda \langle\vec{x};\lambda| , \hat{x}_i \,\Big] = \frac{1}{2}i A_{ij}\frac{\partial}{\partial x_j}\Big(\,|\vec{x};\lambda\rangle *_\lambda \langle\vec{x};\lambda|\,\Big).
\end{equation}
This important distinctive feature of the noncommutative situation thus impacts, for instance, the continuity equation for the probability current density
as expressed in the $*_\lambda$-representation.

Consider the abstract time dependent Schr\"odinger equation,
\begin{equation}
i\hbar\frac{d}{dt}|\psi,t\rangle = \hat{H}\,|\psi,t\rangle,\qquad
-i\hbar\langle\psi,t| \frac{\stackrel{\leftarrow}{d}}{dt} = \langle\psi,t|\,\hat{H},
\end{equation}
with the quantum Hamiltonian operator, say in the typical nonrelativistic situation of a single particle of mass $\mu$, expressed as,
\begin{equation}
\hat{H}=\frac{1}{2\mu}\,\hat{\vec{p}}\,^2\,+\,\hat{V}_q\left(\hat{x}_i\right),\qquad
\hat{V}_q(\hat{x}_i)=\frac{1}{2}\left(\hat{V}(\hat{x}_i) + \hat{V}^\dagger(\hat{x}_i)\right).
\end{equation}
Note that because of the noncommutativity of the coordinate operators $\hat{x}_i$ and the requirement that the Hamiltonian be self-adjoint (or at least hermitian),
the quantum potential energy now needs to be chosen in terms of $\hat{V}_q(\hat{x}_i)$ as defined above, rather than simply by $\hat{V}(\hat{x}_i)$ which
generally is such that $\hat{V}^\dagger(\hat{x}_i)\ne \hat{V}(\hat{x}_i)$ because of configuration space noncommutativity.

For a unitary quantum time evolution generated by $\hat{H}$ the normalisation of any time dependent normalised abstract quantum state remains conserved,
namely,
\begin{equation}
1=\langle\psi,t|\psi,t\rangle = \int_{(\infty)}d^N\vec{x}\,\rho_\lambda(\vec{x},t),\quad
\rho_\lambda(\vec{x},t)=\psi^*_\lambda(\vec{x},t) *_\lambda \psi_\lambda(\vec{x},t),\quad
\psi_\lambda(\vec{x},t) \equiv \langle\vec{x};\lambda | \psi,t\rangle.
\end{equation}
Hence, $\rho_\lambda(\vec{x},t)=\psi^*_\lambda(\vec{x},t)\, *_\lambda\, \psi_\lambda(\vec{x},t)$ still represents the probability density of the state $|\psi,t\rangle$.
As follows from the Schr\"odinger equation, its local conservation equation thus reads,
\begin{equation}
\frac{\partial}{\partial t}\rho_\lambda(\vec{x},t) \, + \, \frac{i}{\hbar}
\langle\psi,t|\,\Big[\,|\vec{x};\lambda\rangle *_\lambda\,\langle\vec{x};\lambda|\, , \, \hat{H}\,\Big]\,|\psi,t\rangle =0.
\end{equation}
The kinetic energy term of the Hamiltonian provides the following contribution to the second term on the l.h.s.~of this identity,
\begin{equation}
\frac{i}{\hbar}\langle\psi,t|\,\Big[\,|\vec{x};\lambda\rangle *_\lambda \langle\vec{x};\lambda|\, , \,\frac{1}{2\mu}\hat{\vec{p}}\,^2\,\Big]\,|\psi,t \rangle
= \vec{\nabla}_{\vec{x}}\cdot\vec{J}_{\lambda,K}(\vec{x},t),
\label{eq:continuity-2}
\end{equation}
where the kinetic probability current density $\vec{J}_{\lambda,K}(\vec{x},t)$, which coincides with its expression in the commutative configuration space case,
is defined by, 
\begin{equation}
\vec{J}_{\lambda,K}(\vec{x},t) = \frac{\hbar}{2i \mu}\,\psi^*_\lambda(\vec{x},t)\, *_\lambda \stackrel{\leftrightarrow}{\nabla}_{\vec{x}} \psi_\lambda(\vec{x},t),\qquad
\left(\stackrel{\leftrightarrow}{\nabla}_{\vec{x}}\right)_i=\left(\frac{\stackrel{\rightarrow}{\partial}}{\partial x_i}\right)\,-\,
\left(\frac{\stackrel{\leftarrow}{\partial}}{\partial x_i}\right).
\end{equation}
However in contradistinction to the commutative case, the quantum potential energy term of the Hamiltonian now also provides an additional nonvanishing
contribution to the total probability current density, since one now has,
\begin{equation}
\frac{i}{\hbar}\Big[\,|\vec{x};\lambda\rangle *_\lambda \langle\vec{x};\lambda|\, , \, \hat{x}_i\,\Big]
=-\frac{1}{2\hbar}\,A_{ij}\,\frac{\partial}{\partial x_j}\Big(\,|\vec{x};\lambda\rangle *_\lambda \langle \vec{x};\lambda|\,\Big).
\end{equation}
By extension, and through a power series expansion representation of $\hat{V}_q(\hat{x}_i)$, it thus follows that the contribution to the second term
on the l.h.s.~of the identity (\ref{eq:continuity-2}) of the quantum potential energy is given again by the divergence of a current density, in the form of,
\begin{equation}
\langle\psi,t | \,\frac{i}{\hbar}\,\Big[\,|\vec{x};\lambda\rangle *_\lambda \langle\vec{x};\lambda |\, , \, \hat{V}_q(\hat{x}_i)\,\Big]\,|\psi,t\rangle
= \frac{\partial}{\partial x_i}\,J_{\lambda,V,i}(\vec{x},t),
\end{equation}
with,
\begin{equation}
\frac{\partial}{\partial x_i}\,J_{\lambda,V,i}(\vec{x};t) = \oint_0\frac{du}{2i\pi}\,\frac{1}{u^2}\,
\langle\psi,t|\,\hat{V}_q\left(\hat{x}_i+u\frac{\partial}{\partial x_j}\left(-\frac{A_{ij}}{2\hbar}|\vec{x};\lambda\rangle *_\lambda\langle\vec{x};\lambda|\,\right)\right)\,
|\psi,t\rangle.
\end{equation}
Here in accordance with the residues theorem, $u$ is a parameter in the complex plane and $\oint_0$ denotes an arbitrary closed contour in the complex plane
encircling once the origin with the trigonometric orientation. Quite obviously, the potential energy probability current density $\vec{J}_{\lambda,V}(\vec{x},t)$,
being directly proportional to $A_{ij}$, vanishes identically in the commutative limit of a vanishing conjugate magnetic field $A_{ij}$, as it should.

In conclusion in the case of noncommutativity in configuration space induced by a nonvanishing conjugate magnetic field $A_{ij}$,
in the $*_\lambda$-representation the conservation of quantum probability is expressed in terms of the continuity equation,
\begin{equation}
\frac{\partial}{\partial t}\rho_\lambda(\vec{x},t) + \vec{\nabla}_{\vec{x}}\cdot\vec{J}_\lambda(\vec{x},t) = 0.
\end{equation}
This time, the total probability current density $\vec{J}_\lambda(\vec{x},t)$ is comprised of two contributions, namely the usual kinetic energy one,
together now with a potential energy one on account of configuration space noncommutativity which induces a Lorentz-like force generated by
the conjugate magnetic field $A_{ij}$, thus,
\begin{equation}
\vec{J}_\lambda(\vec{x},t) = \vec{J}_{\lambda,K}(\vec{x},t) + \vec{J}_{\lambda,V}(\vec{x},t).
\end{equation}

\subsection{The noncommutative configuration space enveloping algebra representation}

Following the same line of argumentation as that presented in section~\ref{Subsection.2.10} in the commutative case based on the momentum space
$|\vec{p}\,\rangle$-representation of the Heisenberg algebra, in the noncommutative case it readily follows once again that there exists a one-to-one map
between abstract quantum states $|\varphi\rangle$ in the abstract quantum Hilbert space and the operators $\hat{\varphi}(\hat{\vec{x}}\,)$ in the enveloping algebra
of the now noncommuting coordinate operators $\hat{x}_i$, constructed in terms of the momentum wave representation
$\tilde{\varphi}(\vec{p}\,)=\langle\vec{p}\,|\varphi\rangle$of the state $|\varphi\rangle$ and the following direct correspondence,
\begin{equation}
\hat{\varphi}(\hat{\vec{x}}\,)=\int_{(\infty)}\frac{d^N\vec{p}}{(2\pi\hbar)^{N/2}}\,e^{\frac{i}{\hbar}\vec{p}\cdot\hat{\vec{x}}}\,\tilde{\varphi}(\vec{p}\,),\qquad
\tilde{\varphi}(\vec{p}\,)=\langle\vec{p}\,|\varphi\rangle.
\end{equation}
Conversely the abstract state $|\varphi\rangle$ is reconstructed through the following representations,
\begin{equation}
|\varphi\rangle = \int_{(\infty)}d^N\vec{p}\,|\vec{p}\,\rangle\,\tilde{\varphi}(\vec{p}\,)
=\int_{(\infty)}d^N\vec{p}\frac{d^N\vec{x}}{(2\pi\hbar)^{N/2}}\,e^{\frac{i}{\hbar}\vec{p}\cdot\hat{\vec{x}}}\,|\vec{x};\lambda\rangle\,\tilde{\varphi}(\vec{p}\,)
=\int_{(\infty)}d^N\vec{x}\,\hat{\varphi}(\hat{\vec{x}}\,)\,|\vec{x};\lambda\rangle.
\end{equation}
This enveloping algebra representation of quantum states relies on the identity (\ref{eq:p-identity}) which allows to represent momentum
eigenstates $|\vec{p}\,\rangle$ in terms of configuration space integrals over the localised and normalisable
states $|\vec{x};\lambda\rangle$ in the noncommutative case as well.

Furthermore the positive definite inner product over the enveloping algebra that represents the inner product of abstract quantum states $|\varphi_1\rangle$
and $|\varphi_2\rangle$ is thus also given by the configuration space doubly integrated matrix element for the $|\vec{x};\lambda\rangle$ states
of the $\hat{x}_i$-enveloping algebra composite operator $\hat{\varphi}^\dagger_1(\hat{\vec{x}}\,)\,\hat{\varphi}_2(\hat{\vec{x}}\,)$,
\begin{equation}
\langle\varphi_1|\varphi_2\rangle = \int_{(\infty)}d^N\vec{x}_1\,d^N\vec{x}_2\,\langle\vec{x}_1;\lambda|\,\hat{\varphi}^\dagger_1(\hat{\vec{x}}\,)\,
\hat{\varphi}_2(\hat{\vec{x}}\,)\,|\vec{x}_2;\lambda\rangle,
\label{eq:inner-2}
\end{equation}
as is the case for the commutative situation (see (\ref{eq:inner})).

However there remains still to establish under which conditions this enveloping algebra representation of quantum states also provides a representation
of the abstract commutation relations of the operators $(\hat{x}_i,\hat{p}_i,\mathbb{I})$. In the commutative case this representation is provided by the
expressions in (\ref{eq:X-1}) and (\ref{eq:P-1}). In particular that for the operators $\hat{P}_i$ involves a partial derivative relative to the coordinate operator $\hat{x}_i$
of the operator $\hat{\varphi}(\hat{\vec{x}}\,)$ that represents the state $|\varphi\rangle$. However in the noncommutative case such a representation
is not well defined anymore since the operators $\hat{x}_i$ then do not commute among themselves. Hence the situation has to be considered anew, based on the
momentum space representations of the abstract operators $\hat{x}_i$ and $\hat{p}_i$ acting on abstract quantum states.

As it turns out (see hereafter), the existence of a representation of the abstract momentum operators $\hat{p}_i$ for the enveloping algebra representation
requires that the real antisymmetric matrix $A_{ij}$ representing the conjugate magnetic field be invertible. Hence the noncommutative configuration space
enveloping algebra representation of the quantum system is available only provided that the configuration space is even dimensional with $N=2n$ and $n\ge 1$
and that  none of the (complex) eigenvalues of $A_{ij}$ is vanishing. Let us henceforth assume to be working under such specific conditions.

Consider first the situation for the abstract momentum operators $\hat{p}_i$, for which $\langle\vec{p}\,|\hat{p}_i| \varphi\rangle=p_i\,\tilde{\varphi}(\vec{p}\,)$.
Hence one has,
\begin{equation}
\hat{P}_i\left(\hat{\varphi}(\hat{\vec{x}}\,)\right)=\int_{(\infty)}\frac{d^N\vec{p}}{(2\pi\hbar)^{N/2}}\,e^{\frac{i}{\hbar}\vec{p}\cdot\hat{\vec{x}}}\,p_i\,\tilde{\varphi}(\vec{p}\,).
\end{equation}
In order to express this linear operator acting within the $\hat{x}_i$-enveloping algebra solely in terms of the noncommuting coordinate operators, let us
revisit the identity,
\begin{equation}
e^{-\frac{i}{\hbar}\vec{p}\cdot\hat{\vec{x}}}\,\hat{x}_i\,e^{\frac{i}{\hbar}\vec{p}\cdot\hat{\vec{x}}}=\hat{x}_i\,-\,\frac{1}{\hbar} A_{ij}\,p_j,\qquad
\hat{x}_i\,e^{\frac{i}{\hbar}\vec{p}\cdot\hat{\vec{x}}}
=e^{\frac{i}{\hbar}\vec{p}\cdot\hat{\vec{x}}}\,\hat{x}_i\,-\,\frac{1}{\hbar}\,A_{ij}\,p_j\,e^{\frac{i}{\hbar}\vec{p}\cdot\hat{\vec{x}}},
\label{eq:identity-twice}
\end{equation}
which thus implies,
\begin{equation}
-\frac{1}{\hbar} A_{ij}\,p_j\,e^{\frac{i}{\hbar}\vec{p}\cdot\hat{\vec{x}}} = \Big[\,\hat{x}_i\, , \, e^{\frac{i}{\hbar}\vec{p}\cdot\hat{\vec{x}}}\,\Big].
\end{equation}
Therefore provided the matrix $A_{ij}$ is invertible one has,
\begin{equation}
p_i\,e^{\frac{i}{\hbar}\vec{p}\cdot\hat{\vec{x}}} = \Big[ -\hbar\left(A^{-1}\right)_{ij} \hat{x}_j\, ,  e^{\frac{i}{\hbar}\vec{p}\cdot\hat{\vec{x}}}\,\Big].
\end{equation}
Consequently one finally finds, under that specific condition, for the representation of the abstract momentum operators
in the $\hat{x}_i$-enveloping algebra representation,
\begin{equation}
\hat{P}_i\left(\hat{\varphi}(\hat{\vec{x}}\,)\right)
=\int_{(\infty)}\frac{d^N\vec{p}}{(2\pi\hbar)^{N/2}}\,\Big[ -\hbar\left(A^{-1}\right)_{ij} \hat{x}_j\, ,  e^{\frac{i}{\hbar}\vec{p}\cdot\hat{\vec{x}}}\,\Big]\tilde{\varphi}(\vec{p}\,)
= -\hbar(A^{-1})_{ij} \left[\hat{x}_j, \hat{\varphi}(\hat{\vec{x}}\,)\right].
\end{equation}

Let us now turn to the situation for the noncommutative coordinate operators $\hat{x}_i$, for which
$\langle\vec{p}\,|\hat{x}_i|\varphi\rangle=\left(i\hbar\frac{\partial}{\partial p_i}-\frac{1}{2\hbar} A_{ij} p_j\right)\,\tilde{\varphi}(\vec{p}\,)$.
Hence one has,
\begin{equation}
\hat{X}_i\left(\hat{\varphi}(\hat{\vec{x}}\,)\right)
=\int_{(\infty)}\frac{d^N\vec{p}}{(2\pi\hbar)^{N/2}}\,e^{\frac{i}{\hbar}\vec{p}\cdot\hat{\vec{x}}}
\,\left(i\hbar\frac{\partial}{\partial p_i}-\frac{1}{2\hbar} A_{ij} p_j\right)\,\tilde{\varphi}(\vec{p}\,).
\end{equation}
In order to complete this evaluation by integration by parts for the term in $\partial/\partial p_i$, for a fixed value of the index $i$ let us denote
by $j\ne i$ the other values of the running index $j$, so that (no summation over $i$ in this expression),
\begin{equation}
\vec{p}\cdot\hat{\vec{x}}=p_i\hat{x}_i\,+\,\sum_{j\ne i} p_j\hat{x}_j.
\end{equation}
Then using the BCH formula one finds (in this expression the index $i$ is not summed over),
\begin{equation}
e^{\frac{i}{\hbar}\vec{p}\cdot\hat{\vec{x}}}=e^{\frac{i}{2\hbar^2}\sum_{j\ne i}A_{ij}p_i p_j}\,e^{\frac{i}{\hbar}p_i\hat{x}_i}\,e^{\frac{i}{\hbar}\sum_{j\ne i}p_j\hat{x}_j}.
\end{equation}
Therefore one has,
\begin{equation}
-i\hbar\frac{\partial}{\partial p_i}\,e^{\frac{i}{\hbar}\vec{p}\cdot\hat{\vec{x}}}
=\left(\hat{x}_i+\frac{1}{2\hbar} \sum_{j\ne i}A_{ij} p_j\right)\,e^{\frac{i}{\hbar}\vec{p}\cdot\hat{\vec{x}}},\quad
\left(-i\hbar\frac{\partial}{\partial p_i}\,-\,\frac{1}{2\hbar}A_{ij} p_j\right)\,e^{\frac{i}{\hbar}\vec{p}\cdot\hat{\vec{x}}}=\hat{x}_i\,e^{\frac{i}{\hbar}\vec{p}\cdot\hat{\vec{x}}}.
\end{equation}
Thus finally,
\begin{equation}
\hat{X}_i\left(\hat{\varphi}(\hat{\vec{x}}\,)\right)
=\int_{(\infty)}\frac{d^N\vec{p}}{(2\pi\hbar)^{N/2}}\,\hat{x}_i\,e^{\frac{i}{\hbar}\vec{p}\cdot\hat{\vec{x}}}\,\tilde{\varphi}(\vec{p}\,)
=\hat{x}_i\,\hat{\varphi}(\hat{\vec{x}}\,).
\end{equation}
Note well that the above analysis implies that this simple representation of the noncommuting coordinate operators on the $\hat{x}_i$-enveloping algebra is
such that $\hat{X}_i$ acts by left multiplication by the operator $\hat{x}_i$ of the operator $\hat{\varphi}(\hat{\vec{x}}\,)$ representing the abstract quantum state
$|\varphi\rangle$, rather than by right multiplication. One could expect that an alternative representation where $\hat{x}_i$ would be acting by right multiplication
may be achieved as well, using the second relation in (\ref{eq:identity-twice}), but then accompanied by an explicit contribution in the conjugate magnetic field
$A_{ij}$, since,
\begin{equation}
\left(-i\hbar\frac{\partial}{\partial p_i}\,-\,\frac{1}{2\hbar}A_{ij} p_j\right)\,e^{\frac{i}{\hbar}\vec{p}\cdot\hat{\vec{x}}}=\hat{x}_i\,e^{\frac{i}{\hbar}\vec{p}\cdot\hat{\vec{x}}}
=e^{\frac{i}{\hbar}\vec{p}\cdot\hat{\vec{x}}}\,\hat{x}_i\,-\,\frac{1}{\hbar} A_{ij}\, p_j\, e^{\frac{i}{\hbar}\vec{p}\cdot\hat{\vec{x}}}.
\end{equation}
However based on that identity one then finds again for the operator $\hat{X}_i$,
\begin{equation}
\hat{X}_i\left(\hat{\varphi}(\hat{\vec{x}}\,)\right)=\hat{\varphi}(\hat{\vec{x}}\,)\,\hat{x}_i\,-\frac{1}{\hbar}\,A_{ij}\,\hat{P}_j\left(\hat{\varphi}(\hat{\vec{x}}\,)\right)
=\hat{\varphi}(\hat{\vec{x}}\,)\hat{x}_i\,+\,\left[\hat{x}_i,\hat{\varphi}(\hat{\vec{x}}\,)\right]=\hat{x}_i\,\hat{\varphi}(\hat{\vec{x}}\,).
\end{equation}

In conclusion, on the $\hat{x}_i$-enveloping algebra representation of operators $\hat{\varphi}(\hat{\vec{x}}\,)$
the abstract algebra of operators $(\hat{x}_i,\hat{p}_i)$ is realised by the following linear operators only involving the noncommuting coordinate operators $\hat{x}_i$,
\begin{equation}
\hat{X}_i\left(\hat{\varphi}(\hat{\vec{x}}\,)\right)=\hat{x}_i\,\hat{\varphi}(\hat{\vec{x}}\,),\qquad
\hat{P}_i\left(\hat{\varphi}(\hat{\vec{x}}\,)\right)=-\hbar(A^{-1})_{ij}\left[\hat{x}_j, \hat{\varphi}(\hat{\vec{x}}\,)\right].
\end{equation}
Furthermore and as is to be expected, it may readily be checked that these operators do indeed generate the same algebra as that of the abstract operators
$(\hat{x}_i,\hat{p}_i,\mathbb{I})$, with,
\begin{equation}
\left[\hat{X}_i,\hat{X}_j\right]=iA_{ij}\,\mathbb{I}_q,\qquad
\left[\hat{X}_i,\hat{P}_j\right]=i\hbar\,\delta_{ij}\,\mathbb{I}_q,\qquad
\left[\hat{P}_i,\hat{P}_j\right]=0,
\end{equation}
where $\mathbb{I}_q$ denotes the identity operator on the enveloping algebra, $\mathbb{I}_q\left(\hat{\varphi}(\hat{\vec{x}}\,)\right)=\hat{\varphi}(\hat{\vec{x}}\,)$.

With the above choices of inner product in (\ref{eq:inner-2}) and of these actions of the operators $\hat{X}_i$ and $\hat{P}_i$, its has thus been established
that the enveloping algebra of the noncommuting coordinate operators $\hat{x}_i$ with operators $\hat{\varphi}(\hat{\vec{x}}\,)$
provides an explicit representation of the abstract algebra of the operators $(\hat{x}_i,\hat{p}_i,\mathbb{I})$ and its Hilbert space of abstract quantum states
$|\varphi\rangle$.

Furthermore, the reconstruction of the abstract quantum state $|\varphi\rangle$ represented by the operator $\hat{\varphi}(\hat{\vec{x}}\,)$ is achieved
in terms of the overcomplete ensemble of localised and normalisable abstract quantum states $|\vec{x};\lambda\rangle$ labelled by all points of
the corresponding classical and commutative configuration space even though the coordinate operators are now noncommuting and thus not
diagonalisable, whatever the choice of real positive definite symmetric matrix of length squared coefficients $\lambda_{ij}$.
As a matter of fact these states $|\vec{x};\lambda\rangle$ are a specific subclass of canonical quantum phase space
coherent states whose definition involves the coefficients $\lambda_{ij}$, thus restricted to configuration space but still generating through
the $*_\lambda$-product the whole abstract Hilbert space on which the
algebra of abstract operators $(\hat{x}_i,\hat{p}_i,\mathbb{I})$ is realised with a positive definite sesquilinear inner product.
This one-to-one map between abstract quantum states $|\varphi\rangle$ and $\hat{x}_i$-enveloping algebra quantum operators $\hat{\varphi}(\hat{\vec{x}}\,)$
is provided by the configuration space integral representations,
\begin{equation}
|\varphi\rangle = \int_{(\infty)}d^N\vec{x}\ \hat{\varphi}(\hat{\vec{x}}\,)\,|\vec{x};\lambda\rangle,\qquad
\langle\varphi_1|\varphi_2\rangle = \int_{(\infty)}d^N\vec{x}_1\,d^N\vec{x}_2\,\langle\vec{x}_1;\lambda|\,\hat{\varphi}^\dagger_1(\hat{\vec{x}}\,)\,
\hat{\varphi}_2(\hat{\vec{x}}\,)\,|\vec{x}_2;\lambda\rangle,
\end{equation}
while the inverse relation involves the momentum eigenbasis $|\vec{p}\,\rangle$,
\begin{equation}
\hat{\varphi}(\hat{\vec{x}}\,)=\int_{(\infty)}\frac{d^N\vec{p}}{(2\pi\hbar)^{N/2}}\,e^{\frac{i}{\hbar}\vec{p}\cdot\hat{\vec{x}}}\,\langle\vec{p}\,|\varphi\rangle.
\end{equation}

Finally let us briefly consider the time dependent Schr\"odinger equation with an abstract Hamiltonian operator, say once again of the nonrelativistic form,
\begin{equation}
\hat{H}=\frac{1}{2\mu}\hat{\vec{p}}\,^2\,+\,\hat{V}_q(\hat{\vec{x}}\,).
\end{equation}
On the $\hat{x}_i$-enveloping algebra this operator is thus represented as,
\begin{eqnarray}
\hat{\mathbb{H}}\left(\hat{\varphi}(\hat{\vec{x}})\,\right) &=& \frac{1}{2\mu}\hat{P}_i\left(\hat{P}_i\left(\hat{\varphi}(\hat{\vec{x}}\,)\right)\right)\,+\,
\hat{V}_q\left(\hat{X}_i\right)\left(\hat{\varphi}(\hat{\vec{x}}\,)\right) \nonumber \\
&=& \frac{\hbar^2}{2\mu}\left(A^{-1}\right)_{ij}\left(A^{-1}\right)_{ik}\Big[\hat{x}_j,\Big[\hat{x}_k,\hat{\varphi}(\hat{\vec{x}}\,)\Big]\Big]\,+\,
\hat{V}_q(\hat{\vec{x}}\,)\,\hat{\varphi}(\hat{\vec{x}}\,).
\end{eqnarray}
Hence in this representation the time dependent Schr\"odinger equation is expressed as, for any time dependent abstract quantum state $|\psi,t\rangle$
and its operator representation $\hat{\psi}(\hat{\vec{x}},t)$,
\begin{equation}
i\hbar\frac{d}{dt}\hat{\psi}(\hat{\vec{x}},t)
=\frac{\hbar^2}{2\mu}\left(A^{-1}\right)_{ij}\left(A^{-1}\right)_{ik}\Big[\hat{x}_j,\Big[\hat{x}_k,\hat{\psi}(\hat{\vec{x}},t)\Big]\Big]\,+\,
\hat{V}_q(\hat{\vec{x}}\,)\,\hat{\psi}(\hat{\vec{x}},t).
\end{equation}

Given that the conserved inner product of a normalised quantum state $|\psi,t\rangle$ is expressed as follows in terms of its $\hat{x}_i$-enveloping algebra
representation,
\begin{equation}
1=\langle\psi,t|\psi,t\rangle
=\int_{(\infty)}d^N\vec{x}_1\,d^N\vec{x}_2\,\langle\vec{x}_1;\lambda|\,\hat{\psi}^\dagger(\hat{\vec{x}},t)\,\hat{\psi}(\hat{\vec{x}},t)\,|\vec{x}_2;\lambda\rangle,
\end{equation}
indicates that the probability density of the quantum system is now represented by the following normalised bi-local density in
configuration space,
\begin{equation}
\rho_\lambda(\vec{x}_1,\vec{x}_2,t)=\langle\vec{x}_1;\lambda|\,\hat{\psi}^\dagger(\hat{\vec{x}},t)\,\hat{\psi}(\hat{\vec{x}},t)\,|\vec{x}_2;\lambda\rangle,\qquad
\int_{(\infty)}d^N\vec{x}_1\,d^N\vec{x}_2\,\rho_\lambda(\vec{x}_1,\vec{x}_2,t)=1,
\end{equation}
or even by the normalised probability density operator $\hat{\rho}(\hat{\vec{x}},t)$ defined as,
\begin{equation}
\hat{\rho}(\hat{\vec{x}},t)=\hat{\psi}^\dagger(\hat{\vec{x}},t)\,\hat{\psi}(\hat{\vec{x}},t),\qquad {\rm with}\quad
\rho_\lambda(\vec{x}_1,\vec{x}_2,t)=\langle\vec{x}_1;\lambda |\, \hat{\rho}(\hat{\vec{x}},t)\,|\vec{x}_2;\lambda\rangle.
\end{equation}
The conservation of probability under a unitary quantum time evolution then translates into specific equations for either
$d\hat{\rho}(\hat{\vec{x}},t)/dt$ or $\partial\rho_\lambda(\vec{x}_1,\vec{x}_2,t)\partial t$, realising in the enveloping algebra representation
the equivalent of an ordinary continuity equation, of which the consideration and details are not included herein.

\subsection{Quantum and classical Hilbert spaces}

To proceed further with the above general analysis, let us now specifically restrict to a configuration space of even dimension, $N=2n$ with $n\ge 1$,
and to a situation when the conjugate magnetic field $A_{ij}$ defines an invertible antisymmetric matrix, ${\rm det}\,A_{ij}\ne 0$. Then in order to connect
with the heuristic construction by F. Scholtz and his collaborators of quantum mechanics over the noncommutative euclidean plane,
it proves useful to consider an alternative coordinate basis in the configuration and momentum spaces such that the conjugate magnetic field is then
represented by a $2\times 2$ block diagonalised antisymmetric $2n\times 2n$ matrix.

As is well known, any $2n\times 2n$ real antisymmetric matrix may be block diagonalised through some O($2n$) orthogonal transformation into
$2\times 2$ antisymmetric blocks. When the matrix is non-degenerate as is now the assumed situation, the off-diagonal entries of these $2\times 2$ blocks
are nonvanishing.

Thus given the constant coefficients $A_{ij}$, there exists a O($2n$) transformation $R_{(\alpha,a),i}$ with $\alpha,\beta=1,2,\cdots,n$ and $a,b=1,2$,
such that $R^{-1}=R^T$, namely,
\begin{equation}
R_{(\alpha,a),i}\,R_{(\beta,b),i}=\delta_{\alpha\beta}\,\delta_{ab},\qquad
R_{(\alpha,a),i}\,R_{(\alpha,a),j} = \delta_{ij},\quad
{\rm det}\,R_{(\alpha,a),i}=\pm 1,
\end{equation}
and leading to the $2\times 2$ block diagonalisation of the matrix $A_{ij}$ in the form of,
\begin{equation}
\Theta_{(\alpha,a),(\beta,b)} = R_{(\alpha,a),i}\,A_{ij}\,R_{(\beta,b),j} = \theta_\alpha\,\delta_{\alpha\beta}\,\epsilon_{ab},\quad
\left(\Theta^{-1}\right)_{(\alpha,a),(\beta,b)}=-\frac{1}{\theta_\alpha}\,\delta_{\alpha\beta}\,\epsilon_{ab},\quad \theta_\alpha>0,
\end{equation}
where the entries $\theta_\alpha>0$ are strictly positive, while $\epsilon_{ab}$ is the 2-index antisymmetric symbol with $\epsilon_{12}=+1$.
Furthermore when requiring that the values $\theta_\alpha$ are ordered in increasing or decreasing order relative to $\alpha=1,2,\cdots,n$,
the O($2n$) orthogonal transformation $R_{(\alpha,a),i}$ is then uniquely determined.

Applying the same linear transformation to the coefficients $\lambda_{ij}$ defining the quantum coherent states $|\vec{x};\lambda\rangle$, one has the
real positive definite symmetric matrix,
\begin{equation}
\Lambda_{(\alpha,a),(\beta,b)}=R_{(\alpha,a),i}\,\lambda_{ij}\,R_{(\beta,b),j}.
\end{equation}
Likewise for the coordinate and momentum operators $\hat{x}_i$ and $\hat{p}_i$, and the associated commuting variables $x_i$ and $p_i$
one has the quantum operators $(\hat{x}_{\alpha,a},\hat{p}_{\alpha,a})$ and variables $(x_{\alpha,a},p_{\alpha,a})$,
\begin{equation}
\hat{x}_{\alpha,a}=R_{(\alpha,a),i}\,\hat{x}_i,\qquad
\hat{p}_{\alpha,a}=R_{(\alpha,a),i}\,\hat{p}_i,\qquad
x_{\alpha,a}=R_{(\alpha,a),i}\,x_i,\qquad
p_{\alpha,a}=R_{(\alpha,a),i}\,p_i,
\end{equation}
of which the commutation relations are,
\begin{equation}
\left[\,\hat{x}_{\alpha,a},\,\hat{x}_{\beta,b}\,\right]=i\,\theta_\alpha\,\delta_{\alpha\beta}\,\epsilon_{ab}\,\mathbb{I},\qquad
\left[\,\hat{x}_{\alpha,a},\,\hat{p}_{\beta,b}\,\right]=i\hbar\,\delta_{\alpha\beta}\,\delta_{ab}\,\mathbb{I},\qquad
\left[\,\hat{p}_{\alpha,a},\,\hat{p}_{\beta,b}\,\right]=0.
\end{equation}
Similarly for the Fock algebra generators $(\hat{A}_i,\hat{A}^\dagger_i)$ (see (\ref{eq:Fock-2}))
in the presence of the conjugate magnetic field which is now represented by the quantities $\theta_\alpha>0$,
one has the transformed operators,
\begin{equation}
\hat{a}_{\alpha,a}=R_{(\alpha,a),i}\,\hat{A}_i,\qquad
\hat{a}^\dagger_{\alpha,a}=R_{(\alpha,a),i}\,\hat{A}^\dagger_i,
\end{equation}
with the following expressions,
\begin{eqnarray}
\hat{a}_{\alpha,a} &=& \hat{x}_{\alpha,a}+\frac{i}{2\hbar}\left(\Lambda_{(\alpha,a),(\beta,b)} - i \Theta_{(\alpha,a),(\beta,b)}\right)\,\hat{p}_{\beta,b}, \nonumber \\
\hat{a}^\dagger_{\alpha,a} &=& \hat{x}_{\alpha,a}-\frac{i}{2\hbar}\left(\Lambda_{(\alpha,a),(\beta,b)} + i \Theta_{(\alpha,a),(\beta,b)}\right)\,\hat{p}_{\beta,b},
\end{eqnarray}
and their general Fock algebra,
\begin{equation}
\Big[\hat{a}_{\alpha,a},\hat{a}_{\beta,b}\Big]=0,\qquad
\left[\hat{a}_{\alpha,a},\hat{a}^\dagger_{\beta,b}\right]=\Lambda_{(\alpha,a),(\beta,b)}\,\mathbb{I},\qquad
\left[\hat{a}^\dagger_{\alpha,a},a^\dagger_{\beta,b}\right]=0.
\end{equation}

Associated to this $N=2n$ dimensional Fock algebra and the complex variables $Z_i=x_i+i(\lambda_{ij}-i A_{ij})p_j/(2\hbar)$ introduced in (\ref{eq:Zi}),
the transformed complex variables thus now read as well,
\begin{eqnarray}
z_{\alpha,a} &=& R_{(\alpha,a),i}\,Z_i
=x_{\alpha,a}+\frac{i}{2\hbar}\left(\Lambda_{(\alpha,a),(\beta,b)} - i \Theta_{(\alpha,a),(\beta,b)}\right)\,p_{\beta,b}, \nonumber \\
z^*_{\alpha,a} &=& R_{(\alpha,a),i}\,Z^*_i
=x_{\alpha,a}-\frac{i}{2\hbar}\left(\Lambda_{(\alpha,a),(\beta,b)} + i \Theta_{(\alpha,a),(\beta,b)}\right)\,p_{\beta,b}.
\end{eqnarray}
Hence the canonical coherent states $|Z_i;\Omega_\lambda\rangle$ related to the initial algebra $(\hat{x}_i,\hat{p}_i,\mathbb{I})$
and the choice of coefficients $\lambda_{ij}$ in the presence of the conjugate magnetic field $A_{ij}$ are now realised as
\begin{eqnarray}
|Z_i;\Omega_\lambda\rangle &\equiv& |z_{\alpha,a};\Omega_\lambda\rangle \nonumber \\
&=& {\rm exp}\left(z_{\alpha,a}\left(\Lambda^{-1}\right)_{(\alpha,a),(\beta,b)}\,\hat{a}^\dagger_{\beta,b}\,-\,
z^*_{\alpha,a}\left(\Lambda^{-1}\right)_{(\alpha,a),(\beta,b)}\,\hat{a}_{\beta,b}\right)\,|\Omega_\lambda\rangle,
\end{eqnarray}
with in particular
\begin{equation}
|\vec{x};\lambda\rangle=|z_{\alpha,a}=x_{\alpha,a};\Omega_\lambda\rangle = e^{-\frac{i}{\hbar}x_{\alpha,a}\,\hat{p}_{\alpha,a}}\,|\Omega_\lambda\rangle
=e^{-\frac{i}{\hbar}\vec{x}\cdot\hat{\vec{p}}}\,|\Omega_\lambda\rangle.
\end{equation}

Consequently the fact that the localised and normalisable quantum states $|\vec{x};\lambda\rangle$ are eigenstates with eigenvalues $x_i$ of the Fock operators
$\hat{A}_i$,
\begin{equation}
\hat{A}_i|\vec{x};\lambda\rangle
=\left(\hat{x}_i+\frac{i}{2\hbar}\left(\lambda_{ij}-i A_{ij}\right)\,\hat{p}_j\right)\,|\vec{x};\lambda\rangle = x_i\,|\vec{x};\lambda\rangle,
\end{equation}
translates into the fact that these states are eigenstates with eigenvalues $x_{\alpha,a}$ of the Fock operators $\hat{a}_{\alpha,a}$,
\begin{equation}
\hat{a}_{\alpha,a}|\vec{x};\lambda\rangle 
= \left(\hat{x}_{\alpha,a}+\frac{i}{2\hbar}\left(\Lambda_{(\alpha,a),(\beta,b)} - i \Theta_{(\alpha,a),(\beta,b)}\right)\,\hat{p}_{\beta,b}\right)\,|\vec{x};\lambda\rangle
= x_{\alpha, a}\,|\vec{x};\lambda\rangle.
\label{eq:obser}
\end{equation}

In order to consider a specific choice of coefficients $\lambda_{ij}$ or $\Lambda_{(\alpha,a),(\beta,b)}$ for the construction of the states $|\vec{x};\lambda\rangle$
such that the above property may be brought into a form which no longer involves a contribution of the action of the momentum operators $\hat{p}_{\alpha,a}$
on these states, let us consider the $2\times 2$ block antisymmetric symbol represented by the $2n\times 2n$ matrix $E$ of which the entries take the values
\begin{equation}
E_{(\alpha,a),(\beta,b)}=\delta_{\alpha\beta}\,\epsilon_{ab},
\end{equation}
and which is such that
\begin{equation}
\left(E^2\right)_{(\alpha,a),(\beta,b)}=-\delta_{\alpha\beta}\,\delta_{ab}.
\end{equation}
Note that $E$ and $\Theta$ are two $2n\times 2n$ block diagonal matrices that commute with one another, $E\Theta=\Theta E$.
Let us then introduce the associated complex valued $2n\times 2n$ projection matrices in the $(\alpha,a)$ basis,
\begin{equation}
\mathbb{P}_\pm=\frac{1}{2}\left(\mathbb{I}_{2n}\pm i E\right),\qquad
\left(\mathbb{P}_\pm\right)_{(\alpha,a),(\beta,b)}=\frac{1}{2}\delta_{\alpha\beta}\left(\delta_{ab}\pm i \epsilon_{ab}\right),
\end{equation}
where $\mathbb{I}_{2n}$ denotes the $2n\times 2n$ identity matrix in the $(\alpha,a)$ basis,
$\left(\mathbb{I}_{2n}\right)_{(\alpha,a),(\beta,b)}=\delta_{\alpha\beta}\,\delta_{ab}$.
Quite obviously one has,
\begin{equation}
\mathbb{P}_+ + \mathbb{P}_- = \mathbb{I}_{2n},\qquad
\mathbb{P}_+ - \mathbb{P}_- = i E,\qquad
\mathbb{P}^*_\pm=\mathbb{P}_\mp,
\end{equation}
as well as,
\begin{equation}
\mathbb{P}^2_\pm = \mathbb{P}_\pm,\qquad
\mathbb{P}_+ \mathbb{P}_- = 0 = \mathbb{P}_- \mathbb{P}_+ .
\end{equation}

In order to exploit these distinctive features in relation to the observation in (\ref{eq:obser}) aiming to project out the contribution of the action of $\hat{p}_{\alpha,a}$,
let us henceforth now restrict the choice for the coefficients $\lambda_{ij}$, namely $\Lambda_{(\alpha,a),(\beta,b)}$, specifically to the values given by,
\begin{equation}
\Lambda=-E\,\Theta= - \Theta \,E,\qquad
\Lambda_{(\alpha,a),(\beta,b)}=\theta_\alpha\,\delta_{\alpha\beta}\,\delta_{ab}.
\end{equation}
To emphasize this specific choice of values for $\lambda_{ij}$ or $\Lambda_{(\alpha,a),(\beta,b)}$, henceforth the corresponding localised and normalisable
quantum states are to be denoted as $|\vec{x};\theta_\alpha\rangle$ rather than as $|\vec{x};\lambda\rangle$ which applies to a generic and general arbitrary choice
for the coefficients $\lambda_{ij}$. And likewise now for the generic $*_\lambda$-product to be denoted henceforth as $*_\theta$, as well as the
generic Fock vacuum $|\Omega_\lambda\rangle$ to be denoted $|\Omega_\theta\rangle$.

Given that specific choice for $\lambda_{ij}$ one has now explicitly,
\begin{equation}
\lambda_{ij} - i A_{ij} = \sum_{\alpha=1}^n\sum_{a,b=1,2}\theta_\alpha\left(\delta_{ab}-i\epsilon_{ab}\right)\,R_{(\alpha,a),i}\,R_{(\alpha,a),j},
\end{equation}
as well as,
\begin{equation}
\left(2\mathbb{P}_\pm\Theta\right)_{(\alpha,a),(\beta,b)}=\mp\,i\,\theta_\alpha\,\delta_{\alpha\beta}\left(\delta_{ab}\,\pm\,i\epsilon_{ab}\right),
\end{equation}
while,
\begin{eqnarray}
\hat{a}_{\alpha,a} &=&\hat{x}_{\alpha,a}+\frac{1}{\hbar}\,\left(\mathbb{P}_- \Theta\right)_{(\alpha,a),(\beta,b)}\,\hat{p}_{\beta,b}
= \hat{x}_{\alpha,a}+\frac{i}{2\hbar}\theta_\alpha\left(\delta_{ab}-i\epsilon_{ab}\right)\,\hat{p}_{\beta,b}, \nonumber \\
\hat{a}^\dagger_{\alpha,a} &=& \hat{x}_{\alpha,a}+\frac{1}{\hbar}\,\left(\mathbb{P}_+ \Theta\right)_{(\alpha,a),(\beta,b)}\,\hat{p}_{\beta,b}
=\hat{x}_{\alpha,a}-\frac{i}{2\hbar}\theta_\alpha\left(\delta_{ab}+i\epsilon_{ab}\right)\,\hat{p}_{\beta,b},
\end{eqnarray}
with in particular,
\begin{equation}
\hat{a}_{\alpha,a}|\vec{x};\theta_\alpha\rangle
= \left(\hat{x}_{\alpha,a}+\frac{1}{\hbar}\,\left(\mathbb{P}_- \Theta\right)_{(\alpha,a),(\beta,b)}\,\hat{p}_{\beta,b}\right)\,|\vec{x};\theta_\alpha\rangle
= x_{\alpha,a}\,|\vec{x};\theta_\alpha\rangle.
\label{eq:identity-3}
\end{equation}
Note as well that,
\begin{equation}
z_{\alpha,a}=x_{\alpha,a}+\frac{i}{2\hbar}\theta_\alpha\left(\delta_{ab}-i\epsilon_{ab}\right)\,p_{\beta,b},\qquad
z^*_{\alpha,a}=x_{\alpha,a}-\frac{i}{2\hbar}\theta_\alpha\left(\delta_{ab}+i\epsilon_{ab}\right)\,p_{\beta,b}.
\end{equation}
Furthermore the canonical coherent states $|z_{\alpha,a};\Omega_\theta\rangle$ are now realised in the form of,
\begin{equation}
|z_{\alpha,a};\Omega_\theta\rangle
= {\rm exp}\left(\sum_{\alpha=1}^n\sum_{a=1,2}\frac{1}{\theta_\alpha}\left(z_{\alpha,a}\,\hat{a}^\dagger_{\alpha,a}\,-\,z^*_{\alpha,a}\,\hat{a}_{\alpha,a}\right)\right)\,
|\Omega_\theta\rangle,
\end{equation}
with for the configuration space restricted canonical coherent states for the algebra $(\hat{x}_i,\hat{p}_i,\mathbb{I})$,
\begin{equation}
|\vec{x};\theta_\alpha\rangle = |z_{\alpha,a}=x_{\alpha,a};\Omega_\theta\rangle=e^{-\frac{i}{\hbar}x_{\alpha,a}\hat{p}_{\alpha,a}}\,|\Omega_\theta\rangle.
\end{equation}

To project out from the identity (\ref{eq:identity-3}) the contributions of the momentum operators $\hat{p}_{\alpha,a}$, it now suffices to multiply it from the left
with the matrix projector $\mathbb{P}_+$. Consider thus the following projected quantum operators, which solely involve the noncommuting
configuration space coordinate operators,
\begin{eqnarray}
\hat{u}_{\alpha,a} &=& 2\left(\mathbb{P}_+\right)_{(\alpha,a),(\beta,b)}\,\hat{a}_{\beta,b}=2\left(\mathbb{P}_+\right)_{(\alpha,a),(\beta,b)}\,\hat{x}_{\beta,b}
=\hat{x}_{\alpha,a} + i\,\epsilon_{ab}\, \hat{x}_{\alpha,b}, \nonumber \\
\hat{u}^\dagger_{\alpha,a} &=& 2\left(\mathbb{P}_-\right)_{(\alpha,a),(\beta,b)}\,\hat{a}^\dagger_{\beta,b}
=2\left(\mathbb{P}_-\right)_{(\alpha,a),(\beta,b)}\,\hat{x}_{\beta,b}=\hat{x}_{\alpha,a}-i\,\epsilon_{ab}\, \hat{x}_{\alpha,b},
\end{eqnarray}
as well as,
\begin{eqnarray}
\hat{v}_{\alpha,a} &=& 2\left(\mathbb{P}_-\right)_{(\alpha,a),(\beta,b)}\,\hat{a}_{\beta,b}
=2\left(\mathbb{P}_-\right)_{(\alpha,a),(\beta,b)}\,\hat{x}_{\beta,b} + \frac{2}{\hbar}\left(\mathbb{P}_-\Theta\right)_{(\alpha,a),(\beta,b)}\,\hat{p}_{\beta,b} \nonumber \\
& = &\left(\delta_{ab}-i\epsilon_{ab}\right)\left(\hat{x}_{\alpha,b}+\frac{i}{\hbar}\theta_\alpha\,\hat{p}_{\alpha,b}\right) 
= \hat{u}^\dagger_{\alpha,a}+\frac{i}{\hbar}\theta_\alpha\left(\hat{p}_{\alpha,a}-i\epsilon_{ab}\,\hat{p}_{\alpha,b}\right), \nonumber \\
\hat{v}^\dagger_{\alpha,a} &=& 2\left(\mathbb{P}+\right)_{(\alpha,a),(\beta,b)}\,\hat{a}^\dagger_{\beta,b}
=2\left(\mathbb{P}_+\right)_{(\alpha,a),(\beta,b)}\,\hat{x}_{\beta,b} +\frac{2}{\hbar}\left(\mathbb{P}_+\Theta\right)_{(\alpha,a),(\beta,b)}\,\hat{p}_{\beta,b}  \nonumber \\
& = & \left(\delta_{ab}+i\epsilon_{ab}\right)\left(\hat{x}_{\alpha,b}-\frac{i}{\hbar}\theta_\alpha\,\hat{p}_{\alpha,b}\right) 
= \hat{u}_{\alpha,a}-\frac{i}{\hbar}\theta_\alpha\left(\hat{p}_{\alpha,a}+i\epsilon_{ab}\,\hat{p}_{\alpha,b}\right),
\end{eqnarray}
jointly with the corresponding complex valued commuting variables,
\begin{eqnarray}
u_{\alpha,a} &=& 2\left(\mathbb{P}_+\right)_{(\alpha,a),(\beta,b)}\,x_{\beta,b}=x_{\alpha,a} + i\, \epsilon_{ab}\, x_{\alpha,b}, \nonumber \\
u^*_{\alpha,a} &=& 2\left(\mathbb{P}_-\right)_{(\alpha,a),(\beta,b)}\,x_{\beta,b}=x_{\alpha,a} - i\, \epsilon_{ab}\, x_{\alpha,b}, \nonumber \\
v_{\alpha,a}& = & 2\left(\mathbb{P}_-\right)_{(\alpha,a),(\beta,b)}\,z_{\beta,b} = u^*_{\alpha,a}\,+
\,\frac{i}{\hbar}\theta_\alpha\left(p_{\alpha,a}-i\epsilon_{ab}\, p_{\alpha,b}\right),    \nonumber \\
v^*_{\alpha,a} & = &  2\left(\mathbb{P}_+\right)_{(\alpha,a),(\beta,b)}\,z_{\beta,b} = u_{\alpha,a}\,-
\,\frac{i}{\hbar}\theta_\alpha\left(p_{\alpha,a}+i\epsilon_{ab}\, p_{\alpha,b}\right).
\end{eqnarray}
Of course one also has,
\begin{equation}
\hat{a}_{\alpha,a}=\frac{1}{2}\left(\hat{u}_{\alpha,a} + \hat{v}_{\alpha,a}\right),\qquad
\hat{a}^\dagger_{\alpha,a}=\frac{1}{2}\left(\hat{u}^\dagger_{\alpha,a} + \hat{v}^\dagger_{\alpha,a}\right),
\end{equation}
with analogous relation between $u_{\alpha,a}$, $v_{\alpha,a}$, $u^*_{\alpha,a}$ and $v^*_{\alpha,a}$, and $z_{\alpha,a}$ and $z^*_{\alpha,a}$.

For the commutation algebra of the operators $\hat{p}_{\alpha,a}$, $\hat{u}_{\alpha,a}$, $\hat{u}^\dagger_{\alpha,a}$, $\hat{v}_{\alpha,a}$
and $\hat{v}^\dagger_{\alpha,a}$, first one readily finds,
\begin{equation}
\left[\hat{u}_{\alpha,a},\hat{p}_{\beta,b}\right] = i\hbar\delta_{\alpha\beta}\left(\delta_{ab}+i\epsilon_{ab}\right)\mathbb{I},\qquad
\left[\hat{u}^\dagger_{\alpha,a},\hat{p}_{\beta,b}\right]=i\hbar\delta_{\alpha\beta}\left(\delta_{ab}-i\epsilon_{ab}\right)\mathbb{I},
\end{equation}
so that,
\begin{eqnarray}
\left[\hat{u}_{\alpha,a},\left(\delta_{bc}-i\epsilon_{bc}\right)\hat{p}_{\beta,c}\right] &=& 2i\hbar\delta_{\alpha\beta}\left(\delta_{ab}+i\epsilon_{ab}\right)\mathbb{I},\qquad
\left[\hat{u}_{\alpha,a},\left(\delta_{bc}+i\epsilon_{bc}\right)\hat{p}_{\beta,c}\right] = 0, \nonumber \\
\left[\hat{u}^\dagger_{\alpha,a},\left(\delta_{bc}-i\epsilon_{bc}\right)\hat{p}_{\beta,c}\right] &=& 0,\qquad
\left[\hat{u}^\dagger_{\alpha,a},\left(\delta_{bc}+i\epsilon_{bc}\right)\hat{p}_{\beta,c}\right] = 2i\hbar\delta_{\alpha\beta}\left(\delta_{ab}-i\epsilon_{ab}\right)\mathbb{I}.
\end{eqnarray}
One then establishes that,
\begin{equation}
\left[\hat{u}_{\alpha,a},\hat{u}_{\beta,b}\right]=0,\quad
\left[\hat{u}_{\alpha,a},\hat{u}^\dagger_{\beta,b}\right]=2\theta_{\alpha}\,\delta_{\alpha\beta}\left(\delta_{ab}+i\epsilon_{ab}\right)\,\mathbb{I},\quad
\left[\hat{u}^\dagger_{\alpha,a},\hat{u}^\dagger_{\beta,b}\right]=0,
\end{equation}
as well as,
\begin{equation}
\left[\hat{v}_{\alpha,a},\hat{v}_{\beta,b}\right]=0,\quad
\left[\hat{v}_{\alpha,a},\hat{v}^\dagger_{\beta,b}\right]=2\theta_{\alpha}\,\delta_{\alpha\beta}\left(\delta_{ab}-i\epsilon_{ab}\right)\,\mathbb{I},\quad
\left[\hat{v}^\dagger_{\alpha,a},\hat{v}^\dagger_{\beta,b}\right]=0,
\end{equation}
and finally,
\begin{equation}
\Big[\hat{u}_{\alpha,a},\hat{v}_{\beta,b}\Big] = 0,\qquad
\left[\hat{u}_{\alpha,a},\hat{v}^\dagger_{\beta,b}\right] = 0,\qquad
\left[\hat{u}^\dagger_{\alpha,a},\hat{v}_{\beta,b}\right] = 0,\qquad
\left[\hat{u}^\dagger_{\alpha,a},\hat{v}^\dagger_{\beta,b}\right] = 0.
\end{equation}
Consequently all the operators $(\hat{u}_{\alpha,a},\hat{u}^\dagger_{\alpha,a})$ commute with all the operators $(\hat{v}_{\alpha,a},\hat{v}^\dagger_{\alpha,a})$.

However the operators $\hat{u}_{\alpha,a}$ on the one hand, and $\hat{v}_{\alpha,a}$ on the other hand, are not all independent, since one has,
\begin{equation}
i\epsilon_{ab}\,\hat{u}_{\alpha,b} = \hat{u}_{\alpha,a},\qquad
i\epsilon_{ab}\hat{v}_{\alpha,a}=-\hat{v}_{\alpha,a},
\end{equation}
namely,
\begin{equation}
\hat{u}_{\alpha,2}=-i\hat{u}_{\alpha,1},\qquad
\hat{u}^\dagger_{\alpha,2}=i\hat{u}^\dagger_{\alpha,1},\qquad
\hat{v}_{\alpha,2}=i\hat{v}_{\alpha,1},\qquad
\hat{v}^\dagger_{\alpha,2}=-i\hat{v}^\dagger_{\alpha,1},
\end{equation}
and likewise for the associated complex variables $u_{\alpha,a}$, $u^*_{\alpha,a}$, and $v_{\alpha,a}$ and $v^*_{\alpha,a}$.
Yet, the states $|\vec{x};\theta_\alpha\rangle$ are the eigenstates of the quantum operators $\hat{u}_{\alpha,a}$ with eigenvalues $u_{\alpha,a}$,
\begin{equation}
\hat{u}_{\alpha,a}\,|\vec{x};\theta_\alpha\rangle = u_{\alpha,a}\,|\vec{x};\theta_\alpha\rangle,\qquad
u_{\alpha,a}=x_{\alpha,a}+i\epsilon_{ab}\, x_{\alpha,b}.
\end{equation}
Since the operators $\hat{u}_{\alpha,a}$ only involve the configuration space coordinate operators $\hat{x}_{\alpha,a}$ (or $\hat{x}_i$) but not the
momentum ones $\hat{p}_{\alpha,a}$ (or $\hat{p}_i$), clearly this last observation strongly suggests that for the specific choice made for the coefficients
$\lambda_{ij}$ the states $|\vec{x};\theta_\alpha\rangle$ are as well some form of canonical coherent states for the restricted subalgebra
of noncommuting configuration space coordinate operators $\hat{u}_{\alpha,a}$, in tensor product with the subalgebra of noncommuting
operators $\hat{v}_{\alpha,a}$, and this not only as a configuration space restricted class of canonical coherent states for the complete initial quantum
algebra $(\hat{x}_i,\hat{p}_i,\mathbb{I})$.

To make this structure manifest, finally let us consider the independent operators $(\hat{b}_\alpha,\hat{b}^\dagger_\alpha)$
and $(\hat{d}_\alpha,\hat{d}^\dagger_\alpha)$, defined by,
\begin{equation}
\hat{b}_\alpha = \hat{u}_{\alpha,1}=\hat{x}_{\alpha,1}+i\hat{x}_{\alpha,2},\qquad
\hat{b}^\dagger_\alpha=\hat{u}^\dagger_{\alpha,1}=\hat{x}_{\alpha,1} - i \hat{x}_{\alpha,2},
\end{equation}
and,
\begin{eqnarray}
\hat{d}_\alpha &=& \hat{v}_{\alpha,1}=(\hat{x}_{\alpha,1}-i\hat{x}_{\alpha,2})+\frac{i}{\hbar}\theta_\alpha\left(\hat{p}_{\alpha,1}-i\hat{p}_{\alpha,2}\right)
=\hat{b}^\dagger_\alpha+\frac{i}{\hbar}\theta_\alpha\,\hat{p}_{\alpha,-}, \nonumber \\
\hat{d}^\dagger_\alpha &=& \hat{v}^\dagger_{\alpha,1}
=(\hat{x}_{\alpha,1}+i\hat{x}_{\alpha,2})-\frac{i}{\hbar}\theta_\alpha\left(\hat{p}_{\alpha,1}+i\hat{p}_{\alpha,2}\right)
=\hat{b}_\alpha-\frac{i}{\hbar}\theta_\alpha\,\hat{p}_{\alpha,+},
\end{eqnarray}
where,
\begin{equation}
\hat{p}_{\alpha,+}=\hat{p}_{\alpha,1}\,+\,i\,\hat{p}_{\alpha,2}=-\frac{i\hbar}{\theta_\alpha}\left(\hat{b}_\alpha-\hat{d}^\dagger_\alpha\right),\qquad
\hat{p}_{\alpha,-}=\hat{p}_{\alpha,1}\,-\,\hat{p}_{\alpha,2}=\frac {i\hbar}{\theta_\alpha}\left(\hat{b}^\dagger_\alpha-\hat{d}_\alpha\right).
\end{equation}
The complex variables corresponding to $\hat{b}_\alpha$ and $\hat{b}^\dagger_\alpha$ are to be denoted as, respectively,
\begin{equation}
z_\alpha=x_{\alpha,1} + i x_{\alpha,2},\qquad
z^*_\alpha=x_{\alpha,1} - i x_{\alpha,2},
\end{equation}
while for $\hat{p}_{\alpha,\pm}$ one has,
\begin{equation}
p_{\alpha,\pm}=p_{\alpha,1}\pm i p_{\alpha,2},\qquad p^*_{\alpha,\pm}=p_{\alpha,\mp}.
\end{equation}

The inverse relations are such that,
\begin{equation}
\hat{a}_{\alpha,1}=\frac{1}{2}\left(\hat{b}_\alpha+\hat{d}_\alpha\right),\qquad
\hat{a}_{\alpha,2}=-\frac{1}{2}i\left(\hat{b}_\alpha-\hat{d}_\alpha\right),
\end{equation}
\begin{equation}
\hat{a}^\dagger_{\alpha,1}=\frac{1}{2}\left(\hat{b}^\dagger_\alpha+\hat{d}^\dagger_\alpha\right),\qquad
\hat{a}^\dagger_{\alpha,2}=\frac{1}{2}i\left(\hat{b}^\dagger_\alpha-\hat{d}^\dagger_\alpha\right).
\end{equation}
Likewise,
\begin{equation}
\hat{x}_{\alpha,1}=\frac{1}{2}\left(\hat{b}_\alpha+\hat{b}^\dagger_\alpha\right),\qquad
\hat{x}_{\alpha,2}=-\frac{1}{2}i\left(\hat{b}_\alpha-\hat{b}^\dagger_\alpha\right),
\end{equation}
\begin{equation}
\hat{p}_{\alpha,1}=\frac{i\hbar}{2\theta_\alpha}\left(\hat{b}^\dagger_\alpha - \hat{b}_\alpha + \hat{d}^\dagger_\alpha - \hat{d}_\alpha\right),\qquad
\hat{p}_{\alpha,2}=-\frac{\hbar}{2\theta_\alpha}\left(\hat{b}^\dagger_\alpha+\hat{b}_\alpha-\hat{d}^\dagger_\alpha-\hat{d}_\alpha\right),
\end{equation}
\begin{equation}
\hat{p}_{\alpha,+}=\hat{p}_{\alpha,1}\,+\,i\,\hat{p}_{\alpha,2}=-\frac{i\hbar}{\theta_\alpha}\left(\hat{b}_\alpha-\hat{d}^\dagger_\alpha\right),\qquad
\hat{p}_{\alpha,-}=\hat{p}_{\alpha,1}\,-\,\hat{p}_{\alpha,2}=\frac {i\hbar}{\theta_\alpha}\left(\hat{b}^\dagger_\alpha-\hat{d}_\alpha\right).
\end{equation}
Obviously, analogous relations apply to the associated complex variables $z_\alpha$, $p_{\alpha,\pm}$, $z_{\alpha,a}$, $x_{\alpha,a}$
and $p_{\alpha,a}$.

The commutation relations for the operators $(\hat{b}_\alpha,\hat{b}^\dagger_\alpha)$ and $(\hat{d}_\alpha,\hat{d}^\dagger_\alpha)$ are then obtained as,
\begin{eqnarray}
\left[\hat{b}_\alpha,\hat{b}_\beta\right] &=& 0,\qquad
\left[\hat{b}_\alpha,\hat{b}^\dagger_\beta\right]=2\theta_\alpha\,\delta_{\alpha\beta}\,\mathbb{I},\qquad
\left[\hat{b}^\dagger_\alpha,\hat{b}^\dagger_\beta\right]=0, \nonumber \\
\left[\hat{d}_\alpha,\hat{d}_\beta\right] &=& 0,\qquad
\left[\hat{d}_\alpha,\hat{d}^\dagger_\beta\right]=2\theta_\alpha\,\delta_{\alpha\beta}\,\mathbb{I},\qquad
\left[\hat{d}^\dagger_\alpha,\hat{d}^\dagger_\beta\right]=0, \nonumber \\
\left[\hat{b}_\alpha,\hat{d}_\beta\right] &=& 0,\qquad
\left[\hat{b}_\alpha,\hat{d}^\dagger_\beta\right]=0,\qquad
\left[\hat{b}^\dagger_\alpha,\hat{d}_\beta\right]=0,\qquad
\left[\hat{b}^\dagger_\alpha,\hat{d}^\dagger_\beta\right]=0,
\end{eqnarray}
while for those involving the operators $\hat{p}_{\alpha,\pm}$ are,
\begin{eqnarray}
\left[\hat{b}_\alpha,\hat{p}_{\beta,-}\right] &=&i\hbar\delta_{\alpha\beta}\mathbb{I},\quad
\left[\hat{b}^\dagger_\alpha,\hat{p}_{\beta,-}\right]=0,\quad
\left[\hat{b}_\alpha,\hat{p}_{\beta,+}\right]=0,\quad
\left[\hat{b}^\dagger_\alpha,\hat{p}_{\beta,+}\right]=i\hbar\delta_{\alpha\beta}\mathbb{I}, \nonumber \\
\left[\hat{d}_\alpha,\hat{p}_{\beta,-}\right] &=&0,\quad
\left[\hat{d}^\dagger_\alpha,\hat{p}_{\beta,-}\right]=i\hbar\delta_{\alpha\beta}\mathbb{I},\quad
\left[\hat{d}_\alpha,\hat{p}_{\beta,+}\right]=i\hbar\delta_{\alpha\beta}\mathbb{I},\quad
\left[\hat{d}^\dagger_\alpha,\hat{p}_{\beta,+}\right]=0.
\end{eqnarray}
Hence the operator algebra defining the quantised system is that of the tensor product of $2n=N$ commuting Fock algebras, namely,
for each value of $\alpha=1,2,\cdots,n$, the two commuting Fock algebras generated, on the one hand by the operators $(\hat{b}_\alpha,\hat{b}^\dagger_\alpha)$,
and on the other hand by the operators $(\hat{d}_\alpha,\hat{d}^\dagger_\alpha)$, with each of these sectors characterised
by the noncommutativity real parameter $\theta_\alpha>0$ as a normalisation of the Fock algebra nonvanishing commutation relations
$[\hat{b}_\alpha,\hat{b}^\dagger_\alpha]=2\theta_\alpha\mathbb{I}$
and $[\hat{d}_\alpha,\hat{d}^\dagger_\alpha]=2\theta_\alpha\mathbb{I}$ (no summation over $\alpha$).

Consequently, the structure of the full quantum Hilbert space for the noncommutative Euclidean configuration space
consists of the tensor product of a $b$-sector and a $d$-sector, each of these two sectors being itself
the $n$-fold tensor product over $\alpha=1,2,\cdots,n$ of the irreducible Fock spaces associated to the $(\hat{b}_\alpha,\hat{b}^\dagger_\alpha)$ or to the
$(\hat{d}_\alpha,\hat{d}^\dagger_\alpha)$ Fock algebras, respectively.

With this understanding at hand, let us now reconsider the status of the states $|\vec{x};\theta_\alpha\rangle$ which generate the entire quantum Hilbert space
through the $*_\theta$-product, and are characterised by the following properties,
\begin{equation}
\hat{a}_{\alpha,a}\,|\vec{x};\theta_\alpha\rangle = x_{\alpha,a}\,|\vec{x};\theta_\alpha\rangle,\qquad
|\vec{x};\theta_\alpha\rangle = e^{-\frac{i}{\hbar}\vec{x}\cdot\hat{\vec{p}}}\,|\Omega_\theta\rangle,\qquad
\langle\Omega_\theta|\Omega_\theta\rangle = \prod_{\alpha=1}^n\frac{1}{2\pi\theta_\alpha},
\end{equation}
with thus in particular the fact that $|\Omega_\theta\rangle=|\vec{x}=\vec{0};\theta_\alpha\rangle$. Given that one has
$\hat{a}_{\alpha,1}=(\hat{b}_\alpha + \hat{d}_\alpha)/2$ and $\hat{a}_{\alpha,2}=-i(\hat{b}_\alpha-\hat{d}_\alpha)/2$,
while $\hat{a}_{\alpha,a}|\Omega_\theta\rangle=0$, it follows that,
\begin{equation}
\hat{b}_\alpha\,|\Omega_\theta\rangle = 0 , \qquad
\hat{d}_\alpha\,|\Omega_\theta\rangle = 0.
\end{equation}
In other words, the state $|\Omega_\theta\rangle=|\vec{x}=\vec{0};\theta_\alpha\rangle$ is as well a Fock vacuum for all Fock algebras in both the $b$-
and the $d$-sectors of the full quantum Hilbert space. Let us denote as $|\Omega_b\rangle$ and $|\Omega_d\rangle$ the Fock vacua for all Fock
algebras in the $b$- and in the $d$-sectors, respectively, with a nonstandard normalisation chosen to be identical in both sectors and each equal to
$\sqrt{\langle\Omega_\theta|\Omega_\theta\rangle}$, so that,
\begin{equation}
\hat{b}_\alpha|\Omega_b\rangle=0,\qquad
\hat{d}_\alpha|\Omega_d\rangle=0,\qquad
\langle\Omega_b|\Omega_b\rangle = \prod_{\alpha=1}^n\frac{1}{\sqrt{2\pi\theta_\alpha}} = \langle\Omega_d|\Omega_d\rangle.
\end{equation}
In terms of these notations, one thus has for these abstract quantum states,
\begin{equation}
|\Omega_\theta\rangle = |\Omega_b\rangle \otimes |\Omega_d\rangle.
\end{equation}
The normalised Fock states for each of the $b$- or $d$-sectors are therefore to be denoted as, with $k_\alpha,\ell_\alpha\in\mathbb{N}$,
\begin{equation}
|k_\alpha;\Omega_b\rangle=\frac{1}{\sqrt{\langle\Omega_b|\Omega_b\rangle}}\prod_{\alpha=1}^n\frac{1}{\sqrt{(2\theta)^{k_\alpha}\,k_\alpha !}}
\left(\hat{b}^\dagger_\alpha\right)^{k_\alpha}|\Omega_b\rangle,\qquad
\langle k_\alpha;\Omega_b|k'_\alpha;\Omega_b\rangle=\prod_{\alpha=1}^n\delta_{k_\alpha k'_\alpha},
\end{equation}
\begin{equation}
|\ell_\alpha;\Omega_d\rangle=\frac{1}{\sqrt{\langle\Omega_d|\Omega_d\rangle}}\prod_{\alpha=1}^n\frac{1}{\sqrt{(2\theta)^{\ell_\alpha}\,\ell_\alpha !}}
\left(\hat{d}^\dagger_\alpha\right)^{\ell_\alpha}|\Omega_d\rangle,\qquad
\langle \ell_\alpha;\Omega_d|\ell'_\alpha;\Omega_d\rangle=\prod_{\alpha=1}^n\delta_{\ell_\alpha \ell'_\alpha},
\end{equation}
so that their tensor product provides an orthonormalised Fock basis for the entire quantum Hilbert space with the abstract quantum states,
\begin{equation}
|k_\alpha,\ell_\alpha;\Omega_\theta\rangle \equiv |k_\alpha;\Omega_b\rangle \otimes |\ell_\alpha;\Omega_d\rangle.
\end{equation}

Correspondingly one also has canonical quantum coherent states in each of the $b$- and $d$-sectors. Given complex variables $w_{b,\alpha}$ and $w_{d,\alpha}$,
these states are defined according to, respectively,
\begin{equation}
|w_{b,\alpha};\Omega_b\rangle
= \prod_{\alpha=1}^n{\rm exp}\left(\frac{1}{2\theta_\alpha}\left(w_{b,\alpha}\hat{b}^\dagger_\alpha - w^*_{b,\alpha} \hat{b}_\alpha\right)\right)|\Omega_b\rangle,
\end{equation}
\begin{equation}
|w_{d,\alpha};\Omega_d\rangle
= \prod_{\alpha=1}^n{\rm exp}\left(\frac{1}{2\theta_\alpha}\left(w_{d,\alpha}\hat{d}^\dagger_\alpha - w^*_{d,\alpha} \hat{d}_\alpha\right)\right)|\Omega_d\rangle,
\end{equation}
whose tensor product defines in the $(\alpha,a)$ parametrisation the canonical quantum coherent states that generate as an overcomplete ensemble
the entire quantum Hilbert space, to be denoted as,
\begin{equation}
|w_{b,\alpha},w_{d,\alpha};\Omega_\theta\rangle \equiv |w_{b,\alpha};\Omega_b\rangle \otimes |w_{d,\alpha};\Omega_d\rangle.
\end{equation}

Finally, in view of the fact that $|\vec{x};\theta_\alpha\rangle={\rm exp}\,(-i\vec{x}\cdot\hat{\vec{p}}/\hbar)|\Omega_\theta\rangle$,
and in terms of the different quantities introduced above, one works out that,
\begin{equation}
-\frac{i}{\hbar}\vec{x}\cdot\hat{\vec{p}}=-\frac{i}{\hbar}x_{\alpha,a}\hat{p}_{\alpha,a}
=\sum_{\alpha=1}^n\frac{1}{2\theta_\alpha}\left(\left(z_\alpha\hat{b}^\dagger_\alpha - z^*_\alpha\hat{b}_\alpha\right)
+\left(z^*_\alpha\hat{d}^\dagger_\alpha - z_\alpha \hat{d}_\alpha\right)\right),
\end{equation}
so that,
\begin{equation}
|\vec{x};\theta_\alpha\rangle = {\rm exp}\left\{\sum_{\alpha=1}^n\frac{1}{2\theta_\alpha}\left(\left(z_\alpha\hat{b}^\dagger_\alpha - z^*_\alpha\hat{b}_\alpha\right)
+\left(z^*_\alpha\hat{d}^\dagger_\alpha - z_\alpha \hat{d}_\alpha\right)\right)\right\}\,|\Omega_\theta\rangle.
\end{equation}
Hence the localised and normalisable states $|\vec{x};\theta_\alpha\rangle$ provide a specific subset of all canonical coherent states
$|w_{b,\alpha},w_{d,\alpha};\Omega_\theta\rangle$ restricted such that $w_{d,\alpha}=w^*_{b,\alpha}$ with
$w_{b,\alpha}=z_\alpha=x_{\alpha,1}+ix_{\alpha,2}$, namely,
\begin{equation}
|\vec{x};\theta_\alpha\rangle = |z_\alpha,z^*_\alpha;\Omega_\theta\rangle=|z_\alpha;\Omega_b\rangle \otimes |z^*_\alpha;\Omega_d\rangle,\qquad
{\rm with}\ z_\alpha = x_{\alpha,1} + i x_{\alpha,2}.
\end{equation}
This final conclusion thus establishes that indeed as expected, the states $|\vec{x};\theta_\alpha\rangle$ labelled by all points in (the classical and
commutative) Euclidean configuration space are a specific class of canonical quantum coherent states which, through the $*_\theta$-product,
generate the entire quantum Hilbert space.

Having identified these specific structures and characteristics, and by borrowing and extending from the work by F. G. Scholtz {\it et al.} a particular vocabulary
introduced first in the context of quantum mechanics on the noncommutative two dimensional Euclidean plane, let us denote first the entire quantum
Hilbert space with its two commuting Fock $b$- and $d$-sectors as ${\cal H}_q$. The Hilbert space associated to the $b$-sector Fock algebra and its
irreducible representation spanned by the Fock states $|k_\alpha;\Omega_b\rangle$, its Fock vacuum $|\Omega_b\rangle$ and its coherent states
$|w_{b,\alpha};\Omega_b\rangle$ is to be called ``the classical Hilbert space'' and denoted as ${\cal H}_c$. While the Hilbert space associate
to the $d$-sector Fock algebra and its irreducible representation spanned by the Fock states $|\ell_\alpha;\Omega_d\rangle$, its Fock vacuum $|\Omega_d\rangle$
and its coherent states $|w_{d,\alpha};\Omega_d\rangle$ is isomorphic to the dual (or adjoint) of the classical Hilbert space ${\cal H}_c$, and thus to be called
``the dual classical Hilbert space'' and denoted as ${\cal H}^\dagger_c$. The above analysis has thus established that one has the correspondence,
\begin{equation}
{\cal H}_q={\cal H}_c\otimes{\cal H}^\dagger_c.
\end{equation}
Hence any given value for $\vec{x}$ parametrising the Euclidean configuration space determines uniquely the specific quantum coherent state
$|\vec{x};\theta_\alpha\rangle$ in the quantum Hilbert space ${\cal H}_q$, and thereby uniquely as well the specific coherent state $|z_\alpha;\Omega_b\rangle$
in the classical Hilbert space ${\cal H}_c$, and the specific coherent state $|z^*_\alpha;\Omega_d\rangle$ in the dual classical Hilbert space ${\cal H}^\dagger_c$.
Conversely any coherent state $|w_{b,\alpha};\Omega_b\rangle$ in the classical Hilbert space determines uniquely the state $|w_{d,\alpha};\Omega_d\rangle$
with $w_{d,\alpha}=w^*_{b,\alpha}$ in the dual classical Hilbert space, and therefore uniquely as well the quantum coherent state
$|\vec{x};\theta_\alpha\rangle=|w_{b,\alpha};\Omega_b\rangle\otimes|w^*_{b,\alpha};\Omega_d\rangle$ in the quantum Hilbert ${\cal H}_q$
which itself is uniquely associated to the coordinate values $\vec{x}$ such that,
\begin{equation}
x_i=x_{\alpha,a}\,R_{(\alpha,a),i},\qquad
x_{\alpha,1}=\frac{1}{2}(w_{b,\alpha}+w^*_{b,\alpha}),\qquad
x_{\alpha,2}=-\frac{1}{2}i(w_{b,\alpha}-w^*_{b,\alpha}),
\end{equation}
since,
\begin{equation}
\hat{a}_{\alpha,1}=\frac{1}{2}(\hat{b}_\alpha+\hat{d}_\alpha),\qquad
\hat{a}_{\alpha,2}=-\frac{1}{2}i(\hat{b}_\alpha-\hat{d}_\alpha),
\end{equation}
while,
\begin{eqnarray}
\hat{b}_\alpha|w_{b,\alpha};\Omega_b\rangle\otimes|w^*_{b,\alpha};\Omega_d\rangle
& = & w_{b,\alpha}|w_{b,\alpha};\Omega_b\rangle\otimes|w^*_{b,\alpha};\Omega_d\rangle, \nonumber \\
\hat{d}_\alpha|w_{b,\alpha};\Omega_b\rangle\otimes|w^*_{b,\alpha};\Omega_d\rangle
& = & w^*_{b,\alpha}|w_{b,\alpha};\Omega_b\rangle\otimes|w^*_{b,\alpha};\Omega_d\rangle .
\end{eqnarray}
These identifications thus define a one-to-one map between the coherent states $|z_\alpha;\Omega_b\rangle$ in the classical Hilbert space ${\cal H}_c$
and the coherent states $|\vec{x};\theta_\alpha\rangle$ in the quantum Hilbert space ${\cal H}_q$, with $z_\alpha=x_{\alpha,1}+i x_{\alpha,2}$
and $x_{\alpha,a}=R_{(\alpha,a),i}\,x_i$. While each of these ensembles of quantum states generates the entire Hilbert space to which they belong,
either through the projection operator density $|z_\alpha;\Omega_b\rangle\langle z_\alpha;\Omega_b|$ integrated over configuration space
in the first case, or through the $*_\theta$-products $|\vec{x};\theta_\alpha\rangle *_\theta \langle \vec{x};\theta_\alpha|$ likewise integrated
over configuration space in the second case.

As a final consideration, let us now turn to the enveloping algebra representation of the abstract quantum states $|\varphi\rangle$ in the quantum Hilbert
space ${\cal H}_q$. As discussed earlier any such state is represented by an operator in that algebra through the correspondence,
\begin{equation}
\hat{\varphi}(\hat{\vec{x}}\,)=\int_{(\infty)}\frac{d^N\vec{p}}{(2\pi\hbar)^{N/2}}\,e^{\frac{i}{\hbar}\vec{p}\cdot\hat{\vec{x}}}\,\langle\vec{p}\,|\varphi\rangle.
\end{equation}
Through the operator correspondences $\hat{x}_i \leftrightarrow \hat{x}_{\alpha,a} \leftrightarrow (\hat{b}^\dagger_\alpha,\hat{b}_\alpha)$,
any abstract quantum state $|\varphi\rangle$ is thus represented, in the enveloping algebra of the coordinate operators $\hat{x}_i$ or equivalently
eventually of the Fock algebra operators $(\hat{b}_\alpha,\hat{b}^\dagger_\alpha)$, in terms of an operator $\hat{\varphi}(\hat{b}^\dagger_\alpha,\hat{b}_\alpha)$,
possibly written in normal ordered form relative to the Fock vacuum $|\Omega_b\rangle$ in the $b$-sector hence the choice of notation,
\begin{equation}
|\varphi\rangle \longleftrightarrow \hat{\varphi}(\hat{b}^\dagger_\alpha,\hat{b}_\alpha).
\end{equation}

As established earlier, in the enveloping algebra the inner product of abstract quantum states $|\varphi_1\rangle$ and $|\varphi_2\rangle$ is represented as,
\begin{equation}
\langle\varphi_1|\varphi_2\rangle = \int_{(\infty)}d^N\vec{x}_1\,d^N\vec{x}_2\,
\langle\vec{x}_1;\theta_\alpha|\,\hat{\varphi}^\dagger_1(\hat{b}^\dagger_\alpha,\hat{b}_\alpha)
\,\hat{\varphi}_2(\hat{b}^\dagger_\alpha,\hat{b}_\alpha)\,|\vec{x}_2;\theta_\alpha\rangle.
\end{equation}
By implementing the following changes of variables to lead to a polar parametrisation $(\rho_\alpha,\phi_\alpha)$
of configuration space in each of its $\alpha$-sectors,
\begin{equation}
d^N\vec{x}=\prod_{i=1}^{2n}dx_i=\prod_{\alpha=1}^n\prod_{a=1,2}dx_{\alpha,a}=\prod_{\alpha=1}^n d\rho_\alpha\,\rho_\alpha\,d\phi_\alpha,\quad
x_{\alpha,1}=\rho_\alpha\cos\phi_\alpha,\quad
x_{\alpha,2}=\rho_\alpha\sin\phi_\alpha,
\end{equation}
one may establish that,
\begin{equation}
\int_{(\infty)}d^N\vec{x}\,|\vec{x};\theta_\alpha\rangle=
\sum_{\{k_\alpha\}=0}^\infty\left(\prod_{\alpha=1}^n\sqrt{2\pi\theta_\alpha}\right)\,|k_\alpha;\Omega_b\rangle\otimes|k_\alpha;\Omega_d\rangle,
\end{equation}
where,
\begin{equation}
\sum_{\{k_\alpha\}=0}^\infty \equiv \sum_{k_1,k_2,\cdots,k_n=0}^\infty.
\end{equation}
Consequently the above representation of the inner product in the enveloping algebra also acquires the following expression as a trace only over the Fock $b$-sector,
namely within the classical Hilbert space ${\cal H}_c$, through,
\begin{eqnarray}
\langle\varphi_1|\varphi_2\rangle &=& \left(\prod_{\alpha=1}^n\,2\pi\theta_\alpha\right)\sum_{\{k_\alpha,\ell_\alpha\}=0}^\infty
\,\langle k_\alpha;\Omega_b|\otimes\langle k_\alpha;\Omega_d|
\left(\hat{\varphi}^\dagger_1(\hat{b}^\dagger_\alpha,\hat{b}_\alpha)\,\hat{\varphi}_2(\hat{b}^\dagger_\alpha,\hat{b}_\alpha)\right)\,
|\ell_\alpha;\Omega_b\rangle\otimes|\ell_\alpha;\Omega_d\rangle \nonumber \\
& = & \left(\prod_{\alpha=1}^n 2\pi\theta_\alpha\right)\sum_{\{k_\alpha\}=0}^\infty
\langle k_\alpha;\Omega_b|\,\hat{\varphi}^\dagger_1(\hat{b}^\dagger_\alpha,\hat{b}_\alpha)\,\hat{\varphi}_2(\hat{b}^\dagger_\alpha,\hat{b}_\alpha)
\,|k_\alpha;\Omega_b\rangle \nonumber \\
& = & \left(\prod_{\alpha=1}^n 2\pi\theta_\alpha\right)
\,{\rm Tr}_c\left(\hat{\varphi}^\dagger_1(\hat{b}^\dagger_\alpha,\hat{b}_\alpha)\,\hat{\varphi}_2(\hat{b}^\dagger_\alpha,\hat{b}_\alpha)\right),
\end{eqnarray}
where the trace operation ${\rm Tr}_c$ in the classical Hilbert space ${\cal H}_c$ of any operator ${\cal O}(\hat{b}^\dagger_\alpha,\hat{b}_\alpha)$
acting on that Fock space is defined as, since the Fock states $|k_\alpha;\Omega_b\rangle$ are orthonormalised,
\begin{equation}
{\rm Tr}_c\,\hat{\cal O}(\hat{b}^\dagger_\alpha,\hat{b}_\alpha)=
\sum_{\{k_\alpha\}=0}^\infty \langle k_\alpha;\Omega_b|\hat{\cal O}(\hat{b}^\dagger_\alpha,\hat{b}_\alpha)|k_\alpha;\Omega_b\rangle.
\end{equation}

As already established earlier, the abstract operators $(\hat{x}_i,\hat{p}_i)$ acting on the abstract quantum Hilbert space ${\cal H}_q$ are represented
as follows in the enveloping algebra representation, for any state $\hat{\varphi}(\hat{\vec{x}}\,)$,
\begin{equation}
\hat{X}_i(\hat{\varphi})=\hat{x}_i\,\hat{\varphi},\qquad
\hat{P}_i(\hat{\varphi})=-\hbar\left(A^{-1}\right)_{ij}\,\left[\hat{x}_j,\hat{\varphi}\right].
\end{equation}
Correspondingly in the $(\alpha,a)$ parametrisation, one has for the same set of operators,
\begin{equation}
\hat{X}_{\alpha,a}(\hat{\varphi})=\hat{x}_{\alpha,a}\,\hat{\varphi},\qquad
\hat{P}_{\alpha,a}(\hat{\varphi})=\frac{\hbar}{\theta_\alpha}\,\epsilon_{ab}\,\left[\hat{x}_{\alpha,b},\hat{\varphi}\right].
\end{equation}
Furthermore, given the relations $\hat{b}_\alpha=\hat{x}_{\alpha,1}+i\hat{x}_{\alpha,2}$, $\hat{b}^\dagger_\alpha=\hat{x}_{\alpha,1}-i\hat{x}_{\alpha,2}$
and $\hat{p}_{\alpha,\pm}=\hat{p}_{\alpha,1}\pm i \hat{p}_{\alpha,,2}$, and thus the definitions,
\begin{equation}
\hat{B}_\alpha=\hat{X}_{\alpha,1}+i\hat{X}_{\alpha,2},\qquad
\hat{B}^\ddagger_\alpha=\hat{X}_{\alpha,1}- i \hat{X}_{\alpha,2},\qquad
\hat{P}_{\alpha,\pm}=\hat{P}_{\alpha,1}\pm i \hat{P}_{\alpha,2},
\end{equation}
where $\hat{F}^\ddagger$ denotes the adjoint of an operator $\hat{F}(\hat{b}^\dagger_\alpha,\hat{b}_\alpha)$ in the enveloping algebra
relative to the above inner product defined for the operators $\hat{\varphi}(\hat{b}^\dagger_\alpha,\hat{b}_\alpha)$, one finds,
\begin{equation}
\hat{B}_\alpha(\hat{\varphi})=\hat{b}_\alpha\,\hat{\varphi},\qquad
\hat{B}^\ddagger_\alpha(\hat{\varphi})=\hat{b}^\dagger_\alpha\,\hat{\varphi},\quad
\hat{P}_{\alpha,+}(\hat{\varphi})=-\frac{i\hbar}{\theta_\alpha}\,\left[\hat{b}_\alpha,\hat{\varphi}\right],\quad
\hat{P}_{\alpha,-}(\hat{\varphi})=\frac{i\hbar}{\theta_\alpha}\,\left[\hat{b}^\dagger_\alpha,\hat{\varphi}\right],
\end{equation}
and thus as well,
\begin{equation}
\hat{P}_{\alpha,1}\left(\hat{\varphi}\right)=\frac{\hbar}{\theta_\alpha}\left[\hat{x}_{\alpha,2},\hat{\varphi}\right],\qquad
\hat{P}_{\alpha,2}\left(\hat{\varphi}\right)=-\frac{\hbar}{\theta_\alpha}\left[\hat{x}_{\alpha,1},\hat{\varphi}\right],\qquad
\hat{P}_{\alpha,a}\left(\hat{\varphi}\right)=\frac{\hbar}{\theta_\alpha}\,\epsilon_{ab}\,\left[\hat{x}_{\alpha,b},\hat{\varphi}\right].
\end{equation}
Finally, in view of the relations $\hat{d}_\alpha=\hat{b}^\dagger_\alpha+i\theta_\alpha\hat{p}_{\alpha,-}/\hbar$ and
$\hat{d}^\dagger_\alpha=\hat{b}_\alpha-i\theta_\alpha\hat{p}_{\alpha,+}/\hbar$, and thus the following definitions for their representatives
in the enveloping algebra representation,
\begin{equation}
\hat{D}_\alpha=\hat{B}^\ddagger_\alpha+i\frac{\theta_\alpha}{\hbar}\hat{P}_{\alpha,-},\qquad
\hat{D}^\ddagger_\alpha=\hat{B}_\alpha-i\frac{\theta_\alpha}{\hbar}\hat{P}_{\alpha,+},
\end{equation}
one finds,
\begin{equation}
\hat{D}_\alpha(\hat{\varphi})=\hat{\varphi}\,\hat{b}^\dagger_\alpha,\qquad
\hat{D}^\ddagger_\alpha(\hat{\varphi})=\hat{\varphi}\,\hat{b}_\alpha.
\end{equation}
Note how for these operators $(\hat{D}_\alpha,\hat{D}^\ddagger_\alpha)$ the multiplications of $\hat{\varphi}$ by $\hat{b}^\dagger_\alpha$ and $\hat{b}_\alpha$
are from the right, to be contrasted with these multiplications from the left in the case of the operators $(\hat{B}_\alpha,\hat{B}^\ddagger_\alpha)$,
thereby manifesting again the adjoint and dual character of the $d$-Fock sector or dual classical Hilbert space ${\cal H}^\dagger_c$.

Most obviously, any given $\alpha$ sector of the above structure for a fixed value of $\alpha=1,2\cdots,n$ and thus fixed value for the noncommutativity
parameter $\theta_\alpha>0$ corresponds to a single noncommutative Euclidean plane and a conjugate magnetic field whose value
is set by that parameter $\theta_\alpha$. In this case the analysis of this work based on the ordinary principles of canonical quantisation
and having led to conclusions that coincide precisely and exactly with it in the $\hat{x}_i$-enveloping algebra representation, justify {\sl a posteriori}
the heuristic quantum mechanics of the noncommutative Euclidean plane as constructed by
Scholtz {\sl et al.} based solely on the defining configuration space commutation relation $[\hat{x}_1,\hat{x}_2]=i\theta\,\mathbb{I}_c$ with $\theta>0$
(and $\mathbb{I}_c$ being the unit operator on the associated classical Hilbert space).

That construction \cite{Scholtz1,Scholtz2,Scholtz3} relies directly on the fact that this commutation relation defines a Fock algebra
with $\hat{b}=\hat{x}_1+i\hat{x}_2$ and $[\hat{b},\hat{b}^\dagger]=2\theta\,\mathbb{I}_c$, of which the canonical coherent states
correspond to the states $|z_\alpha;\Omega_b\rangle$ of the present
discussion, with $z_\alpha=x_1+ix_2$ (and due account of different choices of normalisation conventions).
These localised and normalisable states play the role of the sharp and non-normalisable
position eigenstates in the commutative case,
replace the classical coordinates of the classical commutative configuration space, and generate the entire so-called classical Hilbert space
through the configuration space
integration of the projection operator density $|z_\alpha;\Omega_b\rangle\langle z_\alpha;\Omega_b|$.

The quantum Hilbert space of quantum states
is then identified with the enveloping algebra of the coordinate operators $(\hat{x}_1,\hat{x}_2)$, namely $\hat{\varphi}(\hat{x}_1,\hat{x}_2)$
or $\hat{\varphi}(\hat{b}^\dagger_\alpha,\hat{b}_\alpha)$ in our notations, which is generated by the configuration space integration of the
operators $|z_\alpha;\Omega_b\rangle*_\theta\langle z_\alpha;\Omega_b|$ that act on the classical Hilbert space spanned by
the states $|z_\alpha;\Omega_b\rangle$, with $*_\theta$ being the star product associated to the parameter $\theta$. Furthermore on the quantum Hilbert
space realised as the enveloping algebra of the coordinate operators $(\hat{x}_1,\hat{x}_2)$, the position operators are realised as the left multiplication
of $\hat{\varphi}$ by the operators $(\hat{x}_1,\hat{x}_2)$, while canonically conjugate momentum operators may be identified to arise and to be realised
in terms of the commutators
of $\hat{\varphi}$ with the coordinate operators $(\hat{x}_1,\hat{x}_2)$ in precisely and exactly the same way as established above. Finally the inner
product of abstract quantum states represented as operators $\hat{\varphi}_1$ and $\hat{\varphi}_2$
within the $(\hat{x}_1,\hat{x}_2)$-enveloping algebra is chosen to be
defined through the trace over the classical Hilbert space---the $b_\alpha$-sector in our notations---of the product $\hat{\varphi}^\dagger_1\hat{\varphi}_2$
of the two operators representing the two quantum states, which is indeed once again in complete agreement with the above conclusion based on standard canonical
quantisation (and due account of different choices of normalisation conventions).

This heuristic construction of a quantum mechanics on a noncommutative Euclidean plane coincides thus
exactly and precisely with the above observations and conclusions that directly derive from a standard canonical quantisation in the presence of a magnetic field
in the phase space sector of the canonical conjugate momenta of a classical dynamics, specifically however, in the configuration space enveloping algebra
representation of the quantum system.

As the analysis of this work has explicitly demonstrated based on a standard canonical quantisation approach, the same structure extends to any
even dimensional Euclidian configuration space in the presence of a nondegenerate conjugate magnetic field in the canonical conjugate momentum
sector of phase space, provided the quantised system and its states is represented through the enveloping algebra of the noncommuting configuration space
position operators that act on a classical Hilbert space spanned by states that play a role akin to position eigenstates without being in fact sharp states,
but are specific finite norm canonical coherent Fock states whose definition is aligned with the values of the conjugate magnetic field.

When represented in the $(\alpha,a)$ parametrisation of the configuration and phase spaces which block diagonalises
the conjugate magnetic field, and factorises the classical and quantum Hilbert spaces into commuting and decoupled sectors for each value of $\alpha$,
each $\alpha$ sector then coincides with the quantum mechanics of the two-dimensional noncommutative Euclidean plane as constructed by
Scholtz {\sl et al.} Yet, based on the canonical quantisation approach, other realisations and representations of the same quantum mechanics on
such a noncommutative Euclidean configuration space of even dimension are also available, some familiar, other less familiar,
thereby offering the potential of complementary methods for explicit calculations relevant to the dynamics of quantum systems
over such noncommutative configuration spaces.

\section{Conclusions}
\label{Sect5}

As would deserve to be more widely known, a conjugate or dual magnetic field in the canonically conjugate momentum sector of a classical phase space
induces noncommutativity in configuration space, whether in terms of nonvanishing classical Poisson brackets, or nonvanishing quantum commutation relations
for the position coordinates or operators of the canonically quantised system. This observation is the basis for the canonical quantisation approach to a formulation
of a quantum mechanics over a noncommutative configuration space that is pursued in the present work. More specifically, this is done herein for an Euclidean
configuration  space of even dimension, and a constant and nondegenerate conjugate magnetic field.

As a preliminary, the quantum mechanics of a commutative configuration space has been revisited first, in order to set up considerations that are directly
relevant to the noncommutative case, but not readily available from standard treatises of quantum mechanics. In particular through the choice of some
real, symmetric and positive definite matrix of constant coefficients of length-squared dimension, a specific class of localised and normalisable
quantum states parametrised by all points in configuration space is introduced in terms of which it is possible to define a specific and related $*$-product
even in the commutative case. These distinguished states turn out to correspond to a specific subclass of canonical quantum coherent states but
restricted to configuration space. Yet through the associated $*$-product and by being overcomplete
these states generate the entire abstract quantum Hilbert space of the
system. The consideration of these states allows for the construction of a specific wave function representation in configuration space
of quantum states different from the ordinary position or momentum wave function ones. Finally, a representation of abstract quantum states
and the abstract Heisenberg algebra in terms of the enveloping algebra of the commuting position operators is thereby identified, with the specific class
of distinguished quantum states playing the role of ordinary configuration space coordinates in the commutative case.

When a constant conjugate magnetic field is introduced, a parallel analysis is developed, to lead to new insight into the understanding of the
quantum mechanics on a noncommutative configuration space. By adjusting the ensemble of distinguished quantum states introduced in the
commutative case to the choice of conjugate magnetic field, canonical quantisation produces in a nontrivial manner an appealing organisation
of the quantum Hilbert space of quantum states as a tensor product of two ensembles of Fock algebras and Fock Hilbert spaces,
one associated to the initially distinguished quantum states that then span a so-called classical Hilbert space, and the other being isomorphic
to the dual of that classical Hilbert space. While in a most natural way, a representation of the quantum states is achieved in terms of
operators in the enveloping algebra of the noncommuting configuration space coordinate operators acting on the states of the classical Hilbert space,
the latter thus playing the role of what are in the commutative case position eigenstates but now being normalisable. However besides
this representation, other representations also arise, of possible practical relevance as well.

By having followed this path towards a quantum mechanics of noncommutative configuration spaces, has also produced and thus justified
{\sl a posteriori} precisely the very heuristic construction of a quantum mechanics on the noncommutative Euclidean plane that F. G. Scholtz and his
collaborators have developed over the years. Combining Dirac's standard principles of canonical quantisation with a conjugate magnetic field
in phase space thus totally justifies that work and its results from this complementary point of view, and provides further understanding of its
particular features, as for instance the apparent emergence of canonically conjugate momenta out of an algebraic structure only based
on the noncommuting quantum position operators in configuration space of which the abstract representation is that of the classical Hilbert space.

The specific structures that organise the quantum mechanics of such noncommutative Euclidean configuration spaces, as they were first identified
by Scholtz {\sl et al.} by pursuing analogies with quantum mechanics on commutative configuration spaces, may suggest pathways towards extensions to other
topologies and geometries than those of Euclidean configuration spaces. Indeed, the doubled tensor structure in terms of a classical Hilbert space
and of its dual is a tantalizing possible avenue, whose exploration is also part of F. G. Scholtz's on-going programme. One may also wonder how other choices
for a conjugate magnetic field, now dependent whether on  configuration or momentum space, or both, would produce specific classes
of quantum mechanics on other types of noncommutative structures inclusive of noncommutative configuration spaces. Or conversely,
for instance, which classes of well known quantum mechanics on other types of noncommutative configuration spaces, say for example 
noncommutative spheres (see for instance \cite{spheres}), could also be understood from a similar approach combining Dirac's ordinary canonical quantisation
based on the correspondence principle and a conjugate magnetic field in phase space.

\section*{Acknowledgements}

The author is most grateful to Professor Frederik Scholtz (University of Stellenbosch, South Africa) for the many, plenty and inspiring discussions
that we have had over the years on the topics of the present paper and on many other subjects. And even more so,
for his always truly most constant welcoming and warm hospitality at the Department of Physics of the University of Stellenbosch
and at the then National Institute for Theoretical Physics (NITheP), now the National Institute for Theoretical \& Computational Sciences (NITheCS) of South Africa.
This work is supported in part by the {\sl Institut Interuniversitaire des Sciences Nucl\'eaires} (IISN, Belgium).

\end{document}